\documentclass[%
 reprint,
superscriptaddress,
nofootinbib,
 amsmath,amssymb,
 aps,
pra,
floatfix, % trying this 20230404
table
]{revtex4-2}

\pdfoutput=1

\usepackage{graphicx}% Include figure files
\usepackage{dcolumn}% Align table columns on decimal point
\usepackage{bm}% bold math
\usepackage[hidelinks]{hyperref}% add hypertext capabilities
\usepackage[size=footnotesize,justification=justified]{subcaption}
\usepackage{tikz}
\usetikzlibrary{quantikz}

\usepackage{makecell}  % allows more complex table formatting

\usepackage{multirow}
\usepackage{ragged2e}
\usepackage{enumitem}

\usepackage{adjustbox}

\newcommand{\ProveIt}{\textbf{Prove-It}}
\newcommand{\ProveItWebsite}{\textbf{Prove-It} website \cite{website:Prove-It}}
\newcommand{\ignore}[1]{}
\begin{document}

\preprint{APS/123-QED}

\title{Verifying Quantum Phase Estimation (QPE) using Prove-It}

\author{Wayne M. Witzel}
\affiliation{%
 Center for Computing Research, Quantum Computer Science, Sandia National Laboratories, Albuquerque, NM 87123
}%
\author{Warren D. Craft}%
\affiliation{%
 Center for Computing Research, Quantum Computer Science, Sandia National Laboratories, Albuquerque, NM 87123
}%
\affiliation{%
 Department of Computer Science, The University of New Mexico, Albuquerque, NM 87131
}%
\author{Robert Carr}%
\affiliation{%
 Department of Computer Science, The University of New Mexico, Albuquerque, NM 87131
}%
\author{Deepak Kapur}%
\affiliation{%
 Department of Computer Science, The University of New Mexico, Albuquerque, NM 87131
}%

\date{April 4, 2023}% It is always \today, today,

\begin{abstract}
The general-purpose interactive theorem-proving assistant called \ProveIt{} was used to verify the Quantum Phase Estimation (QPE) algorithm, specifically claims about its outcome probabilities. 
Prove-It is unique in its ability to express sophisticated mathematical statements, including statements about quantum circuits, integrated firmly within its formal theorem-proving framework.  We demonstrate our ability to follow a textbook proof to produce a formally certified proof, highlighting useful automation features to fill in obvious steps and make formal proving nearly as straightforward as informal theorem proving. Finally, we make comparisons with formal theorem-proving in other systems where similar claims about QPE have been proven.
\end{abstract}

\maketitle

%\tableofcontents

\section{Introduction and Motivation}
\label{sec:introduction}

Quantum devices in a variety of technologies have been developing rapidly.  
Several small quantum processors and test-beds are now cloud-accessible using quantum bits encoded in superconductors~\cite{website:IBM_Quantum, website:Rigetti, website:OxfordQuantumCircuits, website:QuantumInspire_Starmon5}, ions~\cite{website:IonQ, website:Qscout}, neutral atoms~\cite{website:QuEra}, photons~\cite{website:Xanadu,website:Quandela}, and electron spins~\cite{website:QuantumInspire_Spin2}.
Although current devices do not yet have the size and fidelity for practical applications to definitively compete against classical computers, quantum computers are now a reality and their eventual utility seems inevitable.
As a result, there has been widespread and accelerating interest in quantum computers and quantum computing not only in academia and research laboratories, but also in government and industry (as evidenced by the efforts of leaders in the computing industry and computational resource use/allocation such as IBM, Amazon, Intel, Google, Facebook, \textit{etc}).
Such interest is driven largely by the fact that quantum computers are theorized to have an advantage over classical computers for many computationally difficult problems where, remarkably, the computational advantage grows exponentially with the size of the problem. 
One well-known example of such an exponential advantage over the best-known classical algorithm is Shor's factoring algorithm \cite{Shor:1994_shor_factoring_algorithm}. 
In Shor's approach, factoring is reduced to order-finding (finding the smallest, positive $r$ for which $x^r = 1~\textrm{mod}~N$ given $x$ and $N$) and order-finding is reduced to quantum phase estimation (QPE).  It is believed that no efficient classical algorithms for factoring and period-finding exist.  
The hidden Abelian subgroup~\cite{Boneh1995} problem is a generalization that may also be solved efficiently on a quantum computer using QPE.
Furthermore, QPE has important applications for solving systems of linear equations~\cite{Harrow_Hassidim_Lloyd:2009_quantum_alg_linear_eqs} and performing quantum simulations for chemistry applications~\cite{Motta_Rice:2021_quantum_algs_quantum_chem}.

Just as in the case of classical-computing algorithms and software implementing them for classical computers, the reliability and correctness of quantum algorithms and their implementations are vital.
Quantum circuits can be complex and (often) counter-intuitive.
Analogous to classical software, and especially during a time when devices may not exist to test quantum algorithms on a convincing scale, it is critically important to seek rigorous forms of confirmation of both the fundamental concepts of an algorithm and its implementation in a specific device of interest.
The eventual physical instantiation of quantum computations and quantum circuits represents substantial physical, financial, and temporal resources.
As devices become sufficiently advanced for testing purposes, it is furthermore valuable to ensure algorithms are valid and implemented correctly so there is no confusion between faulty programming versus faulty equipment.
This is especially important given the inherent challenge of debugging the implementation of a quantum algorithm running on a quantum computer which can have exponentially many states in the number of qubits, with destructive and probabilistic aspects to any related quantum measurement \cite{Chareton_et_al:2022_formal_methods_for_quantum_programs}.
It can be economically beneficial to be able to explore quantum circuits and the computational implications of quantum circuits with relative ease using tools based on formal analysis.

\subsection{Verifying Quantum Algorithms and Quantum Circuits}

Using formal methods to verify quantum algorithms is not new.
A number of languages, frameworks, and environments have been proposed to support the formal analysis and verification of quantum algorithms and quantum circuits.
Some examples include:
VOQC, a fully \underline{v}erified
\underline{o}ptimizer for \underline{q}uantum \underline{c}ircuits, written using the Coq proof assistant, and a low-level
language called SQIR (a simple quantum intermediate representation) for expressing quantum circuits as programs \cite{Hietala_et_al_verified_optimization:2019, Hietala_et_al_verified_optimizer:2021}; 
QWIRE~\cite{Paykin_Rand_Zdancewic_QWIRE:2017, Rand_Paykin_Zdancewic_Phantom_Types_for_Quantum_Programs:2018, Rand_Paykin_ZDancewic_QWIRE_Practice:2018}, a ``quantum circuit language'' embedded in the Coq proof assistant \cite{Coq:2022, website:Coq}, allowing users to write quantum circuits using high-level abstractions and to prove properties of those circuits using Coq's theorem-proving features; Qbricks~\cite{Chareton_et_al:2021_auto_deductive_verif_framework_Qbricks, website:Qbricks}, an environment for automated formal verification of quantum programs, taking advantage of the characterization of quantum circuits as parameterized path-sums; and Quantum Hoare Logic (QHL)~\cite{Liu_Theorem_Prover_Hoare_Logic:2016, Liu_Formal_Verification_Hoare_Logic:2019, Liu_et_al:2019_quantum_hoare_logic_proof_document, Ying:2011_Floyd_Hoare_logic_for_quantum_programs, Ying_Auto_Verif_Quantum_Programs:2019}, an extension of Hoare logic for reasoning about quantum programs.

\subsection{The \ProveIt{} Theorem-Proving Assistant}
\label{subsec:proveit_theorem_proving_assistant}

This paper reports on a verification of the quantum phase estimation (QPE) algorithm \cite[pp 221--226]{Nielsen_Chuang:2010} using a newly developed theorem-proving assistant called \ProveIt{} ~\cite{Preprint:ProveIt, website:Prove-It}.
The verification effort has been developed simultaneously with the evolution of \ProveIt{}, with the two intertwined activities contributing to each other.
The \ProveIt{} approach is unique in its ability to construct a formal proof from a user's proof sketch using standard mathematical language with precise semantics based on set theory, thus providing considerable flexibility to a user.
We demonstrate this ability in this paper by performing a nearly direct translation of a textbook proof \cite{Nielsen_Chuang:2010} of the QPE algorithm into \ProveIt{}.  
A main objective has been to allow verification in a reasonably accessible way: allowing work by subject matter experts (physicists, engineers, mathematicians, \textit{etc}.), without extensive background in formal methods, to explore and generate proofs. The automation capabilities of the \ProveIt{} system allow the user to skip obvious steps, with \ProveIt{} filling in the gaps.

\begin{figure*}
    \captionsetup{font=footnotesize}
    \begin{subfigure}[t]{0.48\textwidth}
    \vskip 0pt\centering
    \includegraphics[width = 0.98\textwidth]{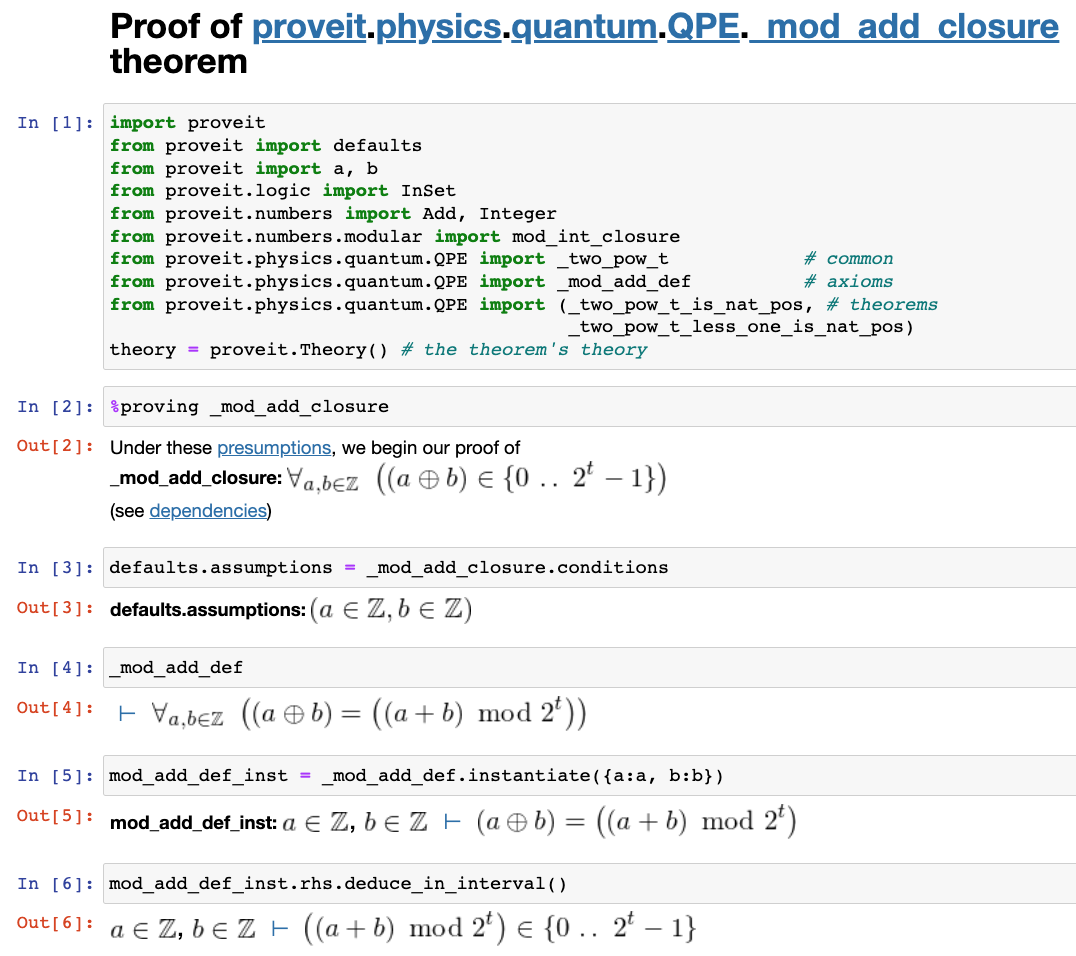}
        %\caption{}
    \end{subfigure}
        \vspace{0.1in}
    \begin{subfigure}[t]{0.48\textwidth}
    \vskip 0pt\centering
    \includegraphics[width = 0.98\textwidth]{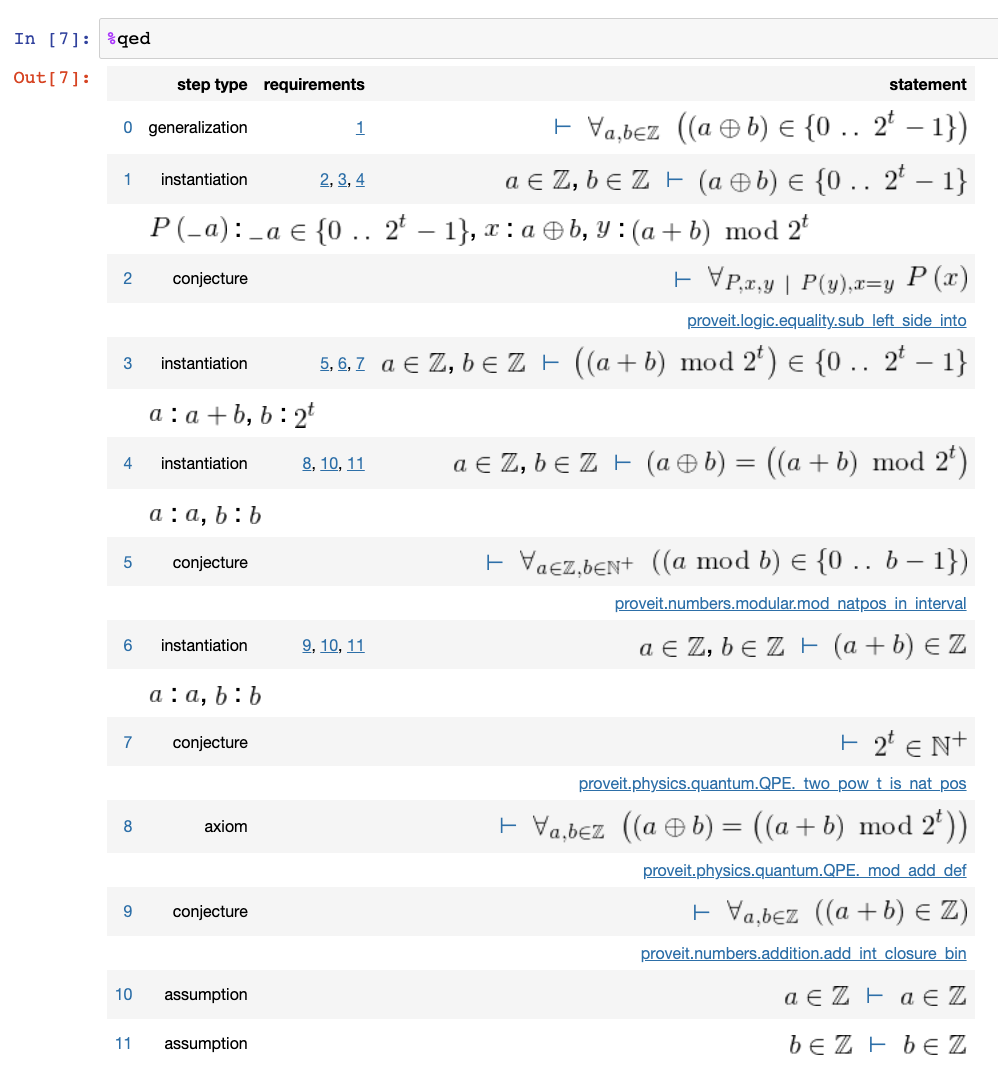}
        %\caption{}
    \end{subfigure}
    \caption{Excerpt from the proof notebook for \texttt{\_mod\_add\_closure} (local QPE) theorem, showing the interactive steps (left) and the \ProveIt{}-generated formal proof (right).
    }
    \label{fig:QPE_thm_mod_add_closure_proof_nb}
\end{figure*}

\ProveIt{} is an open-source, python-based interactive theorem-proving assistant with a Jupyter-notebook-based interface. \ProveIt{} uses \LaTeX{} formatting for output of mathematical expressions, including the mathematical expressions appearing in any eventual formal proof.
User-supplied, interactive proof steps within a Jupyter notebook, similar to steps found in a textbook proof or an informal narrative proof, guide \ProveIt{} to produce a detailed \textit{formal} (but entirely human-readable) proof. A quick example of this process appears in  Fig.~\ref{fig:QPE_thm_mod_add_closure_proof_nb}, showing the interactive steps (left) and eventual \ProveIt{}-generated formal proof (right). This process is discussed in more detail in \S\ref{Sec:interactive_proofs}.

\textsc{html} representations of \ProveIt{} system components and related documentation, including all of the code, definitions, axioms, lemmas, and theorems discussed here in relation to the QPE verification efforts, are available for browsing at \url{http://pyproveit.org}.
The system is available for download and user contributions at \url{https://github.com/PyProveIt/Prove-It}.
The system is organized hierarchically into theory packages.
Each theory package contains axioms and theorems (or lemmas which are treated the same as theorems).
If a theorem has been proven, its name will hyperlink to a ``proof notebook'' containing a sequence of commands, and a detailed formal proof after the \texttt{\%qed} command.  
This proof notebook will also have header information in the top few cells that includes a link to the theorem's ``dependencies,'' which lists all not-yet-proven theorems and axioms required by the formal proof as well as a list of the theorems that directly rely on that particular theorem in their proofs.

\subsection{Organization of the Paper}

\noindent 
The next three sections of this paper will address the following questions: What, essentially, did we prove (\S\ref{Sec:essence})?  What, precisely, did we prove (\S\ref{Sec:dependencies})? and How did we construct our proofs (\S\ref{Sec:interactive_proofs})?  These topics are brought together in a demonstration of one specific proof within the QPE package in \S\ref{sec:theorem_proof_demo}.  In \S\ref{sec:comparisons}, we make comparisons with other theorem-proving approaches that have been applied to the specific QPE problem.
Conclusions and acknowledgments are provided in \S\ref{Sec:Discussion_Conclusion} and \S\ref{Sec:Acknowledgments} respectively.

\S\ref{Sec:essence} provides an overview of \ProveIt{}'s QPE theory package, outlining the step-by-step development of the eventual proof for a set of three useful probability bounds for the output from an ideal (error-free) realization of the quantum algorithm: (1) a probability of $1$ to obtain the phase when it can represented exactly; (2) a probability greater than $4/\pi^2$ for the result to be the closest solution that can be represented; (3) the required number of qubits to obtain any desired precision and minimum probability.  We progress, in this section, from initial assumptions and axiomatic definitions to final theorems. \S\ref{Sec:essence} describes the ``essence'' of the QPE package proofs, listing the steps as named statements within a narrative that describes the reasoning to form our conclusions.

\S\ref{Sec:dependencies} drills down to discuss more precisely just what we've proven.  We explain how \ProveIt{} traces the proven lemmas/theorems back to the axioms and yet-to-be-proven theorems (``conjectures'') on which they depend, and how \ProveIt{} avoids circular dependencies. We also provide specific information about the lemmas/theorems that are proven in our QPE theory package and correspond to the named statements of \S\ref{Sec:essence}.  We list other theory packages that our proofs rely upon and the precise number of axioms and ``conjectures'' (general, fundamental facts that simply have not been proven in our system yet) that were used.  We expect that many of the statements that we call axioms at this time will be transformed into conservative definitions, with proven unique existence, to reduce the dependencies to a small number of true axioms (e.g., the Zermelo–Fraenkel set theory axioms and postulates of quantum mechanics).  At this time, however, we can only claim our proofs to be true conditional upon our current axioms and conjectures.

\S\ref{Sec:interactive_proofs} discusses in more detail just \textit{how} we proved the theorems interactively, discussing convenient methods and automation features that allow a user to skip obvious steps.
We provide a number of examples of using derivation commands to push our QPE-specific proofs along.
For each of the QPE-specific lemmas/theorems (corresponding to named statements of \S\ref{Sec:essence}) we list the number of derivation commands used in its interactive proof as well as the number of steps in its formal proof.

We focus on one specific supporting, QPE-specific proof in \S\ref{sec:theorem_proof_demo} to demonstrate \ProveIt{}'s features in one concrete example.
Then, in \S\ref{sec:comparisons}, we make specific comparisons with other theorem-proving systems that have been used to prove two out of the three high-level QPE-specific theorems that we accomplished.
We close with a more general discussion in
\S\ref{Sec:Discussion_Conclusion} and acknowledgments in \S\ref{Sec:Acknowledgments}.

\section{What, essentially, did we prove?}
\label{Sec:essence}

This section provides an overview of the QPE theory package in \ProveIt{} and helps the reader understand and navigate the flow of named statements (assumptions, definitions, lemmas, and theorems) within this theory package (as axioms and theorems) on the website at \url{pyproveit.org}.
We currently treat definitions as axioms in our system which are asserted without proof, but we eventually intend to make a separate category for conservative definitions which require a proof of unique existence; this change will reduce the number of proof dependencies.

\ProveIt{}'s QPE theory package can be seen as a formalization of Nielsen \& Chuang's \cite{Nielsen_Chuang:2010} standard textbook development and eventual proof of three main claims about the QPE outcome probabilities (also see \cite{Cleve_et_al:1998_quantum_algs_revisited}).
We therefore illustrate how \ProveIt{} provides an accessible way to construct a formal proof in much the same way one goes about constructing an informal proof.
The informal-to-formal transformation is achieved through user-interactive steps that will be discussed in \S\ref{Sec:interactive_proofs}.

In quantum phase estimation (QPE), we have a unitary matrix $U$ such that
\begin{align}
    U \ket{u} = e^{2\pi i \varphi} \ket{u}
\end{align}
\textit{i.e.}, $U$ has eigenvector $\ket{u}$ and associated eigenvalue $e^{2\pi i \varphi}$ --- and we want to estimate the value of the phase $\varphi \in [0, 1)$.
The QPE algorithm is shown in quantum circuit form in Fig.~\ref{fig:QPE_circuit}. The first register has $t$ qubits, with an initial input $\ket{0}_{t}$; the second register has as many qubits as required to hold the eigenvector $\ket{u}$.
A Hadamard is applied to each line in the first register, then the $j$th line in the first register controls the application of $U^{2^j}$ to the second register.
Then we apply an inverse quantum Fourier transform $\text{QFT}^{\dagger}$ on the first register, followed by measurement of the first register, resulting in $\ket{2^t \tilde{\varphi}}$ as the measurement outcome. 
In our diagrams, we treat measurements as projective such that their output corresponds to the collapsed state as a random variable with the appropriate probability distribution.
When the $t$ qubits of the first register are sufficient to represent the phase $\varphi$ exactly, the QPE algorithm produces $\tilde{\varphi} = \varphi$, and thus $\varphi$ can be determined from the final measurement $\ket{2^t \varphi}$ by dividing the measurement by $2^t$.
In the more general case, the estimate $\tilde{\varphi}$ of the actual phase $\varphi$ can be shown to be a good estimate with high probability if $t$ is sufficiently large.  

%%%%%%%%%%%%%%%%%%%%%%%%%%%%%%%%%
% Circuit Diagram:              %
% Full QPE Circuit              %
%%%%%%%%%%%%%%%%%%%%%%%%%%%%%%%%%

\begin{figure*}
    \captionsetup{font=footnotesize}
    \centerline{
    \begin{quantikz}
    % ROW 1
    \lstick[wires=6]{First Register:\\$t$ Qbits} &
    \lstick{$\ket{0}$} & \gate{H}\slice{1} &
        \ctrl{6} & \qw & \ \ldots\ \qw & \qw & \qw\slice{2} & \gate[wires=6, nwires={4}]{\text{QFT}^{\dagger}}\slice{3} & \meter{} & \rstick[wires=6]{$\ket{2^t \tilde{\varphi}}$}\qw\\
    % ROW 2
    &\lstick{$\ket{0}$} & \gate{H} &
        \qw & \ctrl{5} & \ \ldots\ \qw & \qw & \qw & \qw & \meter{} & \qw\\
    % ROW 3
    &\lstick{$\ket{0}$} & \gate{H} &
        \qw & \qw & \ \ldots\ \qw & \qw & \qw & \qw & \meter{} & \qw\\
    % ROW 4
    &\vdots & \vdots &
        \vdots & \vdots & \vdots & \vdots & \vdots & \vdots & \vdots\\
    % ROW 5
    &\lstick{$\ket{0}$} & \gate{H} &
        \qw & \qw & \ \ldots\ \qw & \ctrl{2} & \qw & \qw & \meter{} & \qw\\
    % ROW 6
    &\lstick{$\ket{0}$} & \gate{H} &
        \qw & \qw & \ \ldots\ \qw & \qw & \ctrl{1} & \qw & \meter{} & \qw\\
    % ROW 6
    \lstick[wires=1]{2nd Register:\\$s$ Qbits }
    &\lstick{$\ket{u}$} & \qwbundle[alternate]{} &
        \gate{U^{2^{t-1}}}\qwbundle[alternate]{} & \gate{U^{2^{t-2}}}\qwbundle[alternate]{} & \ \ldots\ \qwbundle[alternate]{} &
        \gate{U^{2^{1}}}\qwbundle[alternate]{} &
        \gate{U^{2^{0}}}\qwbundle[alternate]{} & \qwbundle[alternate]{} & \qwbundle[alternate]{} & \rstick{$\ket{u}$}\qwbundle[alternate]{}
    \end{quantikz}
    }
    \caption{Quantum circuit implementing the QPE algorithm. From initial first-register state $\ket{0}_{t}$ and second register state $\ket{u}$, we apply Hadamards to the first register lines, controlled $U^{2^j}$ to the second register, and an inverse quantum Fourier transform $\text{QFT}^{\dagger}$ to the first register, leading to first-register measurement $\ket{2^t \tilde{\varphi}}$, where $\tilde{\varphi}$ is an estimate of the phase.}
    \label{fig:QPE_circuit}
\end{figure*}

As suggested, then, by the algorithmic circuit in Fig.~\ref{fig:QPE_circuit}, we let $t \in \mathbb{N}^{+}$ and $s \in \mathbb{N}^{+}$ represent the number of qubits in the first and second registers, respectively. We let $U \in \text{U}(2^{s})$ be a unitary matrix of size $2^s \times 2^s$ and $\ket{u} \in \mathbb{C}^{2^s}$ be a unit vector such that $U \ket{u} = e^{2 \pi i \varphi} \ket{u}$. These initial assumptions are captured in the following:
\begin{align}
    &\text{QPE Assumption: }\texttt{\_t\_in\_nat\_pos}
    \quad t \in \mathbb{N}^{+}
    \label{eq:axiom_t_in_nat_pos}\\
    &\text{QPE Assumption: }\texttt{\_s\_in\_nat\_pos}
    \quad s \in \mathbb{N}^{+}
    \label{eq:axiom_s_in_nat_pos}\\
    &\text{QPE Assumption: }\texttt{\_unitary\_U}
    \quad U \in \text{U}(2^{s})
    \label{eq:axiom_unitary_U}\\
    &\text{QPE Assumption: }\texttt{\_u\_ket\_register}
    \quad\ket{u} \in \mathbb{C}^{2^s}
    \label{eq:axiom_u_ket_register}\\
    &\text{QPE Assumption: }\texttt{\_normalized\_ket\_u}
    \quad||\ket{u}|| = 1
    \label{eq:axiom_normalized_ket_u}\\
    &\text{QPE Assumption: }\texttt{\_phase\_in\_interval}
    \quad\varphi
    \in [0, 1)
    \label{eq:axiom_phase_in_interval}\\
    &\text{QPE Assumption: }\texttt{\_eigen\_uu}
    \quad U\ket{u}
    = e^{2 \pi i \varphi} \ket{u}
    \label{eq:axiom_eigen_uu}
\end{align}
\noindent Labels beginning with an underscore \texttt{\_} indicate \textit{local} axioms or theorems in \ProveIt{}, valid only inside a particular package.
We would not want, for example, the assumption $t\in\mathbb{N}^{+}$ to bleed over into other packages.
Also note that descriptors such as ``assumption'', ``definition'', or ``lemma'' are adopted here for narrative purposes but \ProveIt{} makes no such distinctions (assumptions are local axioms, definitions are axioms, and lemmas are theorems).

We also define a specific modular addition which will be useful in the statements of several lemmas:
\begin{align}
    &\text{Local QPE Definition: }\texttt{\_mod\_add\_def}\nonumber\\
    &\forall_{a,b\in\mathbb{Z}}((a \oplus b) = ((a + b) \,\text{mod}\, 2^t)),
    \label{eq:axiom_mod_add_def}
\end{align}
where $t\in\mathbb{N}^{+}$ is again the number of qubits available in the first register.
This definition is local because it is defined for a problem-specific value of $t$.

For convenience we define some simplified representations of pieces of the QPE algorithmic circuit shown in Fig.~\ref{fig:QPE_circuit}.
We represent the portion of the circuit between slices 1 and 2 as the gate $\text{QPE}_1(U, t)$, defined with:
\begin{align}
    &\text{QPE Definition: }\texttt{QPE1\_def}\nonumber\\
    &\raisebox{-0.5\height}{\includegraphics[width = 0.4\textwidth]{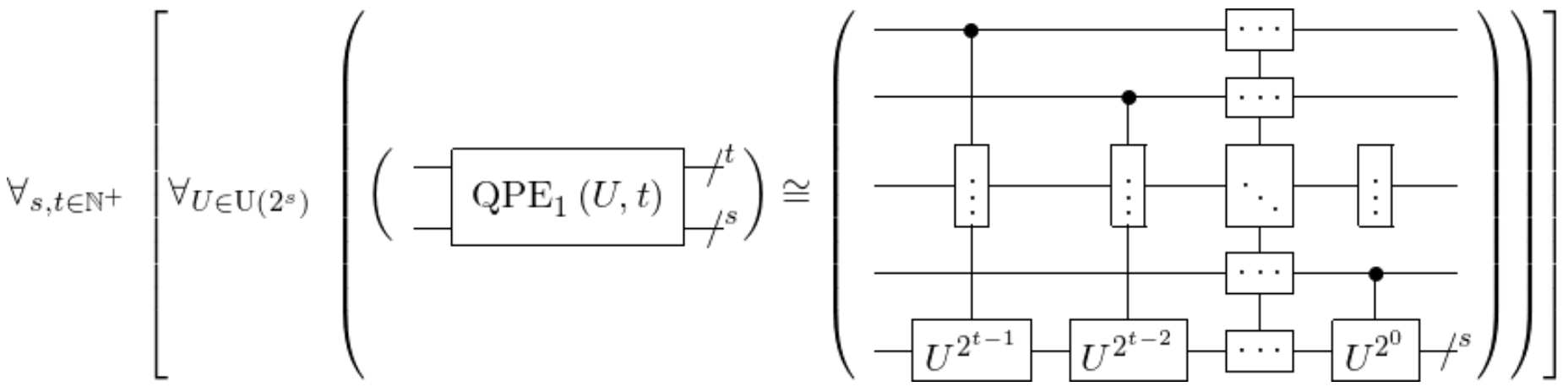}},
    \label{eq:axiom_QPE1_def}
\end{align}
\noindent and the portion of the circuit between slices 1 and 3 as the gate $\text{QPE}(U, t)$, defined with:
\begin{align}
    &\text{QPE Definition: }\texttt{QPE\_def}\nonumber\\
    &\raisebox{-0.5\height}{\includegraphics[width = 0.4\textwidth]{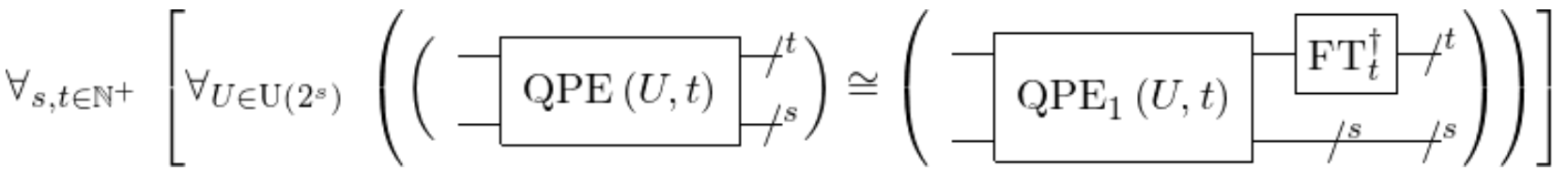}},
    \label{eq:axiom_QPE_def}
\end{align}
\noindent where the symbol $\cong$ denotes circuit equivalence.
Notice that the circuit equivalence \texttt{QPE\_def}~(\ref{eq:axiom_QPE_def}) utilizes $\text{QPE}_1(U, t)$ on its right-hand side. Notice also that \texttt{QPE1\_def}~(\ref{eq:axiom_QPE1_def}) and \texttt{QPE\_def}~(\ref{eq:axiom_QPE_def}) are definitions of the \textit{circuit} components and do not include the inputs to, or outputs of, each circuit.
Moreover, these circuit components do not include the Hadamard gates --- instead, as illustrated presently, we combine the standard initial first-register state $\ket{0}_{t}$ with the subsequent Hadamards to consider the $t$-qubit state $\ket{+}\otimes\ldots\otimes\ket{+}$ as the input to the \texttt{QPE1} component of the QPE circuit.
The \texttt{QPE1\_def} and \texttt{QPE\_def} axiomatic definitions are \textit{global} or ``universal'', valid outside of \ProveIt{}'s QPE package.
The definitions are quantified over $s \in \mathbb{N}^{+}$, $t \in \mathbb{N}^{+}$, and $U \in \text{U}(2^s)$ (as \textit{variables}) instead of holding for the problem-specific $s$, $t$, and $U$ values (as \textit{literals}).

\subsection{QPE Circuit State at Stage 2}

We now prove that the state of the first register at Stage 2 is $\ket{\psi_{t}}$:
\begin{align}
    &\text{Local QPE Theorem: }\texttt{\_psi\_t\_output}\nonumber\\[1ex]
    &\raisebox{-0.5\height}{\includegraphics[width = 0.4\textwidth]{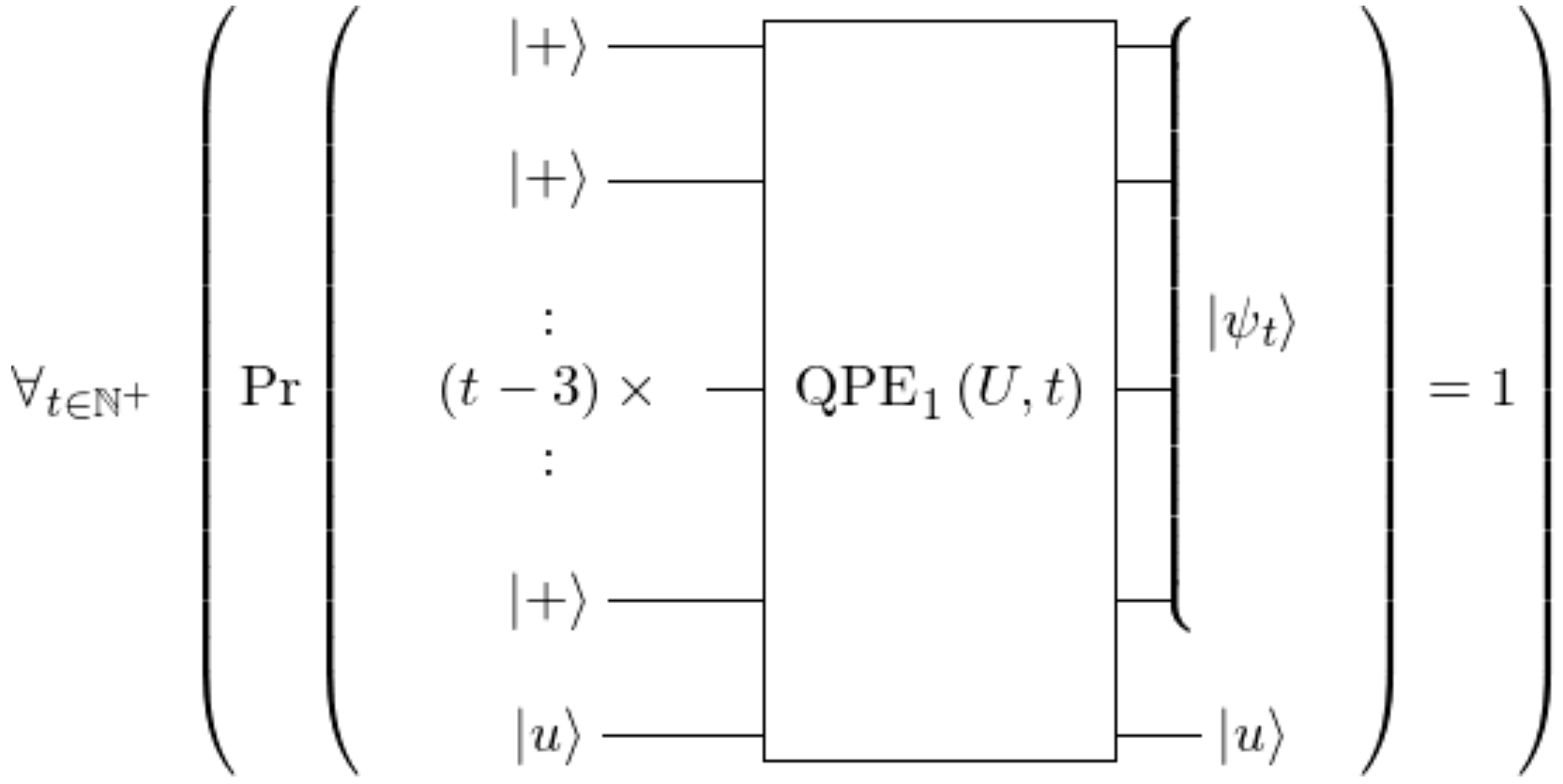}}
    \label{eq:thm_psi_t_output},
\end{align}
where $\ket{\psi_{t}}$ is defined (locally) as
\begin{equation}
\begin{split}
    &\text{Local QPE Definition: }\texttt{\_psi\_t\_def}\\
    &\forall_{t \in \mathbb{N}}~\ket{\psi_{t}}
    =
    \frac{1}{2^{t/2}}
    (\ket{0} + e^{2\pi i (2^{t-1}\varphi)}\ket{1})
    \otimes\\
    &\hspace{0.1in}(\ket{0} + e^{2\pi i (2^{t-2}\varphi)}\ket{1})
    \otimes
    \cdots
    \otimes
    (\ket{0} + e^{2\pi i (2^{0}\varphi)}\ket{1})
    \label{eq:axiom_psi_t_def}.
\end{split}
\end{equation}
Although \texttt{\_psi\_t\_def}~(\ref{eq:axiom_psi_t_def}) quantifies over $t$ as a variable, it is local because it relies upon a problem-specific $\varphi$.  \texttt{\_psi\_t\_output}~(\ref{eq:thm_psi_t_output}) is also local, implicitly relying upon problem-specific assumptions and definitions.
\texttt{\_psi\_t\_output} combined with \texttt{\_psi\_t\_def} correspond to the right-hand side of Nielsen \& Chuang's circuit in their Fig. 5.2, and the left-hand side of their equation (5.20), and is expressed as a \textit{probability} (of 1) of obtaining the first-register state $\ket{\psi_{t}}$ and second-register state $\ket{u}$ from the given inputs to the first stage of the QPE circuit.  
The probability notation may seem odd here, when we are simply making a claim about the deterministic output of a circuit, but quantum theory is intrinsically probabilistic and we will see how this more general framework is useful later.

The proof of \texttt{\_psi\_t\_output}~(\ref{eq:thm_psi_t_output}) relies on the more fundamental ``phase kickback'' phenomenon, captured in \ProveIt{}'s quantum circuit package theorems as:
\begin{align}
    \begin{split}
    &\text{\texttt{quantum.circuit} Theorem: }\texttt{phase\_kickback}\\[1ex]
    &\raisebox{-0.5\height}{\includegraphics[width = 0.4\textwidth]{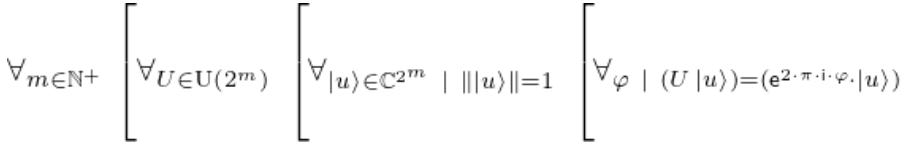}}\\[1ex]
    &\raisebox{-0.5\height}{\includegraphics[width = 0.4\textwidth]{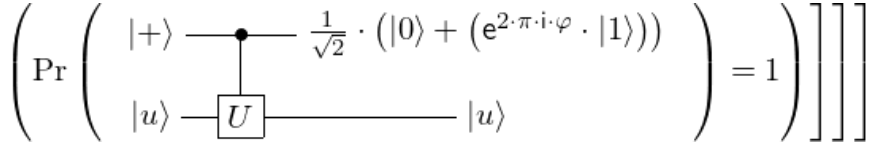}}
    \label{eq:thm_phase_kickback},
    \end{split}
\end{align}
and generalized for application on a register of multiple qubits as
\begin{align}
    &\text{\texttt{quantum.circuit} Theorem: }\nonumber\\
    &\texttt{phase\_kickbacks\_on\_register}\nonumber\\[1ex]
    &\raisebox{-0.5\height}{\includegraphics[width = 0.48\textwidth]{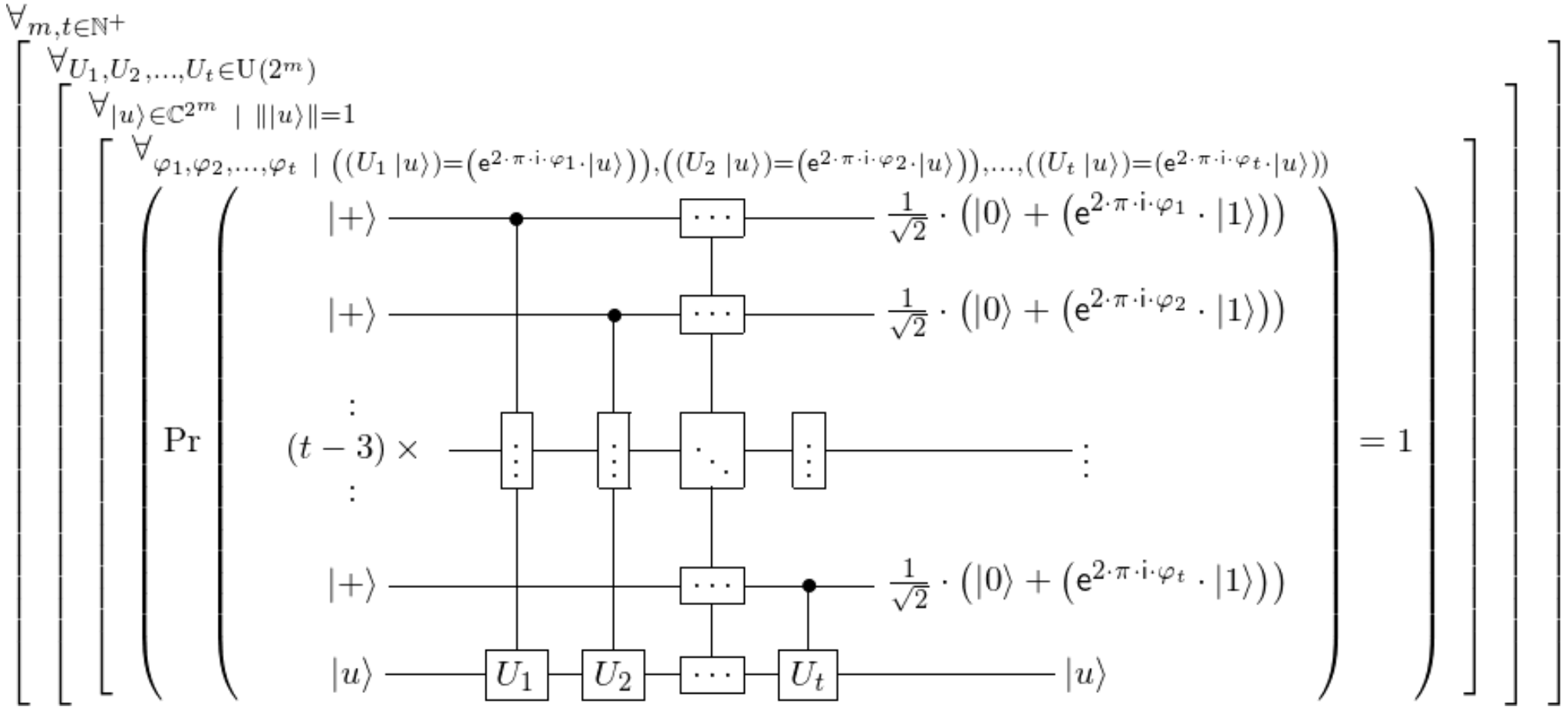}}
    \label{eq:thm_phase_kickbacks_on_register}.
\end{align}

We then prove that the tensor product in \texttt{\_psi\_t\_def}~(\ref{eq:axiom_psi_t_def}) can be rewritten in the compact summation form corresponding to Nielsen \& Chuang's equation (5.20) \cite[pg 222]{Nielsen_Chuang:2010}:\footnote{$\ket{k}_{t}$ denotes a \textit{number ket} (a \texttt{NumKet} object in \ProveIt{}), similar to notation used in Mermin \cite[pg 5]{Mermin_Quantum_Computer_Science:2007} for a multi-Cbit state, where $k$ is the decimal equivalent of a binary sequence within the ket and the subscript $t$ indicates the number of qubits being used in the representation.
The vector $\ket{11010}$ on 5 qubits, for example, can be represented by the number ket $\ket{26}_{5}$.}
\begin{align}
    \begin{split}
    &\text{Local QPE Theorem: }\texttt{\_psi\_t\_formula}\\
    &\forall_{t \in \mathbb{N}}~\ket{\psi_{t}}
    =
    \frac{1}{2^{t/2}}
    \sum_{k = 0}^{2^t - 1} e^{2 \pi i \varphi k} \ket{k}_{t}.
    \label{eq:thm_psi_t_formula}
    \end{split}
\end{align}
The \texttt{\_psi\_t\_formula} theorem is notable for using a proof by induction (on the variable $t$ representing the number of qubits in the first register). For some insight into the transformation from the tensor product in \texttt{\_psi\_t\_def}~(\ref{eq:axiom_psi_t_def}) to the summation formula in \texttt{\_psi\_t\_formula}~(\ref{eq:thm_psi_t_formula}), consider the instantiation and expansion of \texttt{\_psi\_t\_def} for $t=1$ and $t=2$ shown below:
\begin{align}
    \ket{\psi_{1}}
    &=
    \frac{1}{2^{1/2}}
    (\ket{0} + e^{2\pi i \varphi(1)}\ket{1})
    =
    \frac{1}{2^{1/2}}
    \sum_{k = 0}^{1} e^{2 \pi i \varphi k} \ket{k}_{t},\\
    \ket{\psi_{2}}
    &=
    \frac{1}{2^{2/2}}
    \left(
    (\ket{0} + e^{2\pi i \varphi(2)}\ket{1})
    \otimes
    (\ket{0} + e^{2\pi i \varphi(1)}\ket{1})
    \right)\nonumber\\
    &=
    \frac{1}{2}
    \left(
    \ket{00}
    + e^{2\pi i \varphi(1)}\ket{01}
    + e^{2\pi i \varphi(2)}\ket{10}
    + e^{2\pi i \varphi(3)}\ket{11}
    \right)\nonumber\\
    &=
    \frac{1}{2}
    \left(
    \ket{0}_{2}
    + e^{2\pi i \varphi(1)}\ket{1}_{2}
    + e^{2\pi i \varphi(2)}\ket{2}_{2}
    + e^{2\pi i \varphi(3)}\ket{3}_{2}
    \right)\nonumber\\
    &=
    \frac{1}{2}
    \sum_{k = 0}^{3} e^{2 \pi i \varphi k} \ket{k}_{t}.
\end{align}

\subsection{QPE Circuit State at Stage 3}

We next prove that the first register state at stage $3$ is $\ket{\Psi}$:
\begin{align}
    \begin{split}
    &\text{Local QPE Theorem: }\texttt{\_Psi\_output}\\
    &\raisebox{-0.5\height}{\includegraphics[width = 0.4\textwidth]{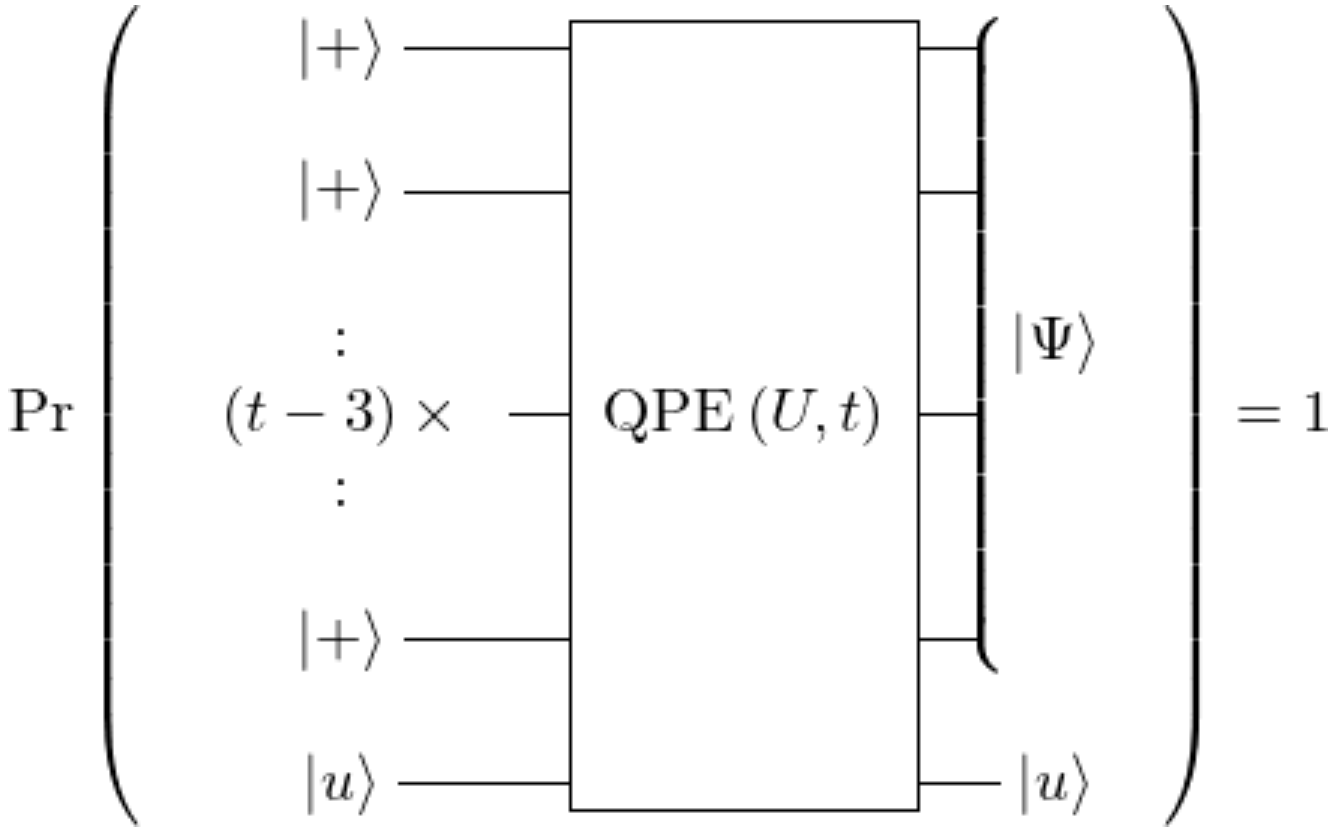}}
    \label{eq:thm_Psi_output},
    \end{split}
\end{align}
where $\ket{\Psi}$ is defined locally from \texttt{\_Psi\_def} as:
\begin{align}
    \begin{split}
    &\text{Local QPE Definition: }\texttt{\_Psi\_def}\\
    &\ket{\Psi}
    =
    \text{FT}^{\dagger}_{t}\ket{\psi_{t}}
    \label{eq:Psi_def},
    \end{split}
\end{align}
and $\text{FT}_{t}^{\dagger}$ represents a $t$-qubit inverse quantum Fourier transform. 
This follows trivially from the definition of the \texttt{QPE} circuit, Definition~\ref{eq:axiom_QPE_def}.
The inverse quantum Fourier transform is an important quantum algorithm of its own, of course, but is treated here as modular ``black box'' function returning a well-defined result.

As with \texttt{\_psi\_t\_output}~(\ref{eq:thm_psi_t_output}), \texttt{\_Psi\_output}~(\ref{eq:thm_Psi_output}) is expressed as a probability --- a probability of $1$ that we obtain at stage 3 the first-register state $\ket{\Psi}$ and second-register state $\ket{u}$ from the given inputs.

\subsection{Probability Space of Measurement Outcomes}

To carefully formalize the QPE process and related measurement probabilities, we take some care in establishing a probability space to formally model the QPE circuit measurement outcomes.

Interpreting the sequence of 0s and 1s obtained in the measurement process as the binary expansion of an integer $m$, we define the sample space $\Omega$ as the set of all possible QPE circuit configurations with the given fixed input state $\ket{+}\otimes\ket{+}\otimes\ldots \otimes\ket{+}\otimes\ket{u}$ and (probabilistic) first-register measurement result $\ket{m}_{t}$, with $m \in \{ 0, 1, \ldots 2^t - 1 \}$:
\begin{align}
    &\text{Local QPE Definition: }\texttt{\_sample\_space\_def}\nonumber\\[1ex]
    % \nonumber\\
    &\raisebox{-0.5\height}{\includegraphics[width = 0.45\textwidth]{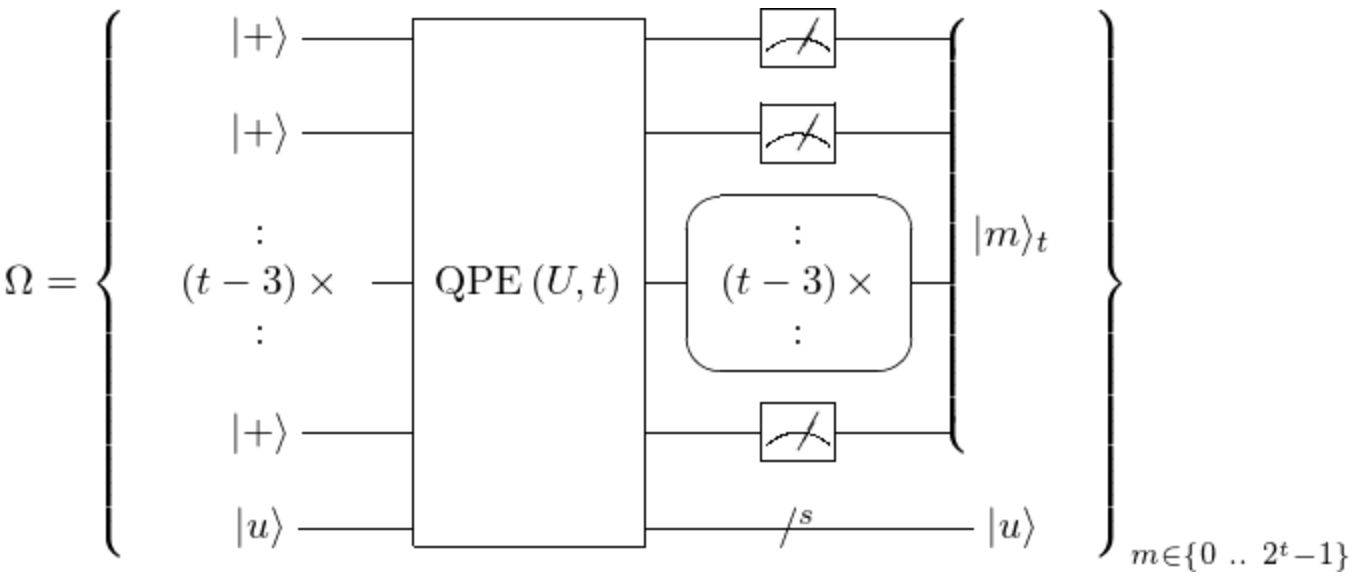}}
    \label{eq:axiom_sample_space_def}
\end{align}
where the right-hand side of \texttt{\_sample\_space\_def}~(\ref{eq:axiom_sample_space_def}) is an example of a \textit{set comprehension} expression in \ProveIt{} --- \textit{i.e.}, the right-hand side of \texttt{\_sample\_space\_def} denotes a \textit{set} of input-output-circuit configurations and the large curly braces are set delimiters.\footnote{Set comprehension expressions in \ProveIt{} look pretty much like standard set comprehension expressions except that domain specifications appear at the end in subscript form. For example, a (sub)set comprehension expression in \ProveIt{} for the positive Reals might appear as $\{x \,|\, x > 0 \}_{x \in \mathbb{R}}$.}

To prove that $\Omega$ has the properties of a sample space, we need to establish that the mapping from integer $m$ to QPE circuit configuration with measurement $\ket{m}_{t}$ is a one-to-one mapping from the set $\{ 0, 1, 2, \ldots, 2^t - 1 \}$ onto $\Omega$:
\begin{align}
    &\text{Local QPE Theorem: }\texttt{\_sample\_space\_bijection}\nonumber\\[1ex]
    % \nonumber\\
    &\raisebox{-0.5\height}{\includegraphics[width = 0.4\textwidth]{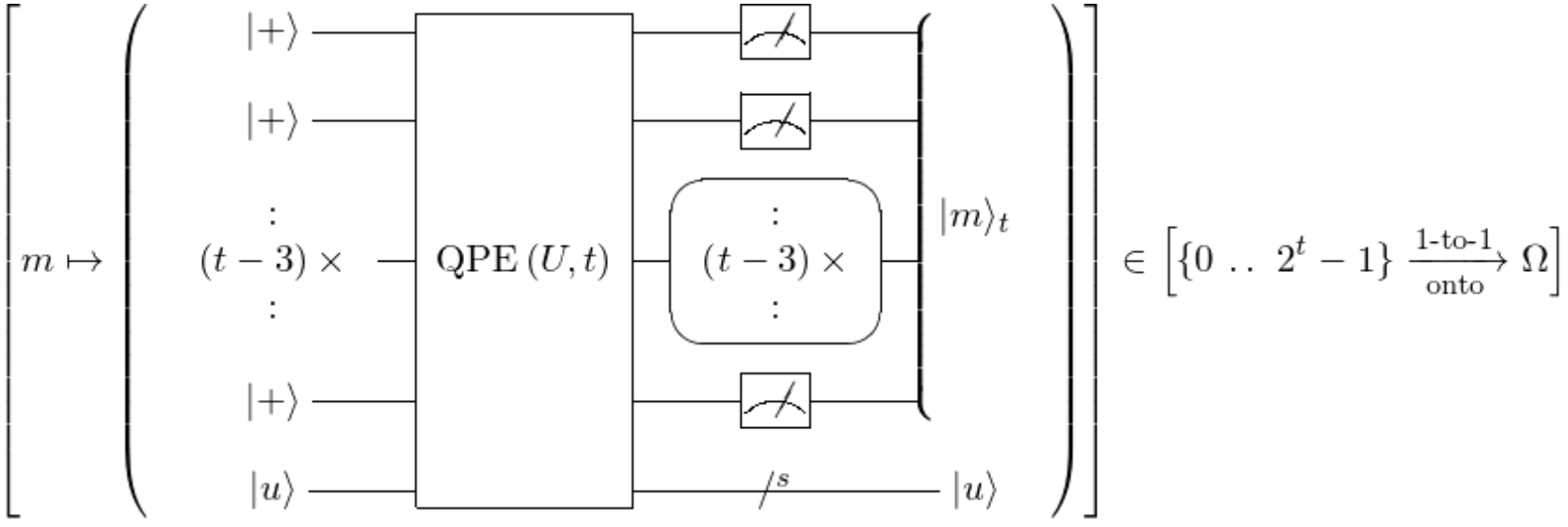}}
    \label{eq:thm_sample_space_bijection}
\end{align}
This means that the values $m \in \{ 0, 1, 2, \ldots, 2^t - 1\}$ induce a partition of the sample space $\Omega$, which is true given that quantum circuits with distinct outputs are distinct from each other.
This is also critical later in evaluating the probability of an \textit{event} consisting of the disjunction $\{ \ket{m_1}_{t}, \ket{m_2}_{t}, \ldots, \ket{m_k}_{t} \}$ of atomic outcomes, allowing us to use the fact that $Pr(\bigvee_i {\ket{m_i}_{t}}) = \sum_i {Pr(\ket{m_i}_{t})}$.

The probability function associated with the probability space is then obtained as a special case of the Born rule (recall above we established $\ket{\Psi}$ as the first-register state of the QPE circuit at stage 3, just before measurement):
\begin{align}
    &\text{Local QPE Theorem: }\texttt{\_outcome\_prob}\nonumber\\[1ex]
    &\raisebox{-0.5\height}{\includegraphics[width = 0.42\textwidth]{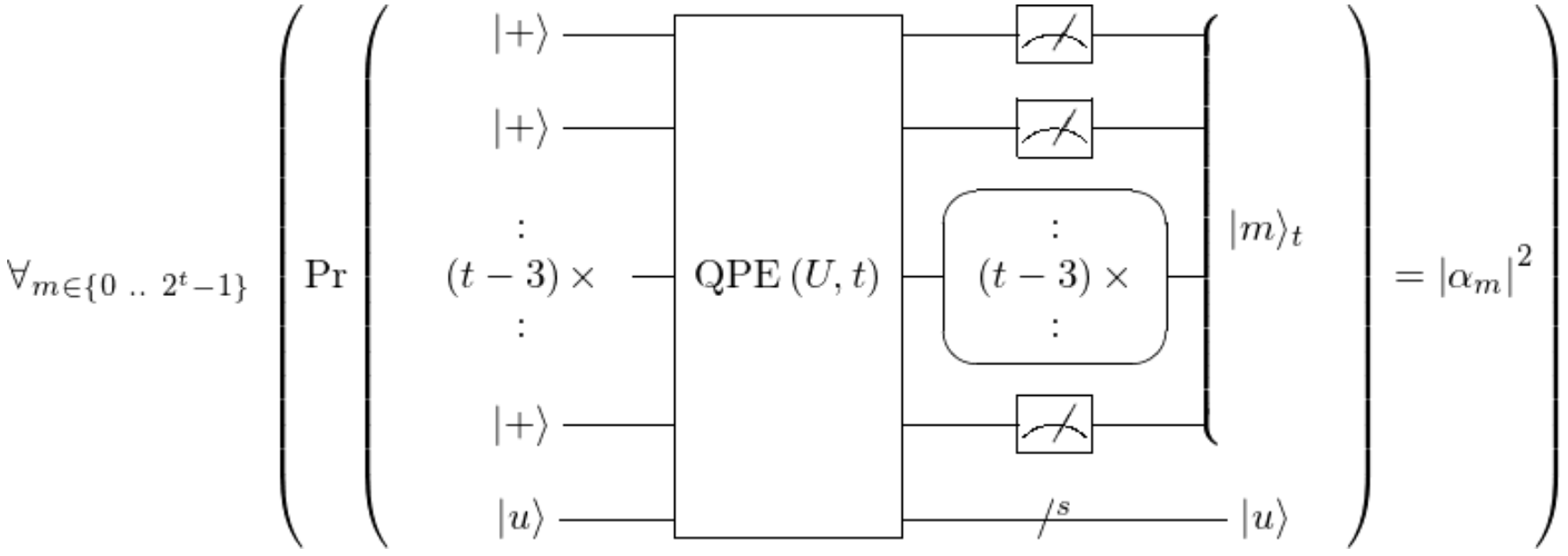}}
    \label{eq:thm_outcome_prob}
\end{align}
where $\alpha_m$ is the probability amplitude corresponding to the measurement outcome $\lvert m \rangle$:
\begin{align}
    \begin{split}
    &\text{Local QPE Definition: }\texttt{\_alpha\_m\_def}\\
    &\forall_{m\in\{0, \ldots, 2^{t}-1\}} \big(\alpha_{m}
    = \braket{m}{\Psi} \big).
    \label{eq:axiom_alpha_m_def}
    \end{split}
\end{align}
We can then prove that the set $\Omega$ defined in \texttt{\_sample\_space\_def}~(\ref{eq:axiom_sample_space_def}) is indeed a sample space:
\begin{align}
    \begin{split}
    &\text{Local QPE Theorem:}\texttt{\_Omega\_is\_sample\_space}\\
    &\Omega\underset{c}{\in}\text{SampleSpaces}
    \label{eq:thm_Omega_is_sample_space}
    \end{split}
\end{align}
meaning that the outcomes specified in $\Omega$ are mutually exclusive and collectively exhaustive, which then also implies that the probabilities of the individual outcomes in $\Omega$ sum to 1.\footnote{The $\in$ notation with the ``underset'' $c$ is used to indicate class membership and distinguish it from standard set membership. Here we are simply taking the collection of all sample spaces to be a class, which is weaker than assuming the collection actually constitutes a set (only sets admit subset comprehension).}

\subsection{Evaluating Probability Amplitude \texorpdfstring{$\alpha_{m}$}{alpha\_m}}

We then prove an initial formula for $\alpha_{m}$. As Nielsen \& Chuang \cite[pg 223, formula (5.23)]{Nielsen_Chuang:2010} point out, applying the inverse Fourier transform to the state in \texttt{\_psi\_t\_formula}~(\ref{eq:thm_psi_t_formula}) produces the double-summation state:
\begin{align}
    &&\ket{\Psi}
    &=\frac{1}{2^t} \sum_{k,l=0}^{2^t - 1} e^{-2\pi i k \ell/2^t} e^{2\pi i \varphi k} \ket{\ell}.
    \label{eq:N&C_5_23}
\end{align}
From the definition of $\alpha_{m}$ in \texttt{\_alpha\_m\_def}~(\ref{eq:axiom_alpha_m_def}) and \texttt{\_psi\_t\_formula}~(\ref{eq:thm_psi_t_formula}), we can derive a formula for the probability amplitude corresponding to measurement outcome $\ket{m}$:
\begin{align}
    &\text{Local QPE Theorem: }\texttt{\_alpha\_m\_evaluation}\nonumber\\[1ex]
    &\forall_{m \in \{0, \ldots, 2^{t}-1\}}~\left(\alpha_{m}
    = \left(\frac{1}{2^{t}} \sum_{k = 0}^{2^{t} - 1} \left(e^{-\frac{2 \pi i k m}{2^{t}}} \cdot e^{2 \pi i \varphi k}\right)\right)\right).
    \label{eq:thm_alpha_m_evaluation}
\end{align}
This follows straightforwardly from the definitions of $\lvert \Psi \rangle$ (\ref{eq:axiom_psi_t_def}), $\lvert \psi_t \rangle$ (\ref{eq:Psi_def}), and 
\begin{align}
  \begin{split}
  &\text{QFT Theorem: }\texttt{invFT\_on\_matrix\_elem}\\[1ex]
  &\forall_{n \in \mathbb{N}^+}~\Big[\forall_{k, l \in \{0~\ldotp \ldotp~2^{n} - 1\}}\\
  &\left(\left({_{n}}\langle l \rvert \thinspace {\mathrm {FT}}^{\dag}_{n} \thinspace \lvert k \rangle_{n}\right) = \left(\frac{1}{2^{\frac{n}{2}}} \cdot \mathsf{e}^{\frac{-\left(2 \cdot \pi \cdot \mathsf{i} \cdot k \cdot l\right)}{2^{n}}}\right)\right)\Big].
  \end{split}
\end{align} 
\texttt{\_alpha\_m\_evaluation}~(\ref{eq:thm_alpha_m_evaluation}) makes explicit an implicit step in Nielsen \& Chuang, intermediate between their expression (5.23) and formula (5.24) \cite[pg 223]{Nielsen_Chuang:2010}.

\subsection{Special Case: The Phase \texorpdfstring{$\varphi$}{phi} Can Be Represented Exactly in \texorpdfstring{$t$}{\textit{t}} Bits}
\label{subsection:special_case}

For the special case when  
the first register of $t$-qubits in the QPE circuit in Fig.~\ref{fig:QPE_circuit} is sufficient to exactly represent the phase $\varphi$ (called ``the ideal case'' by Nielsen \& Chuang \cite[pg 223]{Nielsen_Chuang:2010}), the ``ideal phase condition'' is given by:
\begin{align}
    2^{t} \varphi
    &\in \{0, 1, 2, \ldots, 2^{t}-1\}
    \label{eq:exact_phase_condition}.
\end{align}

Instantiating \texttt{\_alpha\_m\_evaluation}~(\ref{eq:thm_alpha_m_mod_evaluation}) with $m = 2^t \varphi$ under this condition leads to
\begin{align}
    &\text{Local QPE Theorem: }\texttt{\_alpha\_ideal\_case}\nonumber\\[1ex]
    &
    \left(
    (2^{t}\varphi)\in\{0, 1, 2, \ldots, 2^{t}-1\}
    \right)
    \Rightarrow
    \big(\alpha_{(2^{t}\varphi)} = 1\big)
    \label{eq:thm_alpha_ideal_case}
\end{align}
That is, the probability amplitude of the scaled phase $2^{t}\varphi$ is 1. Using \texttt{\_outcome\_prob}~(\ref{eq:thm_outcome_prob}) and \texttt{\_alpha\_ideal\_case}~(\ref{eq:thm_alpha_ideal_case}),
\begin{align}
    &\text{QPE Theorem: }\texttt{qpe\_exact}\nonumber\\[1ex]
    &\raisebox{-0.5\height}{\includegraphics[width = 0.42\textwidth]{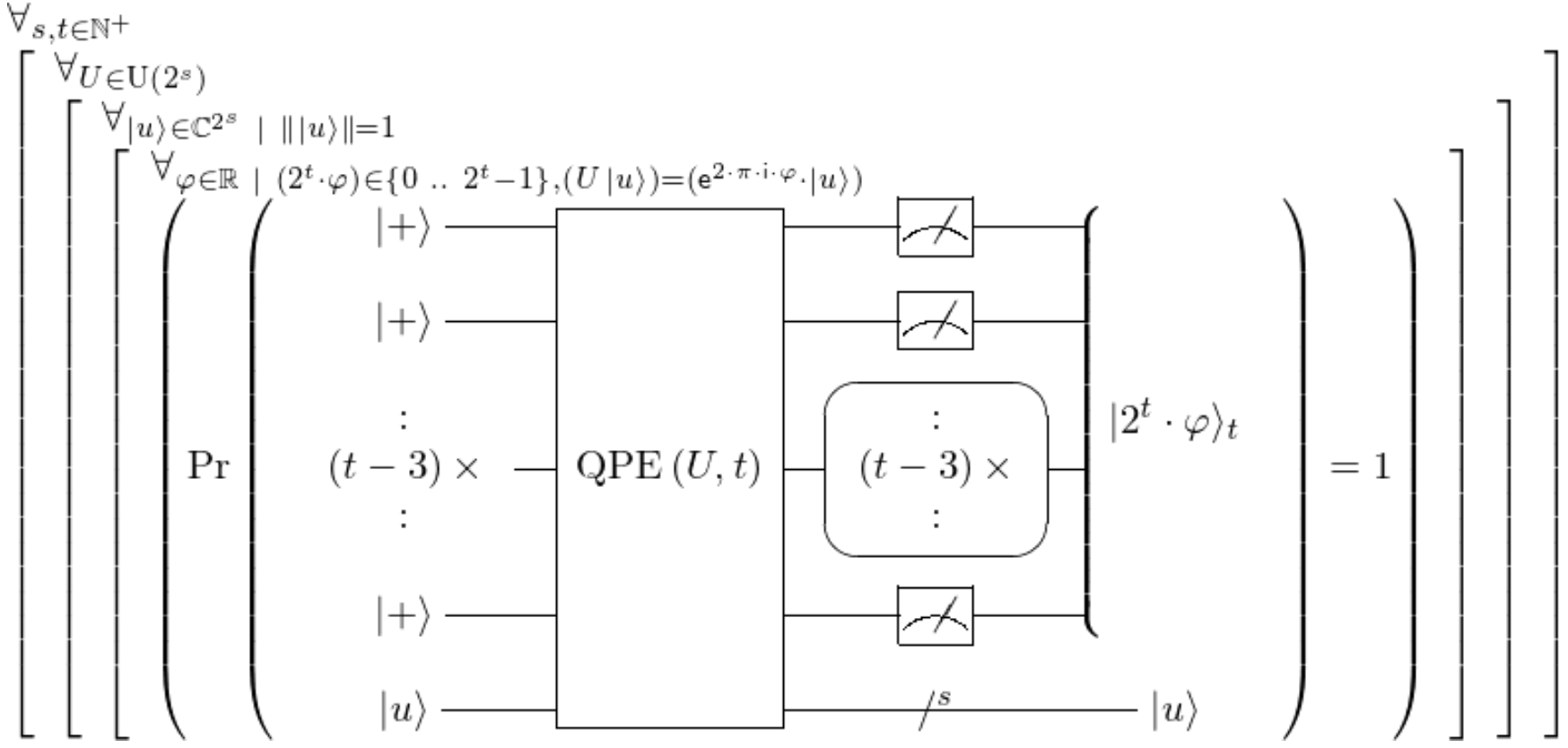}}
    \label{eq:thm_qpe_exact},
\end{align}
which states that, under the standard QPE problem conditions, and under the specific additional ``ideal phase'' condition that $2^t \varphi \in \{ 0, \ldots, 2^t - 1 \}$ (\textit{i.e.} that the phase $\varphi$ can be written exactly with a $t$-bit binary expansion), the first-register measurement in the QPE circuit gives exactly $\ket{2^t \varphi}_{t}$ with a probability of 1. The phase $\varphi$ can then be obtained by dividing the measurement result by $2^t$. 

\subsection{Bounding the Probability for the Best Measurement Outcome 
\texorpdfstring{$b_{r}=\texttt{round}(2^{t}\varphi)$}{b\_r = round(2\^{}t phi)}}
\label{subsect:best_guarantee}

When $2^{t} \varphi$ cannot be represented exactly in the $t$ qubits available in the first register, we can still prove a lower bound for the probability to obtain the best possible measurement outcome, $b_{r} = \texttt{round}(2^{t}\varphi)$.

From \texttt{\_alpha\_m\_evaluation}~(\ref{eq:thm_alpha_m_evaluation}) and using the $2^t$ periodicity of $e^{2 \pi i k m / 2^t}$, we can derive
\begin{align}
    &\text{Local QPE Theorem: }\texttt{\_alpha\_m\_mod\_evaluation}\nonumber\\[1ex]
    &\forall_{m \in \mathbb{Z}}~\left(\alpha_{(m\;\text{mod}\;2^{t})}
    = \left(\frac{1}{2^{t}} \sum_{k = 0}^{2^{t} - 1} \left(e^{-\frac{2 \pi i k m}{2^{t}}} \cdot e^{2 \pi i \varphi k}\right)\right)\right)
    \label{eq:thm_alpha_m_mod_evaluation},
\end{align}
and then rewrite this explicitly as a finite geometric sum:
\begin{align}
    &\text{Local QPE Theorem: }\texttt{\_alpha\_m\_mod\_as\_geometric\_sum}\nonumber\\[1ex]
    &\forall_{m \in \mathbb{Z}}~\left(\alpha_{(m\;\text{mod}\;2^{t})}
    = \left(\frac{1}{2^{t}} \sum_{k = 0}^{2^{t} - 1}
    \left(
    e^{2 \pi i (\varphi - \frac{m}{2^t})}
    \right)^{k}
    \right)\right)
    \label{eq:thm_alpha_m_mod_as_geometric_sum}.
\end{align}

We now define $b_{r}$ and $\delta_{b}$ as follows:
\begin{align}
    \begin{split}
    &\text{Local QPE Definition: }\texttt{\_best\_round\_def}\\
    &b_{r} = \texttt{round}(2^t \varphi);
    \label{eq:axiom_best_round_def}
    \end{split}
    \\[0.1in]
    \begin{split}
    &\text{Local QPE Definition: }\texttt{\_delta\_b\_def}\\
    &\forall_{b\in\mathbb{Z}}\Big(\delta_{b}
    =
    \varphi - \frac{b}{2^t}\Big).
    \label{eq:axiom_delta_b_def}
    \end{split}
\end{align}

\noindent Reducing the finite geometric sum of \texttt{\_alpha\_m\_mod\_as\_geometric\_sum}~(\ref{eq:thm_alpha_m_mod_as_geometric_sum}) and using 
$\theta > 0 \Rightarrow \sin \theta < \theta$ and
$0 \leq \theta \leq \pi/2 \Rightarrow \sin \theta \geq 2 \theta / \pi$ (as illustrated in Fig.~\ref{fig:bound_on_mag_of_1_minus_exp_and_sin}), it follows that
\begin{align}
    &\text{Local QPE Theorem: }\texttt{\_best\_guarantee\_delta\_nonzero}\nonumber\\[1ex]
    &(\delta_{b_r} \ne 0)
    \Rightarrow
    \left(
    |\alpha_{b_{r}\;\text{mod}\;2^t}|^2 > \frac{4}{\pi^2}
    \right).
    \label{eq:thm_best_guarantee_delta_nonzero}
\end{align}
That is: when the $t$ qubits of the first register are insufficient to exactly represent the scaled phase $2^{t} \varphi$, \textit{i.e.}, $(\delta_{b_{r}} \ne 0)$, the squared norm of the probability amplitude of the rounding-based best estimate $b_{r}$ is greater than $4/\pi^2$.
Of course, this doesn't really require that $\delta_{b_r} \ne 0$ since \texttt{qpe\_exact}~(\ref{eq:thm_qpe_exact}) proves that the probability is $1$ when $\delta_{b_r} = 0$.  Therefore, 
\begin{align}
    &\text{Local QPE Theorem: }\texttt{\_best\_guarantee}\nonumber\\[1ex]
    &\left(
    |\alpha_{b_{r}\;\text{mod}\;2^t}|^2 > \frac{4}{\pi^2}
    \right),
    \label{eq:thm_best_guarantee}
\end{align}
which then leads to the general \texttt{qpe\_best\_guarantee} theorem, providing a lower bound on the probability of obtaining $b_{r} = \texttt{round}(2^{t}\varphi)$ as the measurement outcome:
\begin{align}
    &\text{QPE Theorem: }\texttt{qpe\_best\_guarantee}\nonumber\\[1ex]
    &\raisebox{-0.5\height}{\includegraphics[width = 0.42\textwidth]{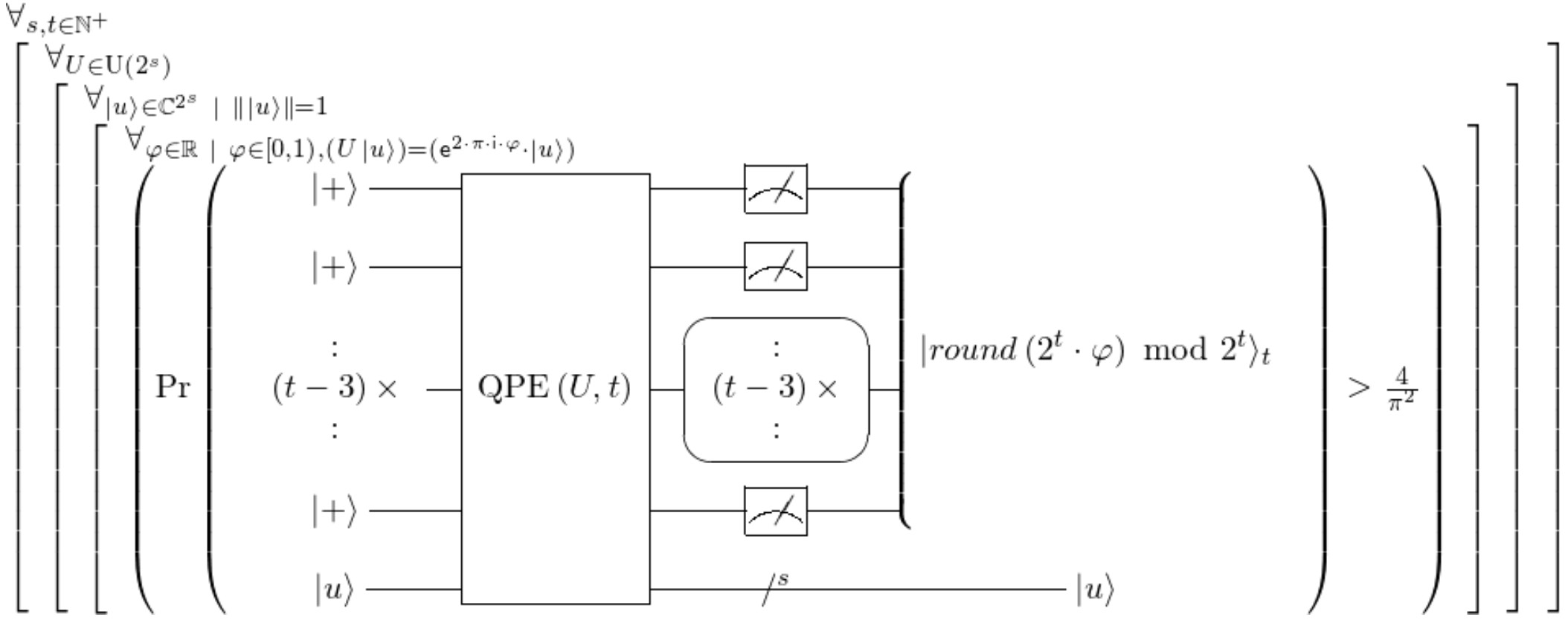}}
    \label{eq:thm_qpe_best_guarantee},
\end{align}
which is consistent with earlier non-formal results \cite{benenti:2004_principles_QCI, Cleve_et_al:1998_quantum_algs_revisited} as well as recent formal proof results in \textit{Coq} \cite{hietala_et_al_proving_quantum_programs:LIPIcs.ITP.2021.21}, but they used a weaker (non-strict) inequality ($\ge$).

\vspace{0.1in}

\subsection{Guaranteeing a Desired Precision}
\label{subsect:general_case}

Closely matching Nielsen \& Chuang's \cite[pg 223]{Nielsen_Chuang:2010} approach in their \S{5.2.1} on ``Performance and Requirements,'' let the integer $b_{\!f} \in \{0, 1, 2, \ldots, 2^t - 1\}$ be such that $b_{\!f}/2^{t} = 0.b_{1}b_{2}{\ldots}b_{t}$ is the best $t$-bit approximation to the phase $\varphi$ which is less than $\varphi$; this is expressed as:
\begin{align}
    \begin{split}
    &\text{Local QPE Definition: }\texttt{\_best\_floor\_def}\\
    &b_{f}
    =
    \lfloor 2^t \varphi \rfloor,
    \label{eq:axiom_best_floor_def}
    \end{split}
\end{align}
giving a difference $\delta_{b_{\!f}} = \varphi - \frac{b_{\!f}}{2^t} > 0$; recall an earlier definition for $\delta$~(\ref{eq:axiom_delta_b_def}), and note that the ``ideal case'' in \S\ref{subsection:special_case} where $b_{\!f} = b_{\!r} = \varphi$ producing $\delta_{b_{\!f}} = 0$ has already been considered.

From the finite geometric sum expression in \texttt{\_alpha\_m\_mod\_as\_geometric\_sum}~(\ref{eq:thm_alpha_m_mod_as_geometric_sum}), we can prove
\begin{align}
    \begin{split}
    &\text{Local QPE Theorem: }\texttt{\_alpha\_summed}\\[1ex]
    &\forall_{\ell \in \{ -2^{t - 1} + 1, \ldots, 2^{t - 1} \}\,|\,\ell \neq 0}\\
    &\hspace{0.1in}\left(\alpha_{b_{\!f}\oplus\ell} = \left(\frac{1}{2^{t}} \cdot \frac{1 - \mathsf{e}^{2 \pi \mathsf{i} \left(2^{t} \delta_{b_{\!f}} - \ell\right)}}{1 - \mathsf{e}^{2 \pi \mathsf{i} \left(\delta_{b_{\!f}} - \frac{\ell}{2^{t}}\right)}}\right)\right),
    \label{eq:thm_alpha_summed}
    \end{split}
\end{align}
which corresponds to expression (5.26) in Nielsen \& Chuang \cite[pg 224]{Nielsen_Chuang:2010} 
with minor differences in notation and the domain for $\ell$ explicitly stated.
\texttt{\_alpha\_summed}~(\ref{eq:thm_alpha_summed}) gives a formula for the probability amplitude corresponding to a measurement outcome that is $\ell$ units away from the desired ``best'' measurement outcome $b_{\!f}$~(\ref{eq:axiom_best_floor_def}).

Proving \texttt{\_alpha\_summed}~(\ref{eq:thm_alpha_summed}) involves instantiating  \texttt{\_alpha\_m\_mod\_as\_geometric\_sum}~(\ref{eq:thm_alpha_m_mod_as_geometric_sum}) with $m=b_{\!f} + \ell$ to produce the equivalent of Nielsen \& Chuang's expression (5.24):
\begin{align}
    \begin{split}
    &
    \ell \in \{ -2^{t - 1} + 1, \ldots, 2^{t - 1} \}\\
    &\hspace{0.1in}\vdash
    \Big[
    \alpha_{b_{\!f}\oplus\ell}
    =
    \frac{1}{2^t}
    \sum_{k=0}^{2^t - 1}
    \left(
    e^{2 \pi i (\varphi - \frac{b_{\!f}+\ell}{2^t})}
    \right)^{k}
    \Big],
    \label{eq:intermediate_alpha_b_plus_ell_sum}
    \end{split}
\end{align}
then substituting $\varphi = \delta_{b_{\!f}} + \frac{b_{\!f}}{2^{t}}$ and simplifying to obtain
\begin{align}
    \begin{split}
    &\ell \in \{ -2^{t - 1} + 1, \ldots, 2^{t - 1} \}\\
    &\hspace{0.1in}\vdash
    \Big[
    \alpha_{b_{\!f}\oplus\ell}
    =
    \frac{1}{2^t}
    \sum_{k=0}^{2^t - 1}
    \left(
    e^{2 \pi i (\delta_{b_{\!f}} - \frac{\ell}{2^t})}
    \right)^{k}
    \Big],
    \label{eq:intermediate_alpha_b_plus_ell_sum_02}
    \end{split}
\end{align}
then reducing the geometric sum.
Reducing the geometric sum requires that the common ratio $e^{2 \pi i (\delta_{b_{\!f}} - \ell/2^t)}$ is not 1, which follows from the fact that $(\delta_{b_{\!f}}-\frac{\ell}{2^{t}})\notin \mathbb{Z}$ unless $\delta_{b_{\!f}} = 0$ which was handled as the special case in \S\ref{subsection:special_case}.

Letting $e$ be ``a positive integer characterizing
our desired tolerance to error'' \cite[pg 224]{Nielsen_Chuang:2010}, we define the probability of a ``successful'' measurement outcome (and its complementary ``failure'') with the following definitions:
\begin{align}
    &\text{Local QPE Definition: }\texttt{\_success\_def}\nonumber\\[1ex]
    &\raisebox{-0.5\height}{\includegraphics[width = 0.42\textwidth]{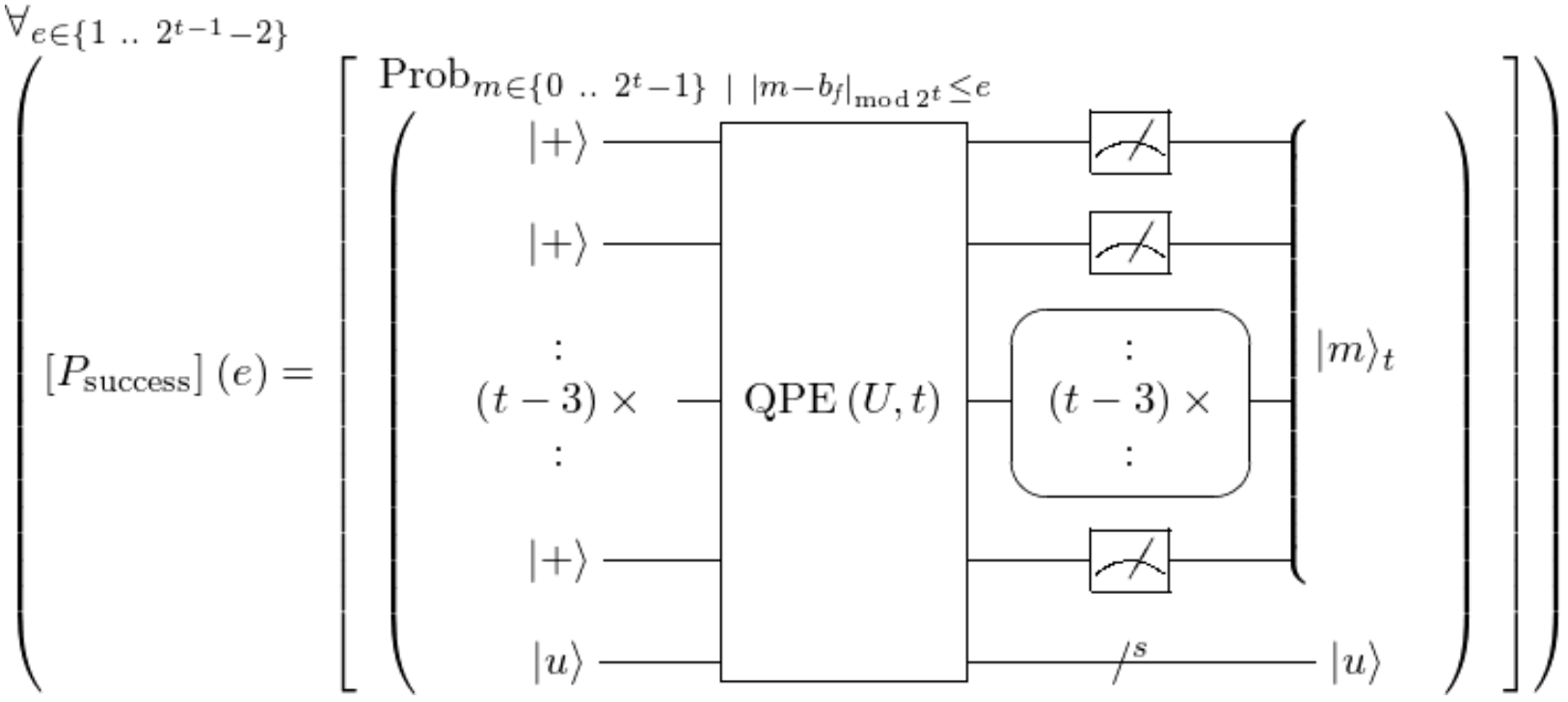}},
    \label{eq:axiom_success_def}
\end{align}
\begin{align}
    &\text{Local QPE Definition: }\texttt{\_fail\_def}\nonumber\\[1ex]
    &\raisebox{-0.5\height}{\includegraphics[width = 0.42\textwidth]{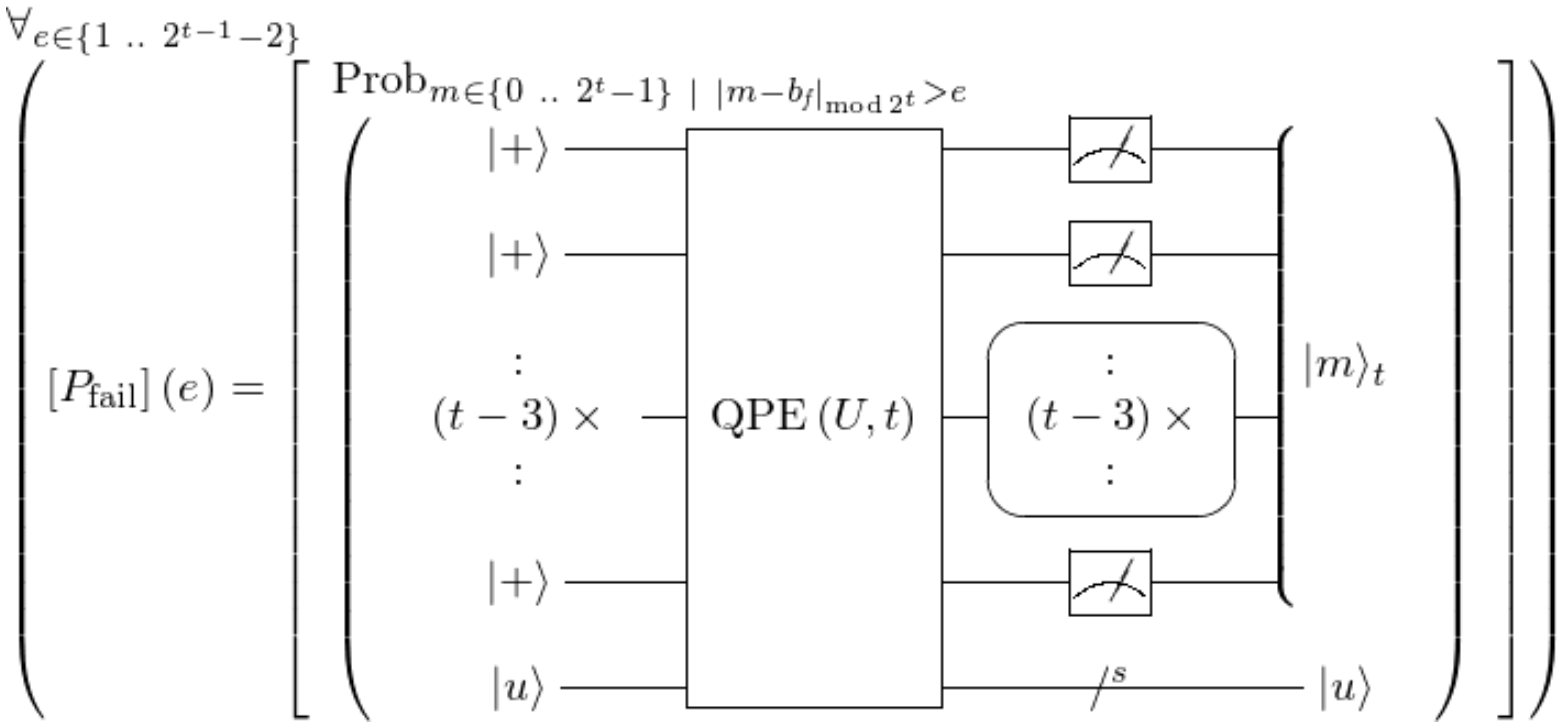}}.
    \label{eq:axiom_fail_def}
\end{align}

\texttt{\_success\_def}~(\ref{eq:axiom_success_def}) states that the probability $[P_{\text{success}}](e)$ of a measurement $m$ that is no more than $e$ units away from the best under-estimate $b_{\!f}$ of $2^t \varphi$ is simply the probability of obtaining the QPE circuit configuration with first-register measurement output $\ket{m}_{t}$ such that $|m - b_{\!f}|_{\text{mod}\,2^t} \le e$. 
The probability of \textit{failing} to obtain a measurement within the desired error tolerance $e$ of $b_{\!f}$ is defined similarly, but now with the condition that $|m - b_{\!f}|_{\text{mod}\,2^t} > e$. Nielsen \& Chuang express the failure probability using the notation $p(|m - b| > e)$ \cite[pg 224, lhs of Eq (5.27)]{Nielsen_Chuang:2010}.
Although the constraint is unstated in Nielsen \& Chuang, the upper bound on $e$ of $2^{t-1}-1$ is important to ensure that the summations that are to imminently appear in \texttt{\_fail\_sum}~ (\ref{eq:thm_fail_sum}) are properly defined.

Recall from earlier that $\alpha_{b_{\!f}\oplus\ell}$ is the probability magnitude of the vector $\ket{b_{\!f} \oplus \ell}_{t}$, where $b_{\!f}$ is the best $t$-bit under-estimate for the scaled phase $2^t \varphi$.
We also defined the ``tolerable error'' $e$ as the number of units away from the best estimate $b$ that would be acceptable as an eventual measurement.
The probability, then, of \textit{failing} to obtain a sufficiently accurate estimate of the phase $\varphi$ is the sum of the squared magnitudes of all possible measurements $\ket{m}_{t}$ such that $|m - b_{\!f}| > e$.
This is proven in  \texttt{\_fail\_sum}~(\ref{eq:thm_fail_sum}):
\begin{align}
    \begin{split}
    &\text{Local QPE Theorem: }\texttt{\_fail\_sum}\\
    &\forall_{e \in \{1~\ldotp \ldotp~2^{t - 1} - 2\}}
    \Biggl(\left[P_{\rm fail}\right]\left(e\right) =\\ &\hspace{0.075in}\left(\left(\sum_{\ell = -2^{t - 1} + 1}^{-\left(e + 1\right)} \left|\alpha_{b_{\!f}\oplus\ell}\right|^{2}\right) + \left(\sum_{\ell = e + 1}^{2^{t - 1}} \left|\alpha_{b_{\!f}\oplus\ell}\right|^{2}\right)\right)\Biggr),
    \label{eq:thm_fail_sum}
    \end{split}
\end{align}
which corresponds to formula (5.27) in Nielsen \& Chuang \cite[pg 224]{Nielsen_Chuang:2010}. 
Part of proving (\ref{eq:thm_fail_sum}) involves showing that the 
$\left|m - b_{\textit{f}}\right|_{\textup{mod}\thinspace 2^{t}} > e$ condition in \texttt{\_fail\_def}~(\ref{eq:axiom_fail_def}) is consistent with the ranges of the two summations over $\ell$ (which takes the role of $m - b_{\!f}$):
\begin{align}
    &\text{Local QPE Theorem: }\texttt{\_fail\_sum\_prob\_conds\_equiv\_lemma}\nonumber\\[1ex]
    &\raisebox{-0.5\height}{\includegraphics[width = 0.5\textwidth]{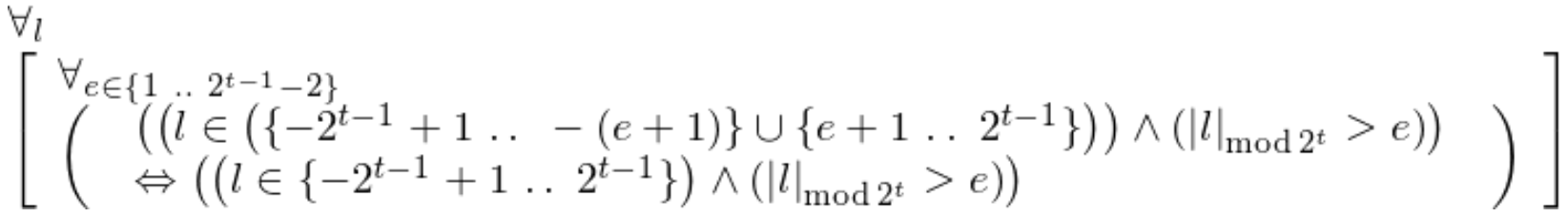}}.
    \label{eq:thm_fail_sum_prob_conds_equiv_lemma}
\end{align}

\begin{figure}[tb]
  \captionsetup{font=footnotesize}
  % \centering
  \includegraphics[width = 0.47\textwidth]{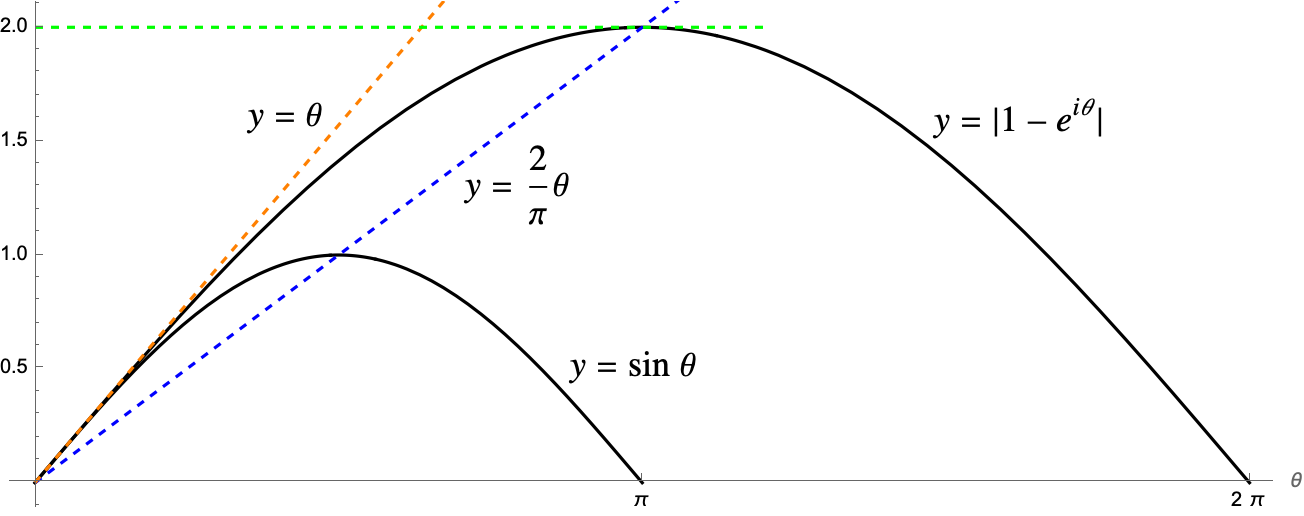}
\caption{Illustration of the bounds $\frac{2}{\pi}\theta \le |1 - e^{i\theta}| \le 2$ for $\theta \in [0, \pi]$ and $\sin{\theta} \ge \frac{2}{\pi}\theta$ for $\theta \in [0, \frac{\pi}{2}]$, utilized in the proof of \texttt{\_alpha\_sqrd\_upper\_bound}~(\ref{eq:thm_alpha_sqrd_upper_bound}), and the bound $\sin\theta < \theta$ for $\theta > 0$ used in the proof of \texttt{\_best\_guarantee\_delta\_nonzero} (\ref{eq:thm_best_guarantee_delta_nonzero}).
}
\label{fig:bound_on_mag_of_1_minus_exp_and_sin}
\end{figure}

From \texttt{\_alpha\_summed}~(\ref{eq:thm_alpha_summed}), we can then establish an upper bound on $|\alpha_{b_{\!f}\oplus\ell}|^2$:
\begin{align}
    &\text{Local QPE Theorem: }\texttt{\_alpha\_sqrd\_upper\_bound}\nonumber\\[1ex]
    &\forall_{l \in \{-2^{t - 1} + 1, \ldots, 2^{t - 1}\}}\left(\left|\alpha_{b_{f}\oplus\ell}\right|^{2} \leq \frac{1}{4 \left(\ell - \left(2^{t} \delta_{b_{f}}\right)\right)^{2}}\right),
    \label{eq:thm_alpha_sqrd_upper_bound}
\end{align}
corresponding to formula (5.29) in Nielsen \& Chuang \cite[pg 224]{Nielsen_Chuang:2010} and making use of the bounds $ \frac{2}{\pi}\theta\le |1 - e^{i\theta}| \le2$ for $\theta \in [0, \pi]$ (as illustrated in Fig.~\ref{fig:bound_on_mag_of_1_minus_exp_and_sin}).

\begin{figure}[tb]
  \captionsetup{font=footnotesize}
  \includegraphics[width = 0.47\textwidth]{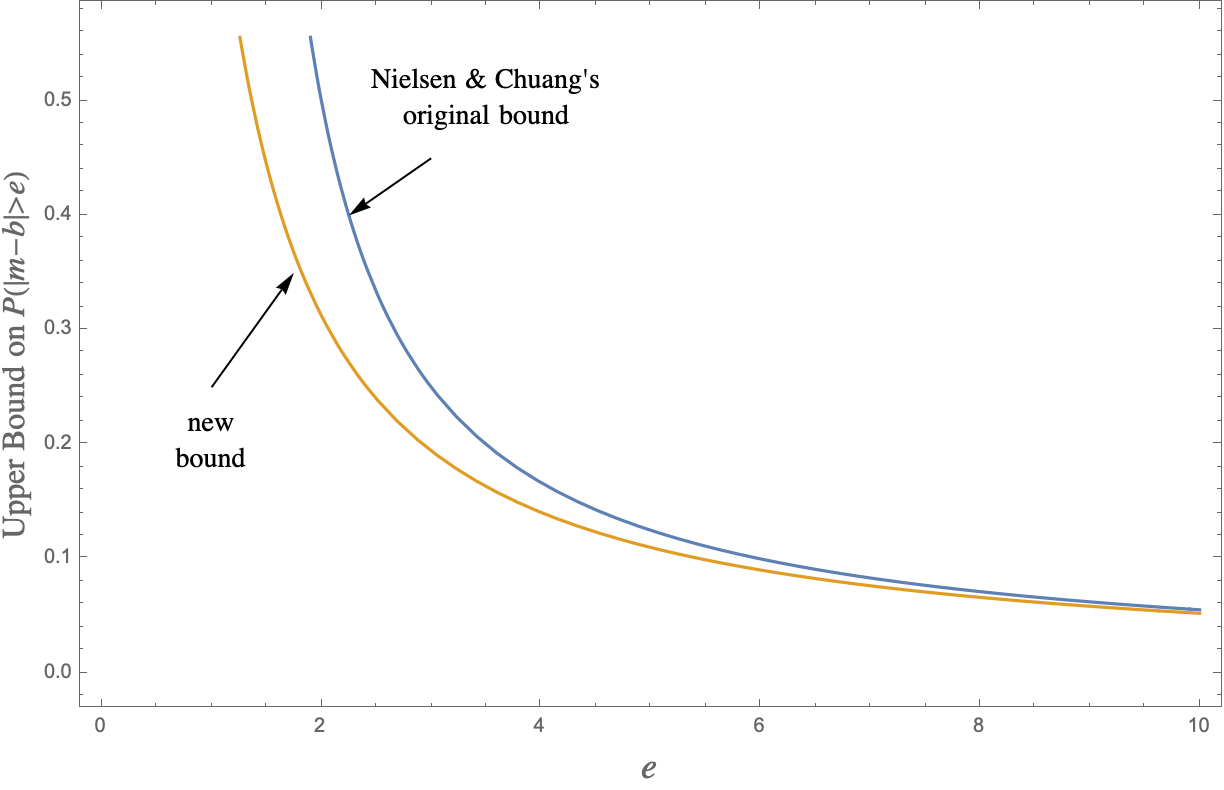}
\caption{Comparison of Nielsen \& Chuang's \cite{Nielsen_Chuang:2010} original upper bound vs. our new, tighter upper bound on the probability of ``failure'' versus the number of unit steps $e$ away from the ``best'' floor-based estimate $b_{f}$.
}
\label{fig:orig_vs_new_bound_on_fail}
\end{figure}

From \texttt{\_fail\_sum}~(\ref{eq:thm_fail_sum}) and \texttt{\_alpha\_sqrd\_upper\_bound}~(\ref{eq:thm_alpha_sqrd_upper_bound}) we then derive an initial upper bound expression for $[P_{\text{fail}}](e)$:
\begin{align}
  \begin{split}
  &\text{Local QPE Theorem: }\texttt{\_failure\_upper\_bound\_lemma}\\[1ex]
  &\forall_{e \in \{1, \dots, 2^{t - 1} - 2\}}
  \Biggl(\left[P_{\rm fail}\right]\left(e\right) \leq\\ &\frac{1}{4} \left(\sum_{\ell = -2^{t - 1} + 1}^{-\left(e + 1\right)} \frac{1}{\left(\ell - \left(2^{t} \delta\right)\right)^{2}}
    +
    \sum_{\ell = e + 1}^{2^{t - 1}} \frac{1}{\left(\ell - \left(2^{t} \delta\right)\right)^{2}}\right)\Biggr),
    \label{eq:thm_failure_upper_bound_lemma}
  \end{split}
\end{align}
corresponding to bound (5.30) in Nielsen \& Chuang \cite[pg 224]{Nielsen_Chuang:2010}. Applying some standard summation manipulations and inequalities to \texttt{\_failure\_upper\_bound\_lemma}~(\ref{eq:thm_failure_upper_bound_lemma}), we then prove
\begin{align}
    &\text{Local QPE Theorem: }\texttt{\_failure\_upper\_bound}\nonumber\\[1ex]
    &\forall_{e \in \{1~\ldotp \ldotp~2^{t - 1} - 2\}}~\left(\left[P_{\rm fail}\right]\left(e\right) \leq \frac{1}{2e} + \frac{1}{4e^2}\right),
    \label{eq:thm_failure_upper_bound}
\end{align}
corresponding to bound (5.34) in Nielsen \& Chuang \cite[pg 224]{Nielsen_Chuang:2010}, but actually a slight improvement over their bound in two different ways: first, the bound holds for $e \ge 1$ whereas their original bound holds only for $e \ge 2$; and second, the bound in \texttt{\_failure\_upper\_bound}~(\ref{eq:thm_failure_upper_bound}) is actually tighter (\textit{i.e.} \textit{less}) than Nielsen \& Chuang's bound, and thus the eventually-derived probability for a ``successful'' measurement is actually \textit{larger}.
See Fig.~\ref{fig:orig_vs_new_bound_on_fail} for a graphical comparison of the original and new bounds.

From \texttt{\_success\_def}~(\ref{eq:axiom_success_def}) and \texttt{\_fail\_def}~(\ref{eq:axiom_fail_def}), we trivially derive
 \begin{align}
    &\text{Local QPE Theorem: }\texttt{\_success\_complements\_failure}\nonumber\\[1ex]
    &\forall_{e \in \{1~\ldotp \ldotp~2^{t - 1} - 2\}}~\left(\left[P_{\rm success}\right]\left(e\right) = \left(1 - \left[P_{\rm fail}\right]\left(e\right)\right)\right).
    \label{eq:thm_success_complements_failure}
\end{align}
We can then show that adopting Nielsen \& Chuang's formula (5.35) \cite[pg 224]{Nielsen_Chuang:2010} as an assumed lower bound on the number of qubits in the first register:
\begin{align}
    \begin{split}
    &\text{QPE Assumption: }\texttt{\_t\_req}\\
    &t \ge
    \left(
    n +
    \left\lceil
    \log_{2}
    \left(
    2 +
    \frac{1}{2 \epsilon}
    \right)
    \right\rceil
    \right)
    \end{split}
    \label{eq:axiom_t_req}
\end{align}
is sufficient to guarantee a successful measurement of phase $\varphi$ to $n$ bits of precision with a probability of  at least $1 - \epsilon$: 
\begin{align}
    &\text{Local QPE Theorem: }\texttt{\_precision\_guarantee}\nonumber\\[1ex]
    &\raisebox{-0.5\height}{\includegraphics[width = 0.45\textwidth]{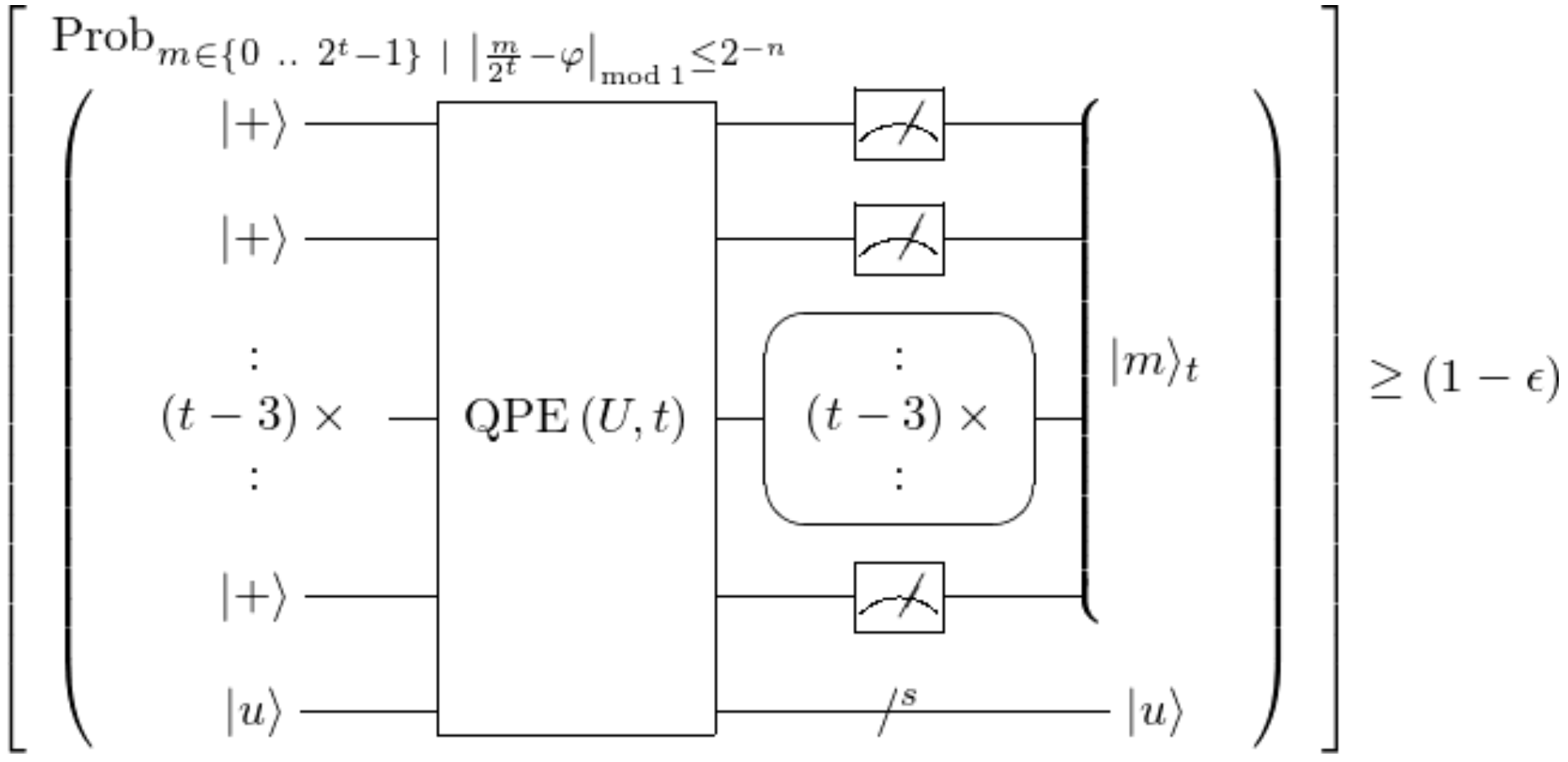}}.
    \label{eq:thm_precision_guarantee}
\end{align}
That local theorem is then generalized (\textit{e.g.} over all failure probabilities $\epsilon\in (0, 1]$ and all conditions satisfying the original QPE problem statement) to produce the universal \texttt{qpe\_precision\_guarantee} theorem:
\begin{align}
    &\text{QPE Theorem: }\texttt{qpe\_precision\_guarantee}\nonumber\\[1ex]
    &\raisebox{-0.5\height}{\includegraphics[width = 0.45\textwidth]{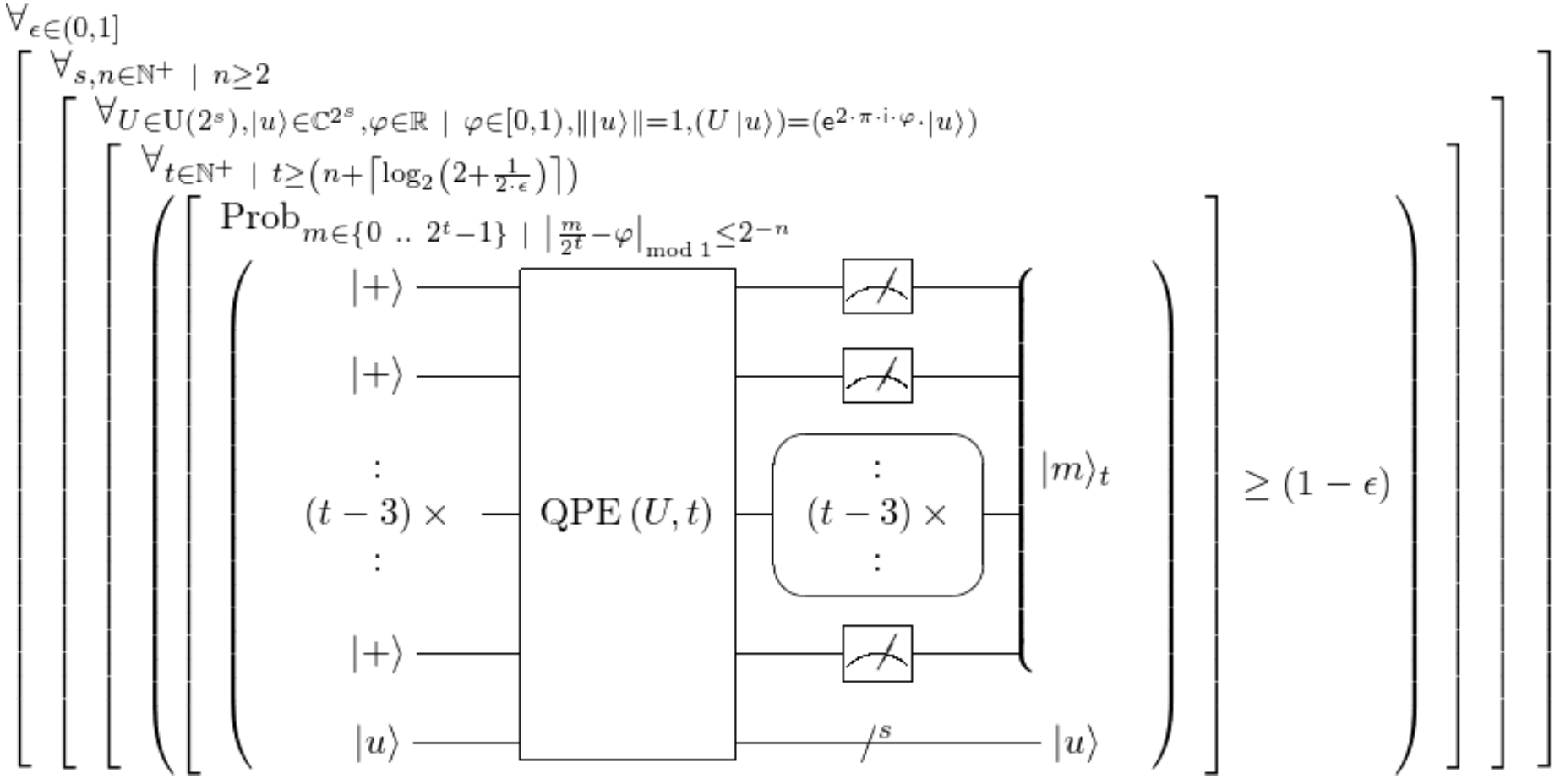}}.
    \label{eq:thm_qpe_precision_guarantee}
\end{align}

\section{What, precisely, did we prove?}
\label{Sec:dependencies}

Using our Prove-It system, we have constructed formal proofs of three universal theorems pertaining to quantum phase estimation:  \texttt{qpe\_exact}~(\ref{eq:thm_qpe_exact}), \texttt{qpe\_best\_guarantee}~(\ref{eq:thm_qpe_best_guarantee}), and \texttt{qpe\_precision\_guarantee}~(\ref{eq:thm_qpe_precision_guarantee}).  Those statements, and an outline of the steps leading to their derivation, are shown in \S\ref{Sec:essence}.  What does it mean to say that we have these formal proofs?  What are the precise claims represented by such proofs and how are these claims verified?

\subsection{Formal Proofs in \ProveIt{}}
\label{subsec:formal_proofs}
A \textit{formal proof} or \textit{proof certificate} in \ProveIt{} is a finite sequence $A_{1}, A_{2}, \ldots, A_{n}$ of formulas, each of which is either an assumption, an axiom, a theorem (proven or unproven), or a \textit{derivation} consisting of a judgment using a rule of inference whose conditions appear among the previous formulas, and ending with $A_{n}$ as the theorem being proved \cite{Hoedel:1995_Intro_Math_Logic, Mendelson:2015}.
This is essentially the same as the definition of a formal proof in mathematical logic except for the
%The difference with a formal proof in mathematical logic is the 
flexibility offered by \ProveIt{} to treat proven or unproven theorems effectively as axioms at the local level while tracking their dependencies and protecting against circular logic on a broader level.
The rules of inference include modus ponens, deduction, instantiation, and generalization (described in detail in Appendix C of \cite{Preprint:ProveIt}).
A formal proof in \ProveIt{} also includes several helpful characteristics of a ``structured proof'' \cite{Lamport:1995_How_to_write_a_proof, Lamport:2012}, including rules of inference and previous-step dependencies being explicitly cited in each step and hyperlinks allowing the user to isolate, zoom into, and explore the details of any step.
These proof characteristics make a formal proof in \ProveIt{} particularly human-readable and human-checkable.

% Combining these two figures inside a single  %
% begin/end{figure} call so they stay together %
% Example formal proof (first few lines)       %
% AND example sub-proof                        %
\begin{figure*}
     % \centering
     \begin{subfigure}[b]{0.55\textwidth}
         % \centering
         \includegraphics[width=\textwidth]{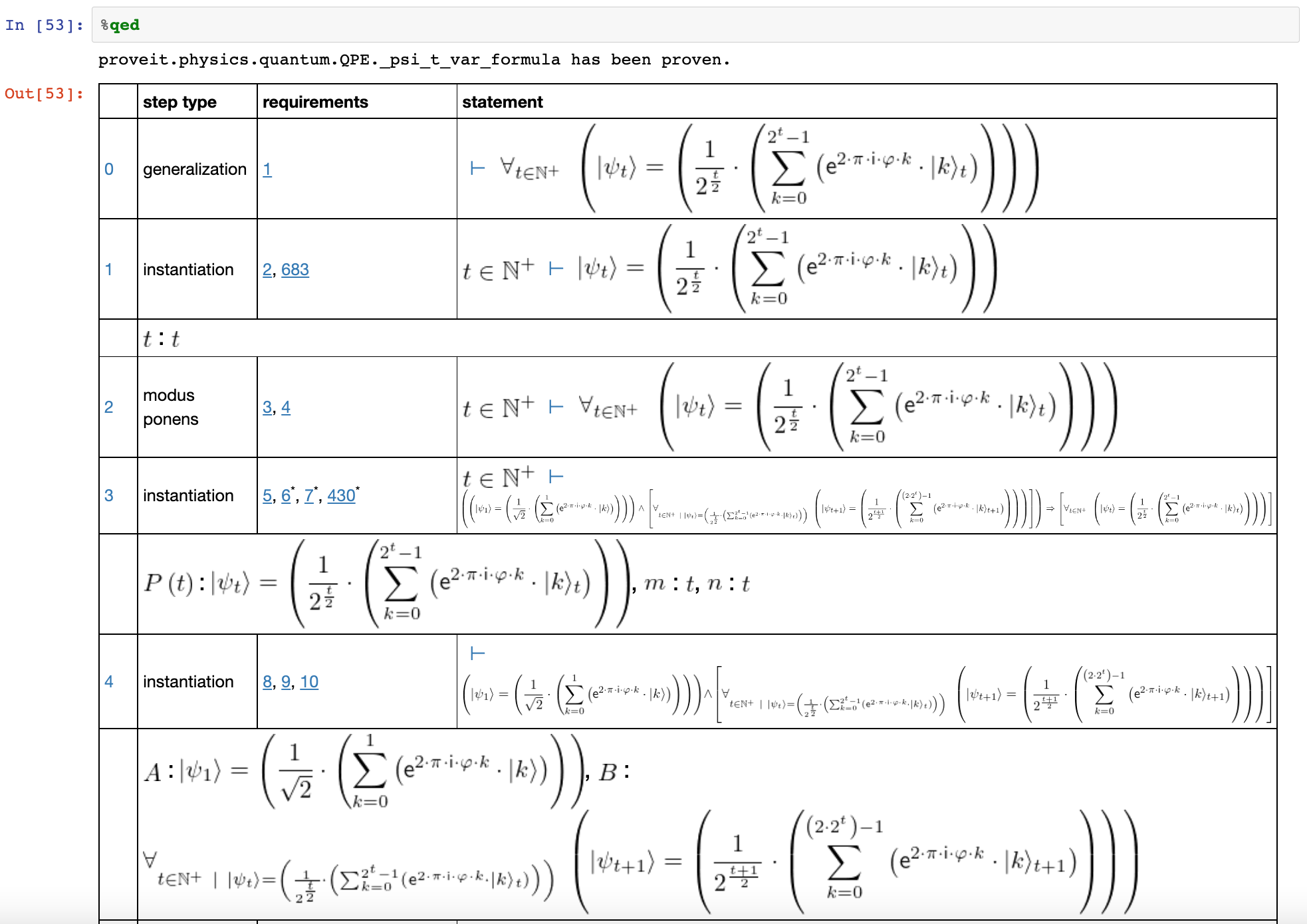}
         \captionsetup{width=10cm}
         \caption{The first few lines of \ProveIt{}'s formal proof output for \texttt{\_psi\_t\_formula}~(\ref{eq:thm_psi_t_formula}), illustrating  both the general structure of a formal proof in \ProveIt{} and the rigorous and explicitly human-checkable details involved in each proof step. Each numbered step (numbers along the left-hand side) includes an explicit listing (underlined blue numbers in 3rd column) of all previous steps on which the current step depends. Instantiation steps explicitly list the instantiations used.}
         \label{fig:example_formal_proof_steps}
     \end{subfigure}

     \vspace{0.2in}

     \begin{subfigure}[b]{0.8\textwidth}
         \centering
         \includegraphics[width=0.8\textwidth]{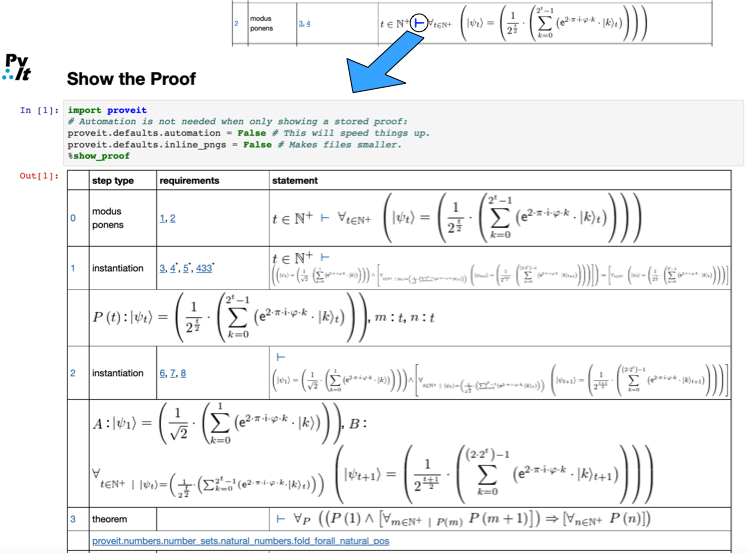}
         \captionsetup{width=10cm}
         \caption{A snapshot of the first few lines of \ProveIt{}'s formal proof of step \texttt{2} in previous Fig.~(\ref{fig:example_formal_proof_steps}), obtained by clicking on the turnstile symbol in that line's judgment (shown circled at the top).}
         \label{fig:example_sub_proof}
     \end{subfigure}
     \captionsetup{width=10cm}
     \caption{Obtaining a sub-proof from within a proof in \ProveIt{}.}
        \label{fig:subproof_from_proof}
\end{figure*}

% \begin{figure*}[p]
%   \captionsetup{font=footnotesize}
%   \begin{subfigure}
%   \includegraphics[width = 0.70\textwidth]{FIGURES/example_formal_proof_steps.png}
%   \caption{The first few lines of \ProveIt{}'s formal proof output for \texttt{\_psi\_t\_formula}~(\ref{eq:thm_psi_t_formula}), illustrating both the general structure of a formal proof in \ProveIt{} and the rigorous and explicitly human-checkable details involved in each proof step. Each numbered step (blue numbers along the left-hand side) includes an explicit listing (note the underlined blue numbers in 3rd column) of all previous steps (higher-numbered to indicate further down the proof tree) on which the current step depends. Instantiation steps explicitly list the instantiations used. 
%   }
%   \label{fig:example_formal_proof_steps}
%   \end{subfigure}
%   \vspace{0.2in}
%   \begin{subfigure}
%   \includegraphics[width = 0.7\textwidth]{FIGURES/example_sub-proof.png}
%   \caption{A snapshot of the first few lines of \ProveIt{}'s formal proof of step \texttt{2} in previous Fig.~(\ref{fig:example_formal_proof_steps}), obtained by clicking on the turnstile symbol in that line's judgment (shown circled at the top).
%   }
%   \label{fig:example_sub_proof}
%   \end{subfigure}
% \end{figure*}

From a theorems page of a theory package in the \ProveItWebsite{}, 
clicking on a theorem name brings up its \emph{proof notebook}.  One can then scroll down to the final cell with the \texttt{\%qed} command.
The output of this cell is the formal proof.
A formal proof produced by \ProveIt{} is often quite lengthy.
Thus the formal proof steps appear in a non-traditional reverse (or top-down) order, which makes it easier for the user to see the main steps in a proof, only working down the tree to ``earlier'' or higher-numbered steps when desired.
As an example, a snapshot of the first few lines of \ProveIt{}'s formal proof of the \texttt{\_psi\_t\_formula} theorem appears in Fig.~\ref{fig:example_formal_proof_steps}.
The steps are numbered in blue along the left-hand side, with \texttt{step 0} being the final theorem statement (and the root of the ``proof tree'' structure of derivations and axioms), with higher-numbered steps representing steps further down into the proof tree and supporting the lower-numbered steps at the top.

When referring to the order of the proof steps we will use the reverse order unless otherwise stated.

Each step explicitly names the type of derivation step being taken and the previous (higher-numbered) steps required for the step. In Fig.~\ref{fig:example_formal_proof_steps}, for example, step \texttt{2} applies \textit{modus ponens} using the results from step \texttt{3} (a logical implication) and step \texttt{4} (the antecedent of the logical implication).
Given the explicit listing of requirements for each step, one \textit{could} manually track through the entire proof to see all the steps that contributed to the conclusion in step \texttt{2}, but the system can actually do all the work for the user: simply clicking on the blue turnstile symbol $\vdash$ in the statement of interest will bring up a new page (a webpage on the system's website or a new Jupyter notebook on a developer's local computer) showing all the detailed steps leading to that particular conclusion, as illustrated in Fig.~\ref{fig:example_sub_proof} for the ``sub-proof'' produced for Step \texttt{2}.

For long formal proofs, hand-verification is tedious and unreliable, of course, but provides some assurance.
\ProveIt{} itself ensures that each derivation step is valid and that there are no circular dependencies, but \ProveIt{} is a large and complicated piece of software that should not be blindly trusted.
As future work, we plan to develop a lightweight, independent checker that is also open-source with detailed documentation to garner more trust.
We aim to make this checker simple enough and with enough detail that other researchers could develop their own and further their confidence in the system and resulting proof certificates.
This will demonstrate that \ProveIt{} satisfies the \textit{de Bruijn} criterion \cite{Barendregt_Barendsen:2002_Autarkic_Computations_in_Formal_Proofs, Barendregt_Wiedijk:2005_challenge_of_computer_mathematics}.

\subsection{Theorem dependencies}
\label{subsec:theorem_dependencies}

Having a proof of a given statement means that we have produced a sequence of small steps that derive the desired statement from more fundamental and accepted statements.
In our case, we derive statements that are specific to the quantum phase estimation algorithm from statements that are more fundamental facts in logic and set theory, number theory, trigonometry, linear algebra, statistics, and quantum mechanics.
The complete list of the dependencies for each proven statement can be viewed from our website \cite{website:Prove-It} by navigating to the theorems page of \texttt{proveit.physics.quantum.QPE} package, clicking on the corresponding theorem name, and then clicking on the ``dependencies'' link.

% ============================== %
% TABLE of Theorem Dependencies  %
% ============================== %
\newcommand\mpwidth{8cm}
\begin{table*}[htb!]
\scriptsize
\begin{tabular}{l|c|c}
\hline
\thead{Theory package} &
\thead{\# dependent\\axioms,\\theorems}
& \thead{Examples} \\
\hline
\texttt{core\_expr\_types} & 9, 22 &
  \begin{minipage}[t]{\mpwidth}{\scriptsize
  $|()| = |(1, 2, \ldots, 0)|$\\[0.0in]
  $\forall_{n\in\mathbb{N}}\big[\forall_{x_{1},\ldots,x_{n},y_{1},\ldots,y_{n}|(x_{1},\ldots,x_{n})=(y_{1},\ldots,y_{n})}$\\$\big(f(x_{1},\ldots,x_{n})=f(y_{1},\ldots,y_{n})\big)\big]$}
  \end{minipage}\\[0.2in]
\hline
\texttt{linear\_algebra} & 4, 23 &
  \begin{minipage}[t]{\mpwidth}{\scriptsize $\forall_{K}\big[\forall_{V\in_{c}\text{VecSpaces}(K)}[\forall_{x,y\in V} (x+y\in V)]\big]$\\[0.0in]
  $\forall_{K}\big[\forall_{V\in_{c}\text{VecSpaces}(K)}[\forall_{a\in K}[\forall_{x\in V} (ax\in V)]]\big]$} \end{minipage}\\[0.2in]
\hline
\texttt{logic.booleans} & 21, 9 &
  \begin{minipage}[t]{\mpwidth}{\scriptsize $\forall_{m\in\mathbb{N}^{+}}\big[\forall_{A_{1}, \ldots, A_{m}|A_{1}, \ldots, A_{m}}\big(A_{1} \land \ldots \land A_{m}\big)\big]$\\[0.0in]
  $\forall_{m\in\mathbb{N}^{+}}\big[\forall_{A_{1}, \ldots, A_{m}|A_{1}, \ldots, A_{m}}\big(A_{1} \lor \ldots \lor A_{m}\big)\big]$} \end{minipage}\\[0.2in]
\hline
\texttt{logic.equality} & 6, 1 &
  \begin{minipage}[t]{\mpwidth}{\scriptsize $\forall_{x}(x=x)$\quad$\forall_{x,y}((y=x)=(x=y))$\quad$\forall_{x,y,z|x=y,y=z}(x=z)$\\[0.0in]
  $\forall_{f,x,y|x=y}(f(x)=f(y))$\quad$\forall_{a,b,c,d | a=b, b=c, c=d}(a=d)$} \end{minipage}\\[0.2in]
\hline
\texttt{logic.sets} & 5, 14 &
  \begin{minipage}[t]{\mpwidth}{\scriptsize $\forall_{A,B}[(A\subseteq B) = (\forall_{x\in A}(x \in B))]$\\[0.0in]
  $\forall_{A,B}[(A \subset B) = (A\subseteq B \land A \not\cong B)]$\\[0.0in]
  $\forall_{f, a}\big([\forall_{x,y\in A | x \ne y}(f(x)\!\!\ne\!\! f(y))] \!\!\Rightarrow\!\!(f\!\! \in\!\! [A\!\! \xrightarrow[\text{onto}]{\text{1-to-1}} f^{\rightarrow}(A)])\big)$
  } \end{minipage}\\
\hline
\texttt{numbers.addition} & 0, 44 & 
  \begin{minipage}[t]{\mpwidth}{\scriptsize $\forall_{A,B\in\mathbb{C}}[A+B \in \mathbb{C}]$\quad$\forall_{A,B\in\mathbb{C}}[A+B = B+A]$\quad$\forall_{a\in\mathbb{C}}[a+0 = a]$\\[0.0in]
  $\forall_{a,x,y\in\mathbb{R}|x<y}[(x+a) < (y+a)]$\quad$\forall_{a,b|a < b}[a-b < 0]$\\[0.0in]
  $\forall_{a, b|a\ne b}[a-b \ne 0]$\quad$\forall_{a,b,c\in\mathbb{C}|(a+b)=c}[c-b = a]$
  } \end{minipage}\\
\hline
\texttt{numbers.division} & 0, 23 & 
  \begin{minipage}[t]{\mpwidth}{\scriptsize $\forall_{a,b\in\mathbb{C}|b\ne 0}[\frac{a}{b} \in \mathbb{C}]$\quad$\forall_{x\in\mathbb{C}|x\ne 0}[\frac{x}{x}=1]$\quad$\forall_{x\in\mathbb{C}|x\ne0}[\frac{0}{x} = 0]$\\[0.0in]
  $\forall_{x,y\in\mathbb{X}|y\ne0}[\frac{-x}{y}=-\frac{x}{y}]$\quad$\forall_{a,x,y\in\mathbb{R}^{+}|x<y}[\frac{a}{x} > \frac{a}{y}]$
  } \end{minipage}\\[0.2in]
\hline
\texttt{numbers.exponentiation} & 0, 41 & 
  \begin{minipage}[t]{\mpwidth}{\scriptsize $\forall_{a\in\mathbb{R}^{+},b,c\in\mathbb{C}}[(a^{b})^{c}=a^{bc}]$
  \;\;
  $\forall_{x\in\mathbb{C}}[x^{1} = x]$
  \;
  $\forall_{a,x,y\in\mathbb{R}|x=y}[x^a = y^a]$
  \\[0.0in]
  $\forall_{x\in\mathbb{C}}[\forall_{n\in\mathbb{N}}(x^{2n}=(-x)^{2n})]$
  \quad
  $\forall_{a,b\in\mathbb{C}}[a^{b}\in\mathbb{C}]$
  \\[0.0in]
  $\forall_{a,x,y\in\mathbb{R}^{+}|a>1, x<y}[a^x < a^y]$
  } \end{minipage}\\[0.2in]
\hline
\texttt{numbers.multiplication} & 0, 33 & 
  \begin{minipage}[t]{\mpwidth}{\scriptsize $\forall_{a,b\in\mathbb{C}}[a\cdot b = b\cdot a]$
  \;\;
  $\forall_{x\in\mathbb{C}}[1\cdot x = x]$
  \;
  $\forall_{a,b\in\mathbb{C}}[a\cdot b\in\mathbb{C}]$
  \\[0.0in]
  $\forall_{a,b\in\mathbb{C}^{\ne 0}}[a\cdot b\in\mathbb{C}^{\ne 0}]$
  \quad
  $\forall_{x,y\in\mathbb{C}}[(-x)\cdot y = -(x\cdot y)]$
  \\[0.0in]
  $\forall_{a,x,y\in\mathbb{R}|x<y, a<0}[a\cdot x > a\cdot y]$
  } \end{minipage}\\[0.2in]
\hline
\texttt{numbers.negation} & 0, 16 & 
  \begin{minipage}[t]{\mpwidth}{\scriptsize $\forall_{a\in\mathbb{C}}[(-a)\in\mathbb{C}]$
  \;\;
  $\forall_{a,b\in\mathbb{C}}[-(a+b)=(-a-b)]$
  \;
  $\forall_{x\in\mathbb{C}}[-(-x)=x]$
  \\[0.0in]
  $\forall_{x\in\mathbb{C}}[(-1)\cdot x = (-x)]$
  \quad
  $\forall_{x,y\in\mathbb{R}|x<y}[(-x)>(-y)]$
  \quad
  $-0 = 0$
  } \end{minipage}\\[0.2in]
\hline
\texttt{numbers.number\_sets} & 2, 83 & 
  \begin{minipage}[t]{\mpwidth}{\scriptsize
  $\forall_{n\in\mathbb{N}}[(n+1)\in\mathbb{N}]$
  \;\;
  $0\in\mathbb{N}$
  \;\;
  $0\in\mathbb{C}$
  \;
  $i\in\mathbb{C}$
  \;
  $\mathbb{R}\subset\mathbb{C}$
  \;
  $\mathbb{Z}\subset\mathbb{C}$
  \\[0.0in]
  $\forall_{a,b\in\mathbb{Z}}[\{a, \ldots, b\}\subset\mathbb{Z}]$
  \quad
  $\forall_{a,b\in\mathbb{Z}}[\forall_{n\in\{a, \ldots, b\}}(n \le b)]$
  \quad
  $\mathbb{N}\subset\mathbb{Z}$
  } \end{minipage}\\[0.2in]
\hline
\texttt{numbers.numerals} & 2, 36 & 
  \begin{minipage}[t]{\mpwidth}{\scriptsize
  $1 = 0 + 1$
  \;\;
  $2 = 1 + 1$
  \;\;
  $0 < 1$
  \;
  $1 < 2$
  \;
  $2\cdot 2 = 4$
  \;
  $3 \in \mathbb{N}$
  \\[0.0in]
  $1\in\mathbb{N}^{+}$
  \quad
  $2\in\mathbb{N}^{+}$
  \quad
  $3\in\mathbb{N}^{+}$
  \quad
  $\forall_{a,b,c}(|(a, b, c)| = |(1, 2, \ldots, 3)|)$
  } \end{minipage}\\[0.2in]
\hline
\begin{minipage}[t]{3.8cm}\raggedright{\texttt{numbers.ordering}, \texttt{numbers.rounding}, \& \texttt{numbers.summation}
  } \end{minipage}
  & 7, 29 & 
  \begin{minipage}[t]{\mpwidth}{\scriptsize
  $\forall_{x,y,z|x<y,y<z}[x < z]$
  \;\;
  $\forall_{x\in\mathbb{R}}\big(\lfloor x \rfloor \le x\big)$
  \;\;
  $\forall_{x\in\mathbb{R}}\big(\texttt{round}(x)\in\mathbb{Z}\big)$
  \\[0.0in]
  $\forall_{x\in\mathbb{R}}((x-\lfloor x \rfloor)<1)$
  \quad
  $\forall_{x\in\mathbb{R}}(\texttt{round}(x)=\lfloor x + \frac{1}{2} \rfloor)$
  \\[0.0in]
  $\forall_{f}\big(\forall_{a,b,c\in\mathbb{Z}}\big(
  \sum_{x=a}^{b}f(x) = \sum_{x=a+c}^{b+c}f(x-c)\big)\big)$
  } \end{minipage}\\[0.2in]
\hline
\begin{minipage}[t]{3.8cm}\raggedright{misc. other\\ \texttt{numbers} subpackages}\end{minipage}
  & 0, 29 & 
  \begin{minipage}[t]{\mpwidth}{\scriptsize
  $\forall_{a,b\in\mathbb{C}}[|a-b| = |b-a|]$
  \;\;
  $\forall_{\theta\in\mathbb{R}}(|e^{i\theta}| = 1)$
  \;\;
  \\[0.0in]
  $\big[x \mapsto \frac{1}{x^{2}}\big]\in\texttt{MonDecFuncs}(\mathbb{R}^{+})$
  \quad
  $\forall_{a,b\in\mathbb{R}^{+}|a \le b}\big((\int_a^{b} \frac{1}{\ell^{2}} d\ell) \le \frac{1}{a}\big)$
  } \end{minipage}\\[0.2in]
\hline
\texttt{physics.quantum} & 3, 44 &
  \begin{minipage}[t]{\mpwidth}{\scriptsize
  $\ket{0}\in\mathbb{C}^{2}$
  \;\;
  $\ket{1}\in\mathbb{C}^{2}$
  \;\;
  $||\ket{0}|| = 1$
  \;\;
  $||\ket{1}|| = 1$
  \;\;
  $\braket{0}{1} = 0$
  \\[0.0in]
  $\forall_{n\in\mathbb{N}^{+}}\big(\texttt{FT}_{n}^{\dagger} \in U(2^{n})\big)$
  \quad
  $\forall_{n\in\mathbb{N}^{+}}\big(C^{n} \in_{c}\texttt{HilbertSpaces}\big)$
  } \end{minipage}\\[0.2in]
\hline
\texttt{statistics} & 1, 7 & 
  \begin{minipage}[t]{\mpwidth}{\scriptsize $\forall_{\Omega\in\text{SampleSpaces}}\left[\forall_{X\in\Omega}\big(\text{Pr}(X)\in[0,1]\big)\right]$} \end{minipage}\\[0.1in]
\hline
\texttt{trigonometry} & 0, 5 &  
  \begin{minipage}[t]{\mpwidth}{\scriptsize
  $\forall_{\theta\in\mathbb{R}^{+}}\big(\sin(\theta)<\theta\big)$
  \quad
  $\forall_{a,b\in\mathbb{R}}\left(|e^{ia}-e^{ib}|=2\sin(\frac{|a-b|}{2})\right)$} \end{minipage}\\
\hline
\end{tabular}
\vspace{0.1in}
\caption{Combined number of axioms and (unproven) theorems in each theory package (along with some examples) upon which the three main QPE theorems, \texttt{qpe\_exact}~(\ref{eq:thm_qpe_exact}), \texttt{qpe\_best\_guarantee}~(\ref{eq:thm_qpe_best_guarantee}), and \texttt{qpe\_precision\_guarantee}~(\ref{eq:thm_qpe_precision_guarantee}), logically depend at the time of this writing.
} \vspace{-0.2in}
\label{tab:dependency_examples}
\end{table*}

% ============================ %
% Table of Proof Steps, for    %
% MAIN and major theorems      %
% ============================ %
\begin{table*}[hbt!]
\scriptsize
% temporarily reset arraystretch
% from default 1 to 1.25 to provide
% more space in rows
\renewcommand{\arraystretch}{1.25}
\begin{tabular}{l|c|c}
\hline
\thead{\footnotesize QPE theorem}
& \thead{\footnotesize\# of nb\\ deriv \\commands}
& \thead{\footnotesize\# steps\\in formal\\proof}
\\
\hline
\hline
\texttt{\_psi\_t\_output}~(\ref{eq:thm_psi_t_output})
& 11
& 752
\\
\hline
\texttt{\_Psi\_output}~(\ref{eq:thm_Psi_output})
& 10
& 411
\\
\hline
\texttt{\_sample\_space\_bijection}~(\ref{eq:thm_sample_space_bijection})
& 4
& 273
\\
\hline
\texttt{\_outcome\_prob}~(\ref{eq:thm_outcome_prob})
& 6
& 392
\\
\hline
\makecell[lt]{
  \texttt{\_Omega\_is\_sample\_space}~(\ref{eq:thm_Omega_is_sample_space})
  $\Omega \underset{c}{\in} \texttt{SampleSpaces}$
}
& 7
& 29
\\
\hline
\makecell[lt]{
  \texttt{\_psi\_t\_formula}~(\ref{eq:thm_psi_t_formula})\\\hspace{0.1in}
  $\forall_{t\in\mathbb{N}^{+}}
  \left(
  \ket{\psi_{t}}
  =
  (
  \frac{1}{2^{t/2}}
  (
  \sum_{k=0}^{2^t - 1}
  e^{2\pi i \varphi k} \ket{k}_{t}
  )
  )
  \right)$
}
& 24
& 685
\\
\hline
\makecell[lt]{
  \texttt{\_alpha\_m\_evaluation}~(\ref{eq:thm_alpha_m_evaluation})\\\hspace{0.1in}
  $\forall_{m\in\{0,\ldots,2^{t}-1\}}
  \left(
  \alpha_{m}
  =
  \frac{1}{2^{t}}
  \sum_{k=0}^{2^{t}-1}
  \left(
  e^{-\frac{2 \pi i k m}{2^{t}}}e^{2 \pi i \varphi k}
  \right)
  \right)$
}
& 8
& 404
\\
\hline
\makecell[lt]{
  \texttt{\_alpha\_ideal\_case}~(\ref{eq:thm_alpha_ideal_case})\hspace{0.1in}
  $
  \left(
  2^{t}\varphi\in\{0,\ldots,2^{t}-1\}
  \right)
  \Rightarrow
  \left(
  \alpha_{2^{t}\varphi}
  =
  1
  \right)$
}
& 2
& 195
\\
\hline
\hline
\rowcolor{lightgray}
\texttt{qpe\_exact}~(\ref{eq:thm_qpe_exact})
& 15
& 161
\\
\hline
\hline
\makecell[lt]{
  \texttt{\_alpha\_m\_mod\_evaluation}~(\ref{eq:thm_alpha_m_mod_evaluation})\\\hspace{0.1in}
  $
  \forall_{m\in\mathbb{Z}}
  \left(
  \alpha_{m\,\text{mod}\,2^{t}}
  =
  \frac{1}{2^{t}}
  \sum_{k=0}^{2^{t}-1}
  \left(
  e^{-\frac{2 \pi i k m}{2^{t}}} e^{2 \pi i \varphi k}
  \right)
  \right)$
}
& 12
& 161
\\
\hline
\makecell[lt]{
  \texttt{\_alpha\_m\_mod\_as\_geometric\_sum}~(\ref{eq:thm_alpha_m_mod_as_geometric_sum})\\\hspace{0.1in}
  $
  \forall_{m\in\mathbb{Z}}
  \left(
  \alpha_{m\,\text{mod}\,2^{t}}
  =
  \frac{1}{2^{t}}
  \sum_{k=0}^{2^{t}-1}
  \left(
  e^{2 \pi i (\varphi - \frac{m}{2^{t}})}
  \right)^{k}
  \right)
  $
}
& 8
& 189
\\
\hline
\makecell[lt]{
  \texttt{\_best\_guarantee\_delta\_nonzero}~(\ref{eq:thm_best_guarantee_delta_nonzero})\hspace{0.1in}
  $
  \left(
  \delta_{b_{r}} \ne 0
  \right)
  \Rightarrow
  \left(
  |\alpha_{b_{r}\,\text{mod}\,2^{t}}|^{2}
  >
  \frac{4}{\pi^{2}}
  \right)
  $
}
& 23
& 625
\\
\hline
\makecell[lt]{
  \texttt{\_best\_guarantee}~(\ref{eq:thm_best_guarantee})\hspace{0.1in}
  $
  |\alpha_{\texttt{round}(2^{t}\varphi)\,\text{mod}\,2^{t}}|^{2}
  >
  \frac{4}{\pi^{2}}
  $
}
& 24
& 259
\\
\hline
\hline
\rowcolor{lightgray}
\texttt{qpe\_best\_guarantee}~(\ref{eq:thm_qpe_best_guarantee})
& 4
& 19
\\
\hline
\hline
\makecell[lt]{
  \texttt{\_alpha\_summed}~(\ref{eq:thm_alpha_summed})\\[-0.075in]\hspace{0.1in}
  $
  \forall_{\ell\in\{-2^{t-1}+1,\ldots,2^{t-1}\}|\ell\ne 0}
  \left(
  \alpha_{b_{\!f}\oplus\ell}
  =
  \frac{1}{2^{t}}
  \cdot
  \frac{1-exp({2 \pi i (2^{t}\delta_{b_{\!f}} - \ell)})}{1 - exp({2 \pi i (\delta_{b_{\!f}} - \frac{\ell}{2^{t}}))}}
  \right)
  $
}
& 12
& 417
\\
\hline
\makecell[lt]{
  \texttt{\_fail\_sum\_prob\_conds\_equiv\_lemma}~(\ref{eq:thm_fail_sum_prob_conds_equiv_lemma})
}
& 12
& 412

\\
\hline
\makecell[lt]{
  \texttt{\_fail\_sum}~(\ref{eq:thm_fail_sum})\\\hspace{0.1in}
  $
  \forall_{e \in \{1, \ldots, 2^{t-1}-2 \} }$
  \\\hspace{0.1in}
  $
  \Big(
  [P_{\text{fail}}](e)
  =
  \sum_{\ell=-2^{t-1}+1}^{-(e+1)}|\alpha_{b_{\!f}\oplus\ell}|^{2}
  +
  \sum_{\ell=e+1}^{2^{t-1}}|\alpha_{b_{\!f}\oplus\ell}|^{2}
  \Big)
  $
}
& 24
& 453
\\
\hline
\makecell[lt]{
  \texttt{\_alpha\_sqrd\_upper\_bound}~(\ref{eq:thm_alpha_sqrd_upper_bound})\\\hspace{0.1in}
  $
  \forall_{\ell\in\{-2^{t-1}+1,\ldots,2^{t-1}\}|\ell\ne 0}
  \left(
  |\alpha_{b_{\!f}\oplus\ell}|^{2}
  \le
  \frac{1}{4(\ell-2^{t}\delta_{b_{\!f}})^{2}}
  \right)
  $
}
& 25
& 706
\\
\hline
\makecell[lt]{
  \texttt{\_failure\_upper\_bound\_lemma}~(\ref{eq:thm_failure_upper_bound_lemma})
}
& 16
& 303
\\
\hline
\makecell[lt]{
  \texttt{\_failure\_upper\_bound}~(\ref{eq:thm_failure_upper_bound})\\\hspace{0.1in}
  $
  \forall_{e\in\{1,\ldots,2^{t-1}-2\}}
  \left(
  [P_{\text{fail}}](e)
  \le
  \frac{1}{2e}
  +
  \frac{1}{4e^{2}}
  \right)
  $
}
& 28
& 547
\\
\hline
\makecell[lt]{
  \texttt{\_success\_complements\_failure}~(\ref{eq:thm_success_complements_failure})\\\hspace{0.1in}
  $
  \forall_{e\in\{1,\ldots,2^{t-1}-2\}}
  \left(
  [P_{\text{success}}](e)
  =
  1 - [P_{\text{fail}}](e)
  \right)
  $
}
& 5
& 70
\\
\hline
\makecell[lt]{
  \texttt{\_precision\_guarantee\_lemma\_01}\\\hspace{0.1in}
  $
  \left(
  1
  - \frac{1}{2 (2^{t-n} - 1)}
  - \frac{1}{4 (2^{t-n} - 1)^{2}}
  \right)
  >
  (1 - \epsilon)
  $
}
& 27
& 589
\\
\hline
\makecell[lt]{
  \texttt{\_precision\_guarantee\_lemma\_02}\\\hspace{0.1in}
  $
  \forall_{m\in\{0,\ldots,2^{t}-1\} \big| |m - b_{\!f}|_{\text{mod}\,2^{t}}\le (2^{t-n}-1)}
  \left(
  |\frac{m}{2^{t}}-\varphi|_{\text{mod}\,1}
  \le
  2^{-n}
  \right)
  $
}
& 20
& 282
\\
\hline
\makecell[lt]{
  \texttt{\_precision\_guarantee}~(\ref{eq:thm_precision_guarantee})
}
& 9
& 139
\\
\hline
\hline
\rowcolor{lightgray}
\texttt{qpe\_precision\_guarantee}~(\ref{eq:thm_qpe_precision_guarantee})
& 2
& 2
\\
\hline
\hline
\end{tabular}
\vspace{0.2in}
\caption{The number of manually-invoked derivation commands (\textit{e.g.}, the calling of specific methods or the instantiation of a theorem to derive a new judgment) in each proof notebook, and the number of steps (nodes in the formal proof tree) appearing in the corresponding formal proof for the three main QPE theorems (highlighted) and major supporting theorems in the QPE theory package. The ordering of the theorems parallels the overview in \S\ref{Sec:essence} with theorems depending only on others higher in the list.} 
\label{tab:proof_steps}
\end{table*}

Statements (axioms and theorems) are organized into theory packages in our system.  
Our universal QPE proofs derive from a number of statements from various packages.
Table~\ref{tab:dependency_examples} lists the number of statements from each theory package (along with some examples of such statements) that, at the time of this writing, were logically required to derive the formal proofs.
The majority of these statements are theorems that have not yet been proven in our system.  Each of these is marked as a ``conjecture'' in the formal proof steps.
It is a convenient feature in our system that conjectures may be posited and used (sometimes through automation) before they are proven, and then noted as unproven in the lists of dependencies.\footnote{This convenience requires no further action on the user's part, but is analogous to the explicit use of the \texttt{Accepted} command in \textit{Coq}.}
As these conjectures are eventually proven, the number of dependencies will decrease, eventually leaving a much smaller number of axiom dependencies.

% ============================ %
% Table of Proof Steps,        %
% for more minor theorems      %
% ============================ %
\begin{table*}[hbt]
\scriptsize
% temporarily reset arraystretch
% from default 1 to 1.5 to provide
% more space in rows
\renewcommand{\arraystretch}{1.5}
\begin{tabular}{l|c|c}
\hline
\thead{QPE theorem}
& \thead{\# of\\ deriv nb\\commands}
& \thead{\# steps\\in formal\\proof}
\\
\hline
\makecell[lt]{
  \texttt{\_two\_pow\_t\_is\_nat\_pos}\hspace{0.1in}
  $
  2^{t}\in\mathbb{N}^{+}
  $
}
& 1
& 7
\\
\hline
\makecell[lt]{
  \texttt{\_two\_pow\_t\_minus\_one\_is\_nat\_pos}\hspace{0.1in}
  $
  2^{t-1}\in\mathbb{N}^{+}
  $
}
& 1
& 20
\\
\hline
\makecell[lt]{
  \texttt{\_psi\_t\_ket\_is\_normalized\_vec}
  $
  \forall_{t\in\mathbb{N}^{+}}
  \left(
  \ket{\psi_{t}}\in\mathbb{C}^{2^{t}}
  \land
  ||\ket{\psi_{t}}|| = 1
  \right)
  $
}
& 4
& 329
\\
\hline
\makecell[lt]{
  \texttt{\_Psi\_ket\_is\_normalized\_vec}
  $
  \ket{\Psi} \in \mathbb{C}^{2^{t}}
  \land
  || \ket{\Psi} || = 1
  $
}
& 4
& 27
\\
\hline
\makecell[lt]{
  \texttt{\_best\_floor\_is\_int}\hspace{0.1in}
  $
  b_{f} \in \mathbb{Z}
  $
}
& 2
& 16
\\
\hline
\makecell[lt]{
  \texttt{\_best\_floor\_is\_in\_m\_domain}\hspace{0.1in}
  $
  b_{f} \in \{0, \ldots, 2^{t}-1\}
  $
}
& 8
& 56
\\
\hline
\makecell[lt]{
  \texttt{\_best\_round\_is\_int}\hspace{0.1in}
  $
  b_{r}\in\mathbb{Z}
  $
}
& 2
& 16
\\
\hline
\makecell[lt]{
  \texttt{\_e\_value\_ge\_two}\hspace{0.1in}
  $
  \left(2^{t-n}-1\right) \ge 2
  $
}
& 8
& 249
\\
\hline
\makecell[lt]{
  \texttt{\_e\_value\_in\_e\_domain}\hspace{0.1in}
  $
  \left(2^{t-n}-1\right) \in \{1,\ldots,2^{t-1}-2\}
  $
}
& 11
& 317
\\
\hline
\makecell[lt]{
  \texttt{\_mod\_add\_closure}
  \hspace{0.1in}
  $
  \forall_{a,b \in \mathbb{Z}}
  \left(
  a \oplus b
  \in \{0, \ldots, 2^{t}-1\}
  \right)
  $
}
& 3
& 12
\\
\hline
\makecell[lt]{
  \texttt{\_phase\_is\_real}
  \hspace{0.1in}
  $
  \varphi \in \mathbb{R}
  $
}
& 1
& 12
\\
\hline
\makecell[lt]{
  \texttt{\_delta\_b\_is\_real}
  \hspace{0.1in}
  $
  \forall_{b\in\mathbb{Z}}
  \left(
  \delta_{b} \in
  \mathbb{R}
  \right)
  $
}
& 3
& 25
\\
\hline
\makecell[lt]{
  \texttt{\_scaled\_delta\_b\_floor\_in\_interval}
  \hspace{0.1in}
  $
  2^{t}\delta_{b_{\!f}} \in
  [0, 1)
  $
}
& 9
& 108
\\
\hline
\makecell[lt]{
  \texttt{\_scaled\_delta\_b\_round\_in\_interval}
  \hspace{0.1in}
  $
  2^{t}\delta_{b_{r}} \in
  [-\frac{1}{2}, \frac{1}{2})
  $
}
& 15
& 167
\\
\hline
\makecell[lt]{
  \texttt{\_delta\_b\_in\_interval}
  \hspace{0.1in}
  $
  \forall_{b\in\{b_{\!f}, b_{r}\}}
  \left(
  \delta_{b}\in
  (-\frac{1}{2}, \frac{1}{2}]
  \right)
  $
}
& 11
& 324
\\
\hline
\makecell[lt]{
  \texttt{\_alpha\_are\_complex}
  \hspace{0.1in}
  $
  \forall_{m\in\{0, \ldots, 2^{t}-1\}}
  \left(
  \alpha_{m} \in \mathbb{C}
  \right)
  $
}
& 2
& 20
\\
\hline
\makecell[lt]{
  \texttt{\_delta\_b\_is\_zero\_or\_non\_int}
  \hspace{0.1in}
  $
  \forall_{b\in\{b_{\!f}, b_{r}\}}
  \left(
  \delta_{b} = 0
  \lor
  \delta_{b} \notin \mathbb{Z}
  \right)
  $
}
& 15
& 150
\\
\hline
\makecell[lt]{
  \texttt{\_scaled\_delta\_b\_is\_zero\_or\_non\_int}
  $
  \forall_{b\in\{b_{\!f}, b_{r}\}}
  \left(
  2^{t}\delta_{b} = 0
  \lor
  2^{t}\delta_{b} \notin \mathbb{Z}
  \right)
  $
}
& 16
& 167
\\
\hline
\makecell[lt]{
  \texttt{\_scaled\_delta\_b\_not\_eq\_nonzeroInt}
  $
  \forall_{b\in\{b_{\!f}, b_{r}\}, \ell\in\mathbb{Z}|\ell\ne 0}
  \left(
  2^{t}\delta_{b} \ne \ell
  \right)
  $
}
& 9
& 31
\\
\hline
\makecell[lt]{
  \texttt{\_delta\_b\_not\_eq\_scaledNonzeroInt}
  $
  \forall_{b\in\{b_{\!f}, b_{r}\}, \ell\in\mathbb{Z}|\ell\ne 0}
  \left(
  \delta_{b} \ne \frac{\ell}{2^{t}}
  \right)
  $
}
& 6
& 56
\\
\hline
\makecell[lt]{
  \texttt{\_delta\_b\_floor\_diff\_in\_interval}
  \hspace{0.1in}
  $
  \forall_{l\in\{-2^{t-1}+1, \ldots, 2^{t-1}\}}
  \left(
  \delta_{b_{\!f}} - \frac{\ell}{2^{t}}
  \in
  [-\frac{1}{2}, \frac{1}{2})
  \right)
  $
}
& 13
& 200
\\
\hline
\makecell[lt]{
  \texttt{\_non\_int\_delta\_b\_diff}\hspace{0.1in}
  $
  \forall_{b\in\{b_{\!f},b_{r}\},\ell\in\{-2^{t-1}+1,\ldots,2^{t-1}\}|\ell\ne 0}
  \left(
  \delta_{b} - \frac{\ell}{2^{t}}
  \notin \mathbb{Z}
  \right)
  $
}
& 14
& 233
\\
\hline
\makecell[lt]{
  \texttt{\_scaled\_abs\_delta\_b\_floor\_diff\_interval}\\\hspace{0.1in}
  $
  \forall_{\ell\in\{-2^{t-1}+1,\ldots,2^{t-1}\}|\ell\ne 0}
  \left(
  \pi
  |\delta_{b_{\!f}} - \frac{\ell}{2^{t}}|
  \in (0, \frac{\pi}{2}]
  \right)
  $
}
& 9
& 93
\\
\hline
\makecell[lt]{
  \texttt{\_pfail\_in\_real}\hspace{0.1in}
  $
  \forall_{e\in\{1,\ldots,2^{t-1}-2\}}
  \left(
  [P_{\text{fail}}](e) \in \mathbb{R}
  \right)
  $
}
& 6
& 44
\\
\hline
\makecell[lt]{
  \texttt{\_phase\_from\_best\_with\_delta\_b}\hspace{0.1in}
  $
  \forall_{b\in\mathbb{Z}}
  \left(
  \varphi
  =
  \frac{b}{2^{t}} + \delta_{b}
  \right)
  $
}
& 3
& 43
\\
\hline
\makecell[lt]{
  \texttt{\_modabs\_in\_full\_domain\_simp}
  $
  \forall_{-2^{t-1}+1,\ldots,2^{t-1}}
  \left(
  |\ell|_{\text{mod}\,2^{t}}
  = |\ell|
  \right)
  $
}
& 6
& 213
\\
\hline
\end{tabular}
\vspace{0.1in}
\caption{The number of manually-invoked derivation commands in the proof notebook, and the number of steps appearing in the formal proof, for each of the more minor theorems in the QPE theory package. These minor theorems typically involve simple algebraic relationships, closure properties, and domain specifications.} 
\label{tab:proof_steps_minor_thms}
\end{table*}

\begin{figure}%[tb]
  \captionsetup{font=footnotesize}
  \includegraphics[width = 0.45\textwidth]{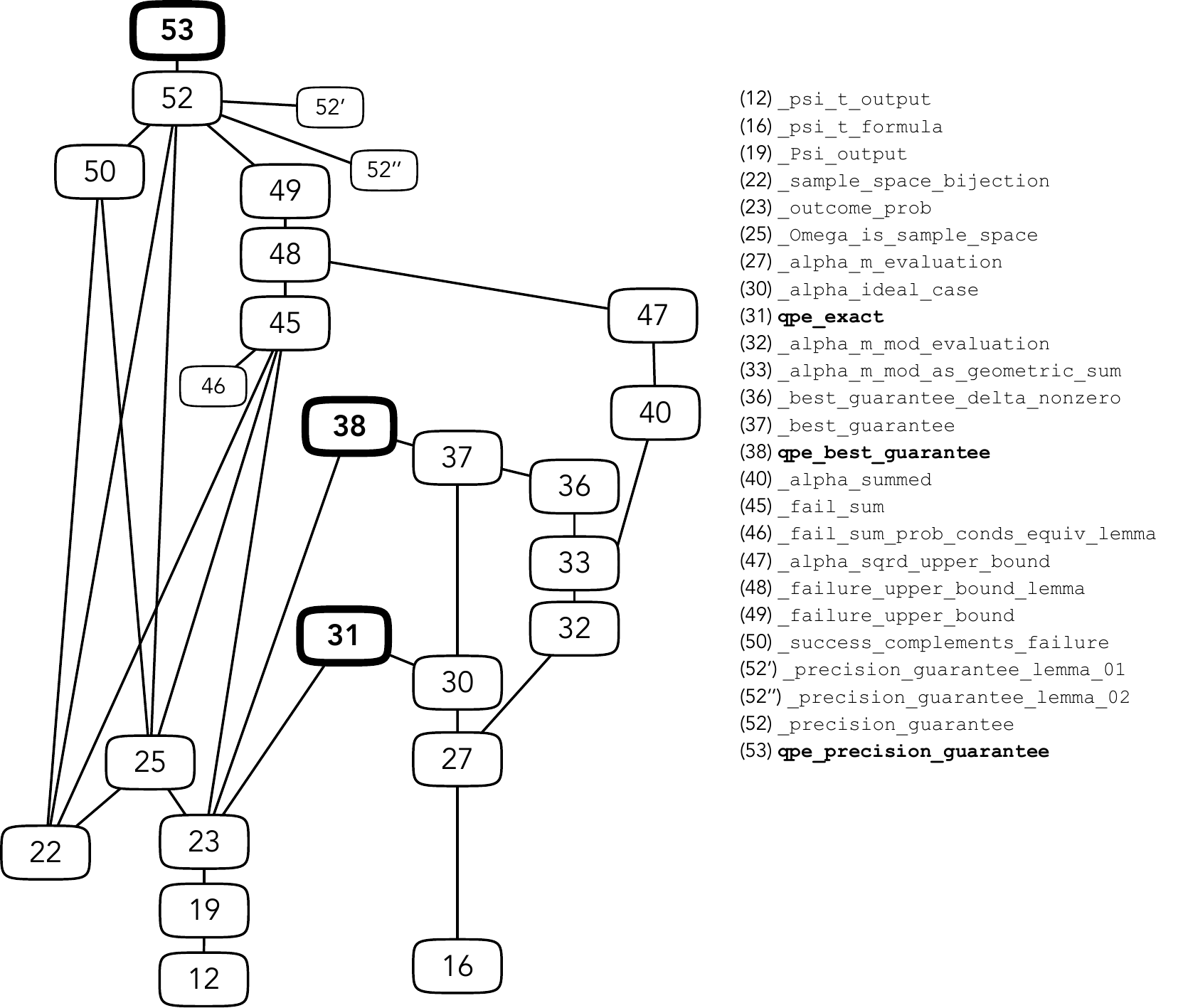}
\caption{Graphical depiction of the dependencies among the ``main'' theorems appearing in Table~\ref{tab:proof_steps}. Numbering corresponds to equation numbers in the text (except for the primed numbers appearing in smaller boxes, for lemmas that don't explicitly appear in the text). The three main universal theorems appear in bold. An upward link means that the higher theorem explicitly depends on the lower theorem.
}
\label{fig:qpe_theorems_dependencies_tree}
\end{figure}

Moreover, many axioms that currently appear as dependencies are merely label definitions and thus each serves as a (conservative) ``extension by definition'' \cite[pp 102--103]{Mendelson:2015}\cite[pp 57--61]{Schoenfield_Math_Logic:1967}.
In a future update of \ProveIt{}, we plan to distinguish conservative definitions from axioms.  After proving the unique existence of the label as it is defined, the definition will no longer be a dependency when used to prove a theorem unless that particular label appears in that particular theorem.  Often in mathematics, convenient notation is used incidentally in the process of constructing a proof, but the validity of the proof is independent of this extraneous notational convenience. We want the dependencies to properly reflect this fact.

Circular dependencies are prevented, both on the local level of a formal proof and the broader level of theorem dependencies.

\subsection{Lengths of the QPE proofs}
\label{subsec:derivation_steps}

As mentioned repeatedly, all of our proofs are viewable on the \ProveItWebsite{}.
All of the theorems in the QPE theory package (both universal and local) are listed in Tables~\ref{tab:proof_steps} and~\ref{tab:proof_steps_minor_thms}, along with the number of user-invoked derivation commands utilized in each Jupyter notebook, and the number of steps in the eventual formal proof produced by \ProveIt{} for that theorem. Table~\ref{tab:proof_steps} includes the three main QPE theorems (shown shaded) and the major supporting theorems, ordered to parallel the presentation in \S\ref{Sec:essence}. Table~\ref{tab:proof_steps_minor_thms} lists more minor theorems. Figure~\ref{fig:qpe_theorems_dependencies_tree} presents a graphical representation of the inter-dependencies among the main and supporting theorems found in Table~\ref{tab:proof_steps}.

The number of steps in a formal proof may be surprising.  A lot of details are required in a formal proof. 
It is also likely possible to substantially reduce the number of formal proof steps by making improvements in the implementation of the automation features that will be described in Sec.~\ref{Sec:interactive_proofs}.
We have focused our efforts on the capabilities of the automation features (to be discussed in \S\ref{Sec:interactive_proofs}) to minimize the number of derivation commands that are needed to generate the proof.
Although the number of formal proof steps listed in Table~\ref{tab:proof_steps} is only a rough estimate of some ideal number of steps, it is interested to note how it compares with the number of derivation commands.  A given derivation command can generate several formal proof steps in order to fill the gaps that result from skipping obvious steps in the sequence of derivation commands.

\subsection{Expression structures and \LaTeX{} formatting}
\label{subsec:expression_structures_and_latex_formatting}

\begin{figure}%[tb]
  \captionsetup{font=footnotesize}
  \includegraphics[width = 0.4\textwidth]{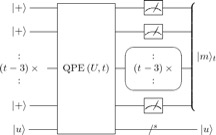}
  \\\vspace{0.2in}
  \includegraphics[width = 0.45\textwidth]{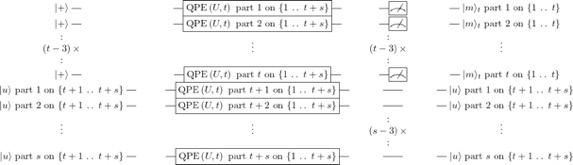}
\caption{An example circuit presentation in \ProveIt{} (top) and a stylized version of its underlying representation in the code (bottom) illustrating its organization as a sequence of rows of circuit components nested within a sequence of columns. See \S\ref{subsec:expression_structures_and_latex_formatting} for discussion.
}
\label{fig:example_circuit_and_representation}
\end{figure}

Expressions in Prove-It have an internal directed acyclic graph (DAG) structure that is critically important to its interpretation and the allowable proof steps.  The \LaTeX{} formatting makes expressions much more human readable, but it is important to note that this formatting is not guaranteed to completely disambiguate its internal structure.  In fact, there is really no safeguard in place to ensure that the formatting is true to its structure.  Therefore, for full end-to-end verification, in addition to checking the derivation steps and theorem dependencies, one should also check the structure of the final derived statement to ensure it is indeed the statement that was intended to be proven, and one should check the structures of the statements that were logically required (\textit{e.g.}, the axioms and conjectures).

Details about Prove-It's core expression types and structure, and their relation with derivation steps, are discussed in Ref.~\cite{Preprint:ProveIt}. \ProveIt{} has been developed particularly with quantum algorithms and quantum circuits in mind. Quantum circuit expressions in this work are good examples of the distinction between \LaTeX{} formatting and internal structures.
\ProveIt{} allows us to present quantum algorithms, and related axioms, definitions, and theorems in quantum circuit form (utilizing Q-Circuit \cite{Eastin_Fammia:2004_Q_Circuit}), all of which provide a more intuitive and ``direct'' understanding for the user, but
which then also have underlying representations in Prove-It’s core in terms of nested tuples or nested arrays of related circuit details, which allow the manipulation of circuits and instantiation of circuit-based axioms and theorems (see Fig.~\ref{fig:example_circuit_and_representation} for an example).

% ==================== %
% HOW DID WE CONSTRUCT %
% OUR PROOFS?          %
% ==================== %
\section{How did we construct our proofs?}
\label{Sec:interactive_proofs}

Our primary objective is to make formal theorem proving as straightforward and natural as it is to develop and present a more informal proof.
So it is vital that our system allows a user to skip obvious steps in a derivation and fill in details automatically.
In this section, we describe some of the system capabilities in this regard and provide examples from interactive proof steps in the QPE theory package.
The system capabilities include:
\begin{enumerate}
  
  \item[(1)] \textit{manual instantiation} of axioms and theorems;
  
  \item[(2)] the availability of a wide variety of \textit{user-invoked derivation} commands (tactics) that perform instantiations indirectly for a higher-level purpose (e.g., distributing, factoring, deducing new bounds, etc.);

  \item[(3)] \textit{goal-driven derivations} that may be invoked directly or indirectly; %capabilities, available as a user-initiated derivation tactic and also widely used ``under the hood'' in many aspects of the automation;
  
  \item[(4)] the use of \textit{canonical forms} to recognize equivalent expressions and guide automated proof steps;
  
  \item[(5)] completely automatic \textit{incidental derivations} performed whenever a judgment is established or assumptions are made;
  
  \item[(6)] and widespread, user-configurable \textit{automatic simplification} of expressions.
  
\end{enumerate}
As the examples will show, most proof steps involve a combination of these capabilities working in concert.

\subsection{Manual instantiation}

Any universally quantified statement (such as an explicitly-imported axiom or theorem) may be instantiated with specific instances of the quantified variables.  This can actually be quite powerful on its own, with an ability to expand ranges (\textit{e.g.}, $x_1 \cdot y_1 + \ldots + x_n \cdot y_n$), automatically attempt to prove required conditions of the quantifiers as needed (see \S\,\ref{subsec:goal-driven_derivations}), and take advantage of automatic simplification of the instantiated formula (see \S\,\ref{subsec:automatic_simplifications}), and otherwise providing an informative error message (specifying, \textit{e.g.}, that some instantiation condition could not be proven).

\begin{figure} %[!htb]
  \captionsetup{font=footnotesize}
  \includegraphics[width = 0.48\textwidth]{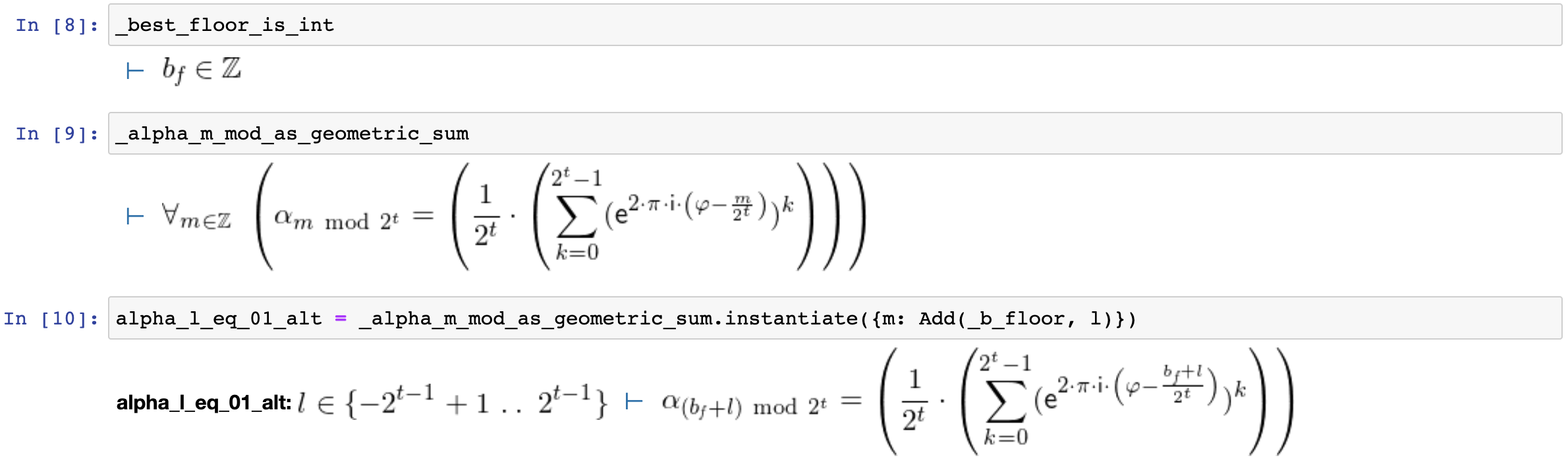}
\caption{Example of manual instantiation excerpted from \texttt{\_alpha\_summed}~(\ref{eq:thm_psi_t_formula}), instantiating \texttt{\_alpha\_m\_mod\_as\_geometric\_sum}~(\ref{eq:thm_alpha_m_mod_as_geometric_sum}) with $m = (b_{\!f}+\ell)$.
}
\label{fig:example_instantiations_alpha_summed}
\end{figure}

An example of manual (\textit{i.e.}, explicitly user-invoked) instantiation is shown in Fig.~\ref{fig:example_instantiations_alpha_summed}, excerpted from the proof notebook for \texttt{\_alpha\_summed}~(\ref{eq:thm_alpha_summed}).
The \texttt{\_alpha\_m\_mod\_as\_geometric\_sum} theorem shown in cell \texttt{[9]} is instantiated in cell \texttt{[10]} with quantified variable $m$ instantiated as the expression $(b_{\!f}+\ell)$, which means that the instantiation process must be able to prove or verify $(b_{\!f}+\ell)\in\mathbb{Z}$.
That required condition is automatically proven from $b_{\!f} \in \mathbb{Z}$ in cell \texttt{[8]} and the assumption $\ell\in\{-2^{t-1}+1,\ldots,2^{t-1}\}$ (which is a ``default'' assumption set earlier in the notebook); this assumption then appears in the judgment produced by the instantiation in cell \texttt{[10]}.

\begin{figure}[!htb]
  \captionsetup{font=footnotesize}
  \includegraphics[width = 0.5\textwidth]{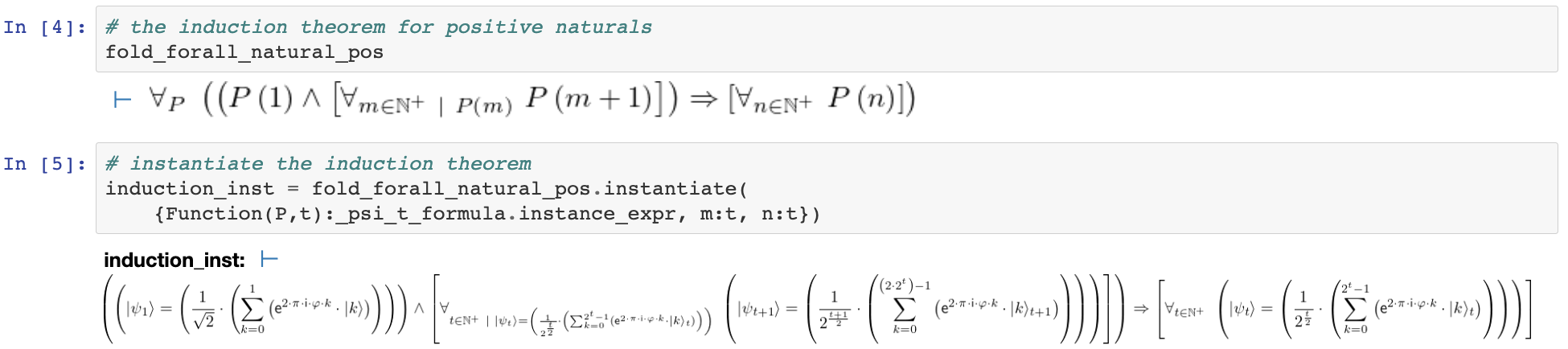}
  \\
  $\vdots$\vspace{0.1in}
  \\
  \includegraphics[width = 0.5\textwidth]{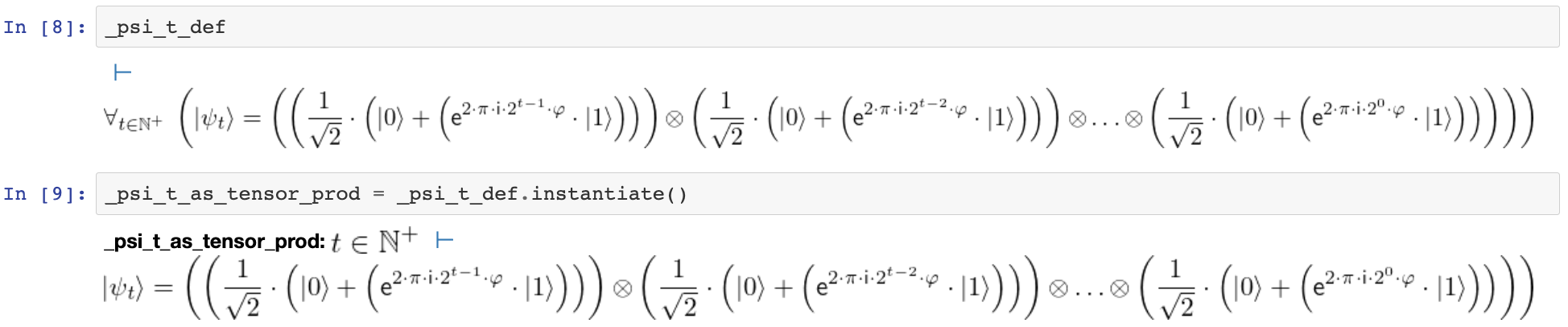}
\caption{Examples of manual instantiations, excerpted from the proof notebook for the \texttt{\_psi\_t\_formula}~(\ref{eq:thm_psi_t_formula}), showing the manual instantiation of the induction theorem for positive naturals (\texttt{cells} \texttt{4} and \texttt{5}) and of the \texttt{\_psi\_t\_def}~(\ref{eq:axiom_psi_t_def}) local QPE definition (\texttt{cells} \texttt{8} and \texttt{9}).
}
\label{fig:example_instantiations_psi_t_formula}
\end{figure}

Two more examples of manual instantiations are shown in Fig.~\ref{fig:example_instantiations_psi_t_formula}, excerpted from the proof notebook for \texttt{\_psi\_t\_formula}~(\ref{eq:thm_psi_t_formula}).
Cells \texttt{[4]} and \texttt{[5]} show the general induction theorem for positive naturals and its instantiation with the specifics for the \texttt{\_psi\_t\_formula} theorem (the variables $m$ and $n$ instantiated with $t$ and the function $P$ instantiated with the summation formula). Cells \texttt{[8]} and \texttt{[9]} respectively show \texttt{\_psi\_t\_def}~(\ref{eq:axiom_psi_t_def}) and its instantiation for $t=1$.

\subsection{User-invoked derivations (commands/tactics)}

\begin{figure}[tb]
  \captionsetup{font=footnotesize}
  \includegraphics[width = 0.5\textwidth]{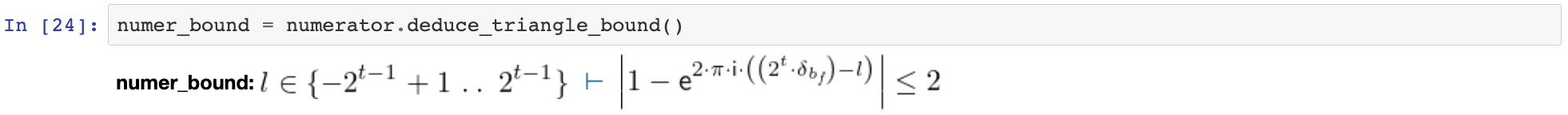}
  \\
  $\vdots$\vspace{0.1in}
  \\
  \includegraphics[width = 0.5\textwidth]{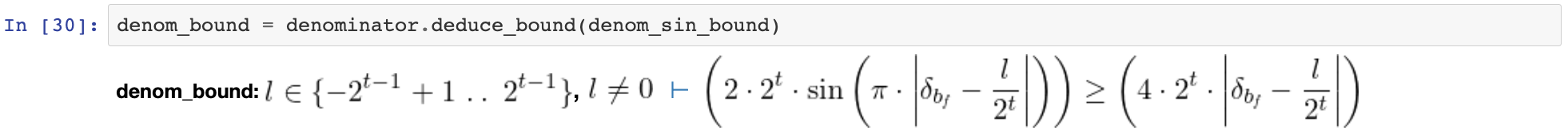}
  \\
  $\vdots$\vspace{0.1in}
  \\
  \includegraphics[width = 0.5\textwidth]{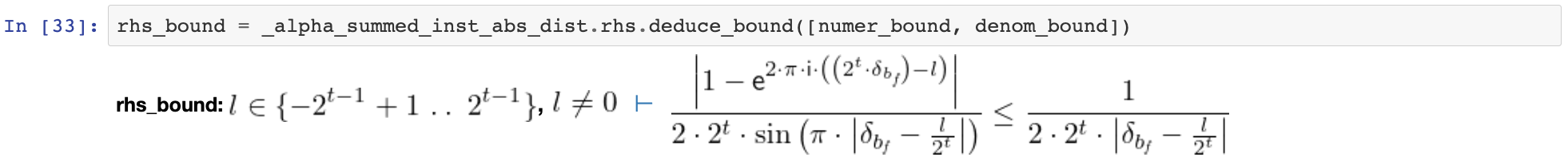}
  \\
  $\vdots$\vspace{0.1in}
  \\
  \includegraphics[width = 0.5\textwidth]{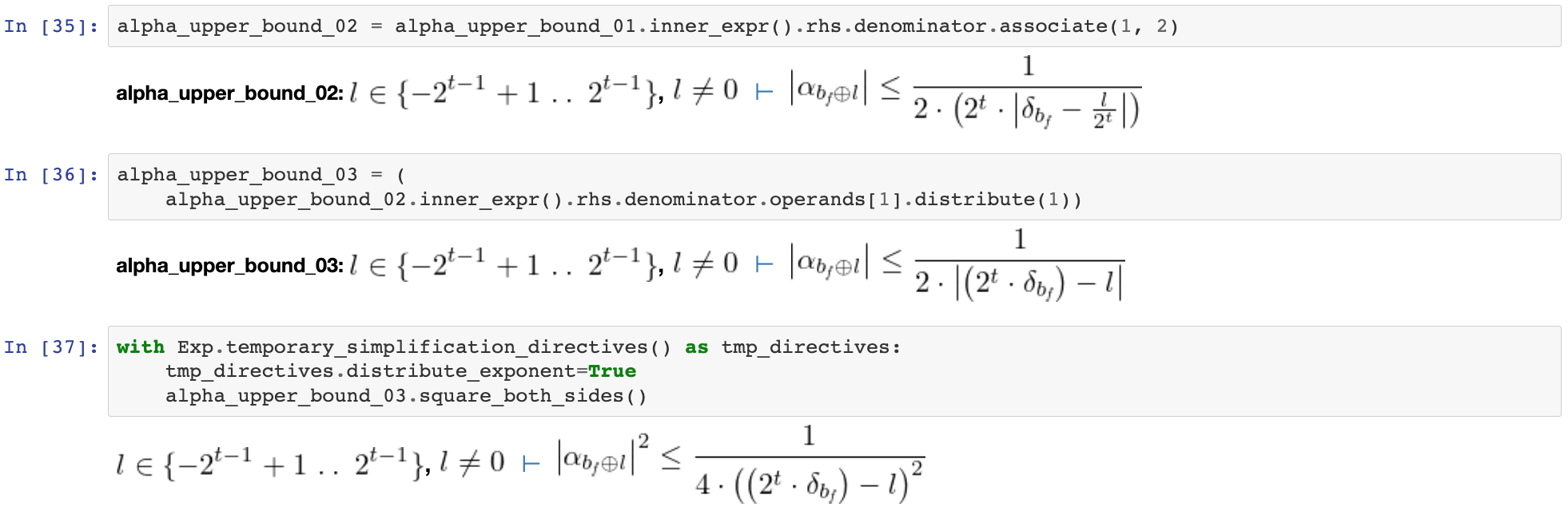}
\caption{Examples of user-invoked derivations commands (tactics), excerpted from the proof notebook for \texttt{\_alpha\_sqrd\_upper\_bound}~(\ref{eq:thm_alpha_sqrd_upper_bound}), showing user-invoked calls to \texttt{deduce\_triangle\_bound()}, \texttt{deduce\_bound()}, \texttt{associate()}, \texttt{distribute()}, and \texttt{square\_both\_sides()} methods.
}
\label{fig:example_tactics}
\end{figure}

A user-invoked derivation typically utilizes an \texttt{Expression} class method that contains the instructions to determine axiom(s)/theorem(s) to invoke and appropriate instantiations of their quantified variables.\footnote{The object-oriented organization of the underlying code also then makes it easy to find desired methods.}
This frees the user from having to find the right axiom(s)/theorem(s), look up which variables need to be instantiated, and figure out exactly how to instantiate them.  Some of these methods are more complicated and powerful, with provability checks, recursion, and/or post-processing of resulting judgments.

Some examples of such user-invoked derivations are shown in Fig.~\ref{fig:example_tactics}, excerpted from the proof notebook for 
\texttt{\_alpha\_sqrd\_upper\_bound}~(\ref{eq:thm_alpha_sqrd_upper_bound}).  The \texttt{NumberOperation} class \texttt{deduce\_bound()} method, employed in cells \texttt{[30]} and \texttt{[33]}, is a particularly powerful method.  It propagates the effects of any number of bounds on sub-expressions.
Under the hood, the \texttt{deduce\_bound()} step in cell \texttt{[33]} also employs goal-driven derivations (see \S\,\ref{subsec:goal-driven_derivations}) to satisfy requirements (\textit{e.g.}, establishing the signs of the numerator and denominator) as well as automatic simplification (see \S\,\ref{subsec:automatic_simplifications}) in canceling out a factor of $2$.

\subsection{Goal-Driven derivations}
\label{subsec:goal-driven_derivations}

A goal-driven derivation is an attempt to prove a particular statement indicated as the goal in advance.
Goal-driven derivations are initiated either directly (manually) or indirectly (e.g., when an instantiation is made and conditions must be satisfied).  
A user may directly invoke the \texttt{.prove()} method on an expression, which instructs \ProveIt{} to attempt to prove that expression, or it may be invoked indirectly through cascading requirements (\textit{e.g.}, the user wants to prove statement $A$ but  statement $B$ must be proven as a prerequisite to deriving $A$).

Each \texttt{Expression} class has its own tactics for trying to prove something of its own type. As a deliberate design choice in \ProveIt{}, we do \emph{not} perform a deep, combinatorial search to find a tactic that works. Thus, simply calling \texttt{.prove()} on a formula of interest will often fail.
As a guiding principle, we only invest in a single tactic for a given goal-driven derivation.
If that fails, an exception is raised.
To determine which one tactic to attempt, we perform efficient provability checks.  The provability checks are neither exhaustive nor fail-proof.
We only check if something is readily (or obviously) provable.
We may assume that expressions are properly typed (\textit{e.g.}, given $a+b$, assuming $a$ and $b$ are numbers) even though an actual proof will have to ensure the types are proper.
These checks simply serve as a guide to choosing a tactic. The goal in \ProveIt{} is to skip obvious steps, not provide fully automated theorem proving.

\begin{figure}[!htb]
  \captionsetup{font=footnotesize}
  \includegraphics[width = 0.47\textwidth]{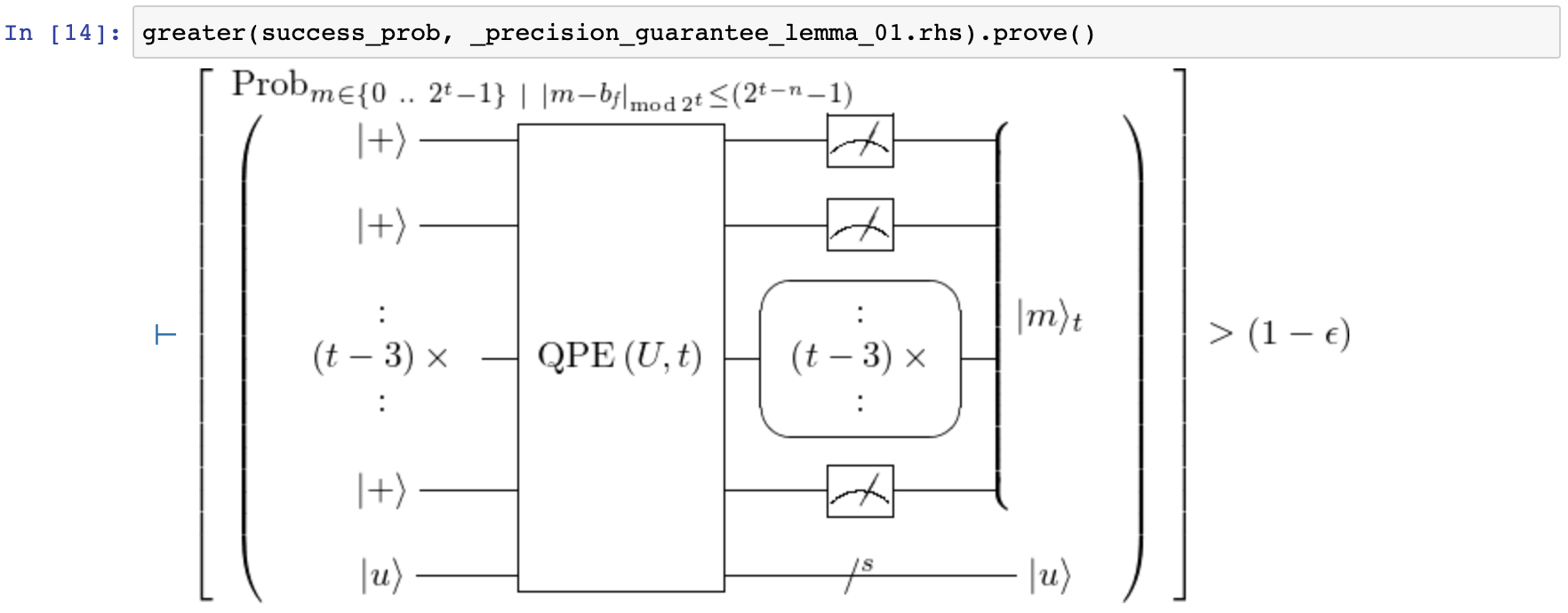}
\caption{Example of a goal-driven derivation, excerpted from the proof notebook for \texttt{\_precision\_guarantee}~(\ref{eq:thm_precision_guarantee}) showing the manual calling of \texttt{.prove()} on a formula of interest.
}
\label{fig:example_goal_drive_deriv}
\end{figure}
 
An indirect (automatic) goal-driven derivation was already encountered in the instantiation step of cell \texttt{[10]} in Fig.~\ref{fig:example_instantiations_alpha_summed}, with the system proving ``under the hood'' that $(b_{\!f}+\ell)\in\mathbb{Z}$ to satisfy the condition of the quantified statement being instantiated.  An example of a directly (manually) invoked goal-driven derivation is shown in Fig.~\ref{fig:example_goal_drive_deriv}, excerpted from the proof notebook for \texttt{\_precision\_guarantee}~(\ref{eq:thm_precision_guarantee}).
In this particular instance, the inequality is proven by automatically recognizing that a simple substitution can be made on the left-hand side of a previously known inequality.
More generally, Prove-It may attempt a bidirectional search among known number-ordering relations ($<$, $\leq$, and $=$) to see if some inequality may be proven via transitivity (e.g., $a < d$ can be proven automatically if we know  $a < b$, $b=c$, and $c \leq d$).
The same bidirectional search algorithm can be utilized in goal-driven derivations of other transitive relations (e.g., subset relations: $\subset$, $\subseteq$, and $\cong$). 

\subsection{Incidental derivations}
\label{subsec:incidental}

Many simple derivations are performed automatically, without user direction, to proactively derive facts which may be useful for later derivations.
When a judgment is proven or assumptions are made, incidental derivations are often made.
These incidental derivations are directed by each \texttt{Expression} class.

Incidental derivations are not obvious to spot, since they do not appear explicitly in an interactive notebook step, but they assist derivations throughout.
For example, proving that $A \vdash x \in \{2 ~\ldotp \ldotp~ 10\}$ will incidentally derive (in the background, so to speak) $A \vdash x \geq 2$ and $A \vdash x \leq 10$ (where $A$ represents any number of assumptions).
When an equality judgment such as $A \vdash x = y$ is established, $A \vdash y =x$ is also automatically derived.
The ``unfolding'' of a definition is a common incidental derivation as well --- for example, $A \vdash \lnot(x=y)$ is automatically derived upon proving that $A \vdash x \ne y$.
Such incidental judgments are then available for deriving other conclusions, both automatic and user-invoked.
As a rule, \ProveIt{} reserves automatic incidental derivations for obvious things and makes sure that incidental derivations can never cascade in a way that expands exponentially.

\begin{figure}[tb]
\captionsetup{font=footnotesize}
  \includegraphics[width = 0.47\textwidth]{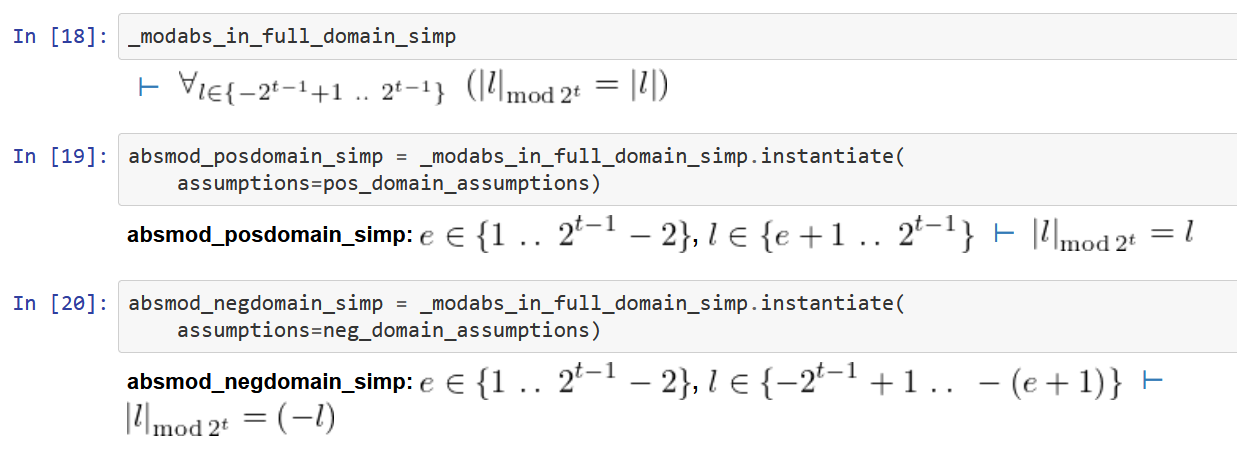}
\caption{Incidental derivation examples from an except of the proof notebook for \texttt{\_fail\_sum}~(\ref{eq:thm_fail_sum})).
}
\label{fig:example_incidental_derivs}
\end{figure}

A pair of related examples of incidental derivation appear in 
cells \texttt{[19]}, \texttt{[20]} of Fig.~\ref{fig:example_incidental_derivs} (excerpted from the proof notebook for \texttt{\_fail\_sum}~(\ref{eq:thm_fail_sum})).
In each example, assumptions generate incidental derivations that enable an instantiation and its simplification.  We assume $e \in \{1~\ldotp \ldotp~2^{t - 1} - 2\}$ in both cases which incidentally derives $e \geq 1$ and $e \leq 2^{t - 1} - 2$.  In cell \texttt{[19]}, we additionally assume $l \in \{e + 1~\ldotp \ldotp~2^{t - 1}\}$ which incidentally derives $l \geq e+1$ and $l \leq 2^{t - 1}$ from which it can automatically derive $l \in \{-2^{t - 1} + 1~\ldotp \ldotp~2^{t - 1}\}$ to satisfy the instantiation condition from \texttt{[18]}, and we can derive $l > 1$ so that $|l|$ simplifies to $l$ (automatic simplification will be discussed in \S\ref{subsec:automatic_simplifications}).  In cell \texttt{[20]}, we additionally assume $l \in \{-2^{t - 1} + 1~\ldotp \ldotp~-\left(e + 1\right)\}$ from which we can also derive $l \in \{-2^{t - 1} + 1~\ldotp \ldotp~2^{t - 1}\}$ but in this case $l < 1$ such that $|l|$ simplifies to $-l$.
Incidental and goal-driven derivations are ubiquitous and often occur together like this in a bi-directional fashion and can be very powerful.

\subsection{Equating expressions with the same canonical forms}

\begin{figure}[tb]
\captionsetup{font=footnotesize}
  \includegraphics[width = 0.47\textwidth]{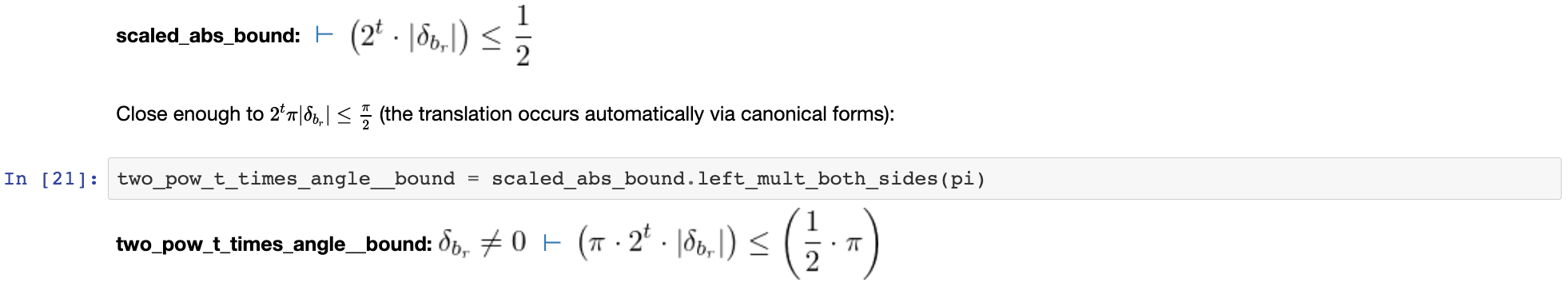}
  \\[-0.1in]
  $\vdots$\vspace{0.05in}
  \\
  \includegraphics[width = 0.47\textwidth]{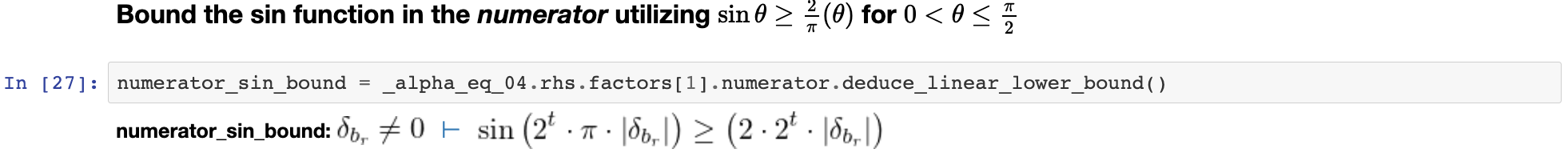}
  \\[-0.075in]
  $\vdots$\vspace{0.025in}
  \\
  \includegraphics[width = 0.47\textwidth]{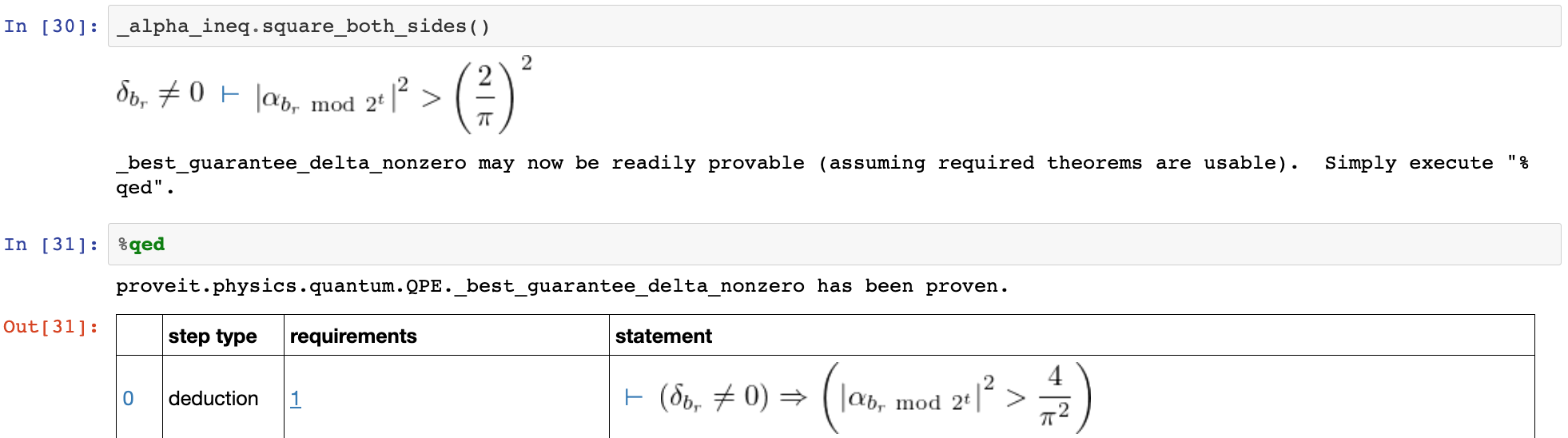}
\caption{Examples of equivalent canonical forms allowing automatic derivation steps, excerpted from the proof notebook for \texttt{\_best\_guarantee\_delta\_nonzero}~(\ref{eq:thm_best_guarantee_delta_nonzero}).
}
\label{fig:example_canonical_forms}
\end{figure}

In mathematics, there are often multiple ways of expressing the same thing with uninteresting or insignificant differences.  In some instances, there is no difference in the inherent meaning and it is simply a style choice.  For example, $\forall_{x_1, \ldots, x_n} P(x_1, \ldots, x_n)$ is no different than $\forall_{y_1, y_2, \ldots, y_n} P(y_1, y_2, \ldots, y_n)$ in its meaning (simply changing the internal $x$ label to $y$ and changing the presentation of the ranges).
Another example is the total-ordering style of a conjunction of transitive relations; for example, $(a < b) \land (b < c)$ can be presented as $a < b < c$ in \ProveIt{} using a style option.
\ProveIt{} offers flexibility in choosing the presentation style for such things while treating them as logically identical.
In other cases, such as $(2 x / 2 + 6)$ \text{vs.} $(2 \cdot 3 + x)$, the inherent meaning is different but it is straightforward to derive the equality.
For the latter scenario, we use canonical forms.
The choice of the canonical form (\textit{e.g.}, whether to use $6 + x$ or $x + 6$) has no bearing on proof generation or presentation, and so we omit details here about how such canonical forms are generated.  The important thing is that we choose consistent conventions to exploit knowledge about canonical form equivalence classes (sets of expressions that have the same canonical form), which then help drive proof automation.

Some simple examples of the use of canonical forms are shown in Fig.~\ref{fig:example_canonical_forms}, excerpted the proof notebook for \texttt{\_best\_guarantee\_delta\_nonzero}~(\ref{eq:thm_best_guarantee_delta_nonzero}). The user-invoked method $\texttt{deduce\_linear\_bound()}$ called in cell \texttt{[27]} needs to know that the angle $\theta = 2^{t}\pi|\delta_{b_{r}}|\le\frac{\pi}{2}$, which the system recognizes through canonical forms as equivalent to the inequality derived in cell \texttt{[21]}. Later, the interactive steps have established a formula in cell \texttt{[30]} involving the expression $\left(\frac{2}{\pi}\right)^{2}$, while the theorem to be proven uses the expression $\frac{4}{\pi^{2}}$.
While $\left(\frac{2}{\pi}\right)^{2}$ and $\frac{4}{\pi^{2}}$ are distinct expression forms, they have the same canonical form. \ProveIt{} determines that one can be readily transformed into the other, and thus no further manual steps are needed in the proof notebook.

\subsection{Automatic simplifications}
\label{subsec:automatic_simplifications}

\begin{figure}[tb]
\captionsetup{font=footnotesize}
  \centering
  \includegraphics[width = 0.47\textwidth]{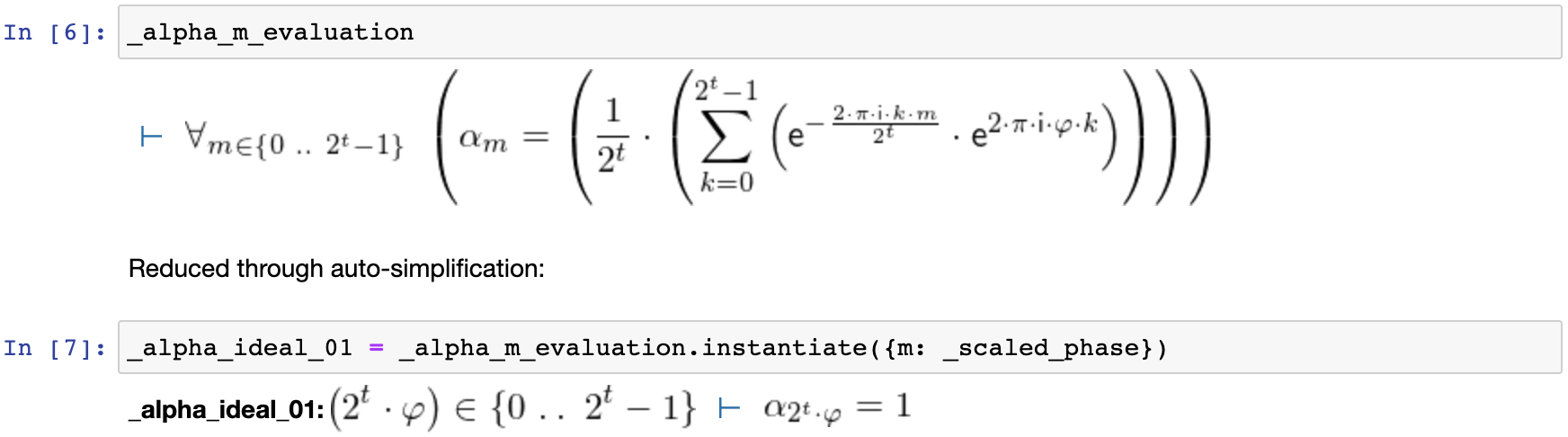}
\caption{An example of automatic simplification, drawn from the proof notebook for \texttt{\_alpha\_ideal\_case}~(\ref{eq:thm_alpha_ideal_case}). 
}
\label{fig:example_auto_simplification_alpha_ideal}
\end{figure}

Why not automatically simplify everything to its canonical form?  This is not always desirable and sometimes counter-productive.  For example, $y + z/x$ may be the desired form even though $z/x + y$ may happen to be the canonical form.  There are certain kinds of simplification that are sensible, but not always.
It makes sense to simplify $y + 2 x / 2$ to $y + x$ (without changing the order in case it is preferable to the user for their purpose). It is not so obvious which of the forms $2x + 2y$ versus $2(x+y)$ should be preferred automatically, even though they have the same canonical form. \ProveIt{}'s solution is to allow fairly straightforward automatic simplifications that are configurable --- that is, the user can manually disable auto-simplifications, or explicitly change default simplification directives for each expression type.

Automatic simplifications are performed as a post-processing step for any derivation command unless it has been disabled.  After implementing a theorem-instantiation step, for example, \ProveIt{} traverses the resulting expression to simplify sub-expressions according to the active simplification directives.  Simplification substitutions are explicitly proven in the formal proof that is constructed.

A fairly dramatic example of such automatic simplification appears in Fig.~\ref{fig:example_auto_simplification_alpha_ideal}, drawn from the proof notebook for \texttt{\_alpha\_ideal\_case}~(\ref{eq:thm_alpha_ideal_case}).
The expression in cell \texttt{[7]} resulting from the manual instantiation of the \texttt{\_alpha\_m\_evaluation} theorem in \texttt{[6]} is greatly reduced through auto-simplification. When $m$ is instantiated as the \texttt{\_scaled\_phase} $2^{t}\varphi$, the product of exponentials in the summand is automatically reduced to $1$, and then the summation of the constant summand $1$ is automatically reduced and eventually canceled with the coefficient $\frac{1}{2^t}$. Such auto-simplification saves many steps throughout our QPE derivations.

\section{Examining a Specific Proof in the QPE package}
\label{sec:theorem_proof_demo}

In previous sections, we have provided scattered examples of both derivation commands fed into \ProveIt{} in the process of proving QPE-specific theorems as well as some formal proof output.  All of the proof constructions and formal output for QPE-specific theorems may be viewed on the \ProveItWebsite{}.  There is too much material there to include in this article.  But to provide a sense for how the pieces come together, this section will focus on \texttt{\_psi\_t\_formula}~(\ref{eq:thm_psi_t_formula}) as one particular example.

A subset of the input/output cells for the Jupyter notebook used to construct the proof of \texttt{\_psi\_t\_formula}
is displayed in Fig.~\ref{fig:example_interactive_steps}.
While many steps are omitted for brevity, these chosen steps show the basic flow of this inductive proof and illustrate how derivation commands (discussed in \S\ref{Sec:interactive_proofs}) are used to move the proof along.
The final step is to issue a \texttt{\%qed} command to produce the formal proof, the first few lines of which are shown in Fig.~\ref{fig:example_formal_proof_12_steps_of_psi_t_formula}.
Note that the formal proof contains all of the fine details to make an airtight proof, each step of which is easy to confirm by human or machine (in principle).
With 43 user-invoked input/output cells to construct the \texttt{\_psi\_t\_formula} proof (which includes 24 distinct derivation commands), a formal proof of 684 steps is generated with all of the fine details to derive the \texttt{\_psi\_t\_formula} from other proven theorems or more fundamental facts.  
We expect the formal proof sizes to reduce to some extent over time as \ProveIt{} routines develop better strategies to generate succinct formal proofs, but the relatively large formal proof size does convey how much \ProveIt{} is doing to fill in the obvious steps omitted in the proof construction.

The readability of both the proof construction input/output cells and the formal proof output offers interesting possibilities for communicating proofs and concepts, clearly and succinctly, among experts.
In Appendix~\ref{appendix:structured_proof}, we explore the potential connections between proofs in \ProveIt{} and structured proofs intended for proof communication using \texttt{\_psi\_t\_formula} as an example.

We can also compare 14 steps in the structured, informal proof of Appendix~\ref{appendix:structured_proof} with the 24 derivation commands used to prove \texttt{\_psi\_t\_formula}.  
This ratio is analogous to what de~Bruijn called a ``loss factor'' \cite{deBruijn:1980_automath} and Wiedijk has called ``the de Bruijn factor'' \cite{Wiedijk:nd_De_Bruijn_Factor}.
Wiedijk reported finding a de Bruijn factor of about 4 across a variety of texts and computer-proof systems.  This comparison suggests a loss factor of roughly $2$ for this particular example (though this ratio really depends upon the level of detail provided in the informal proof).

\newcommand\tempgraphicswidth{0.45}
\begin{figure*}[tp]
\captionsetup{font=footnotesize}
  \includegraphics[width = \tempgraphicswidth\textwidth]{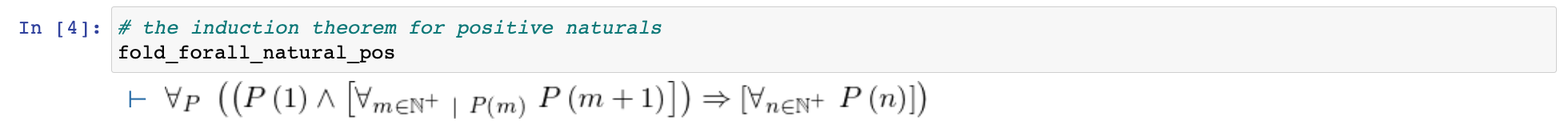}\\
  \includegraphics[width = \tempgraphicswidth\textwidth]{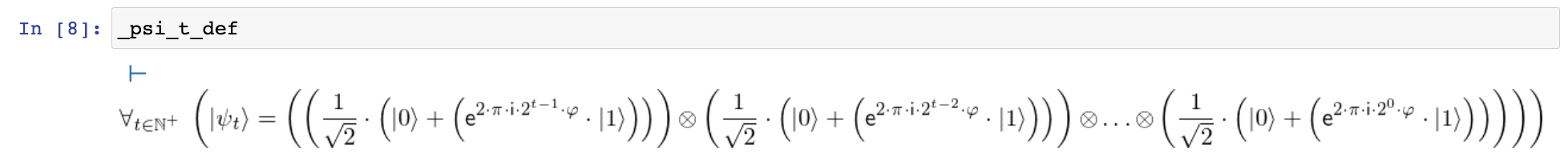}\\
  \includegraphics[width = \tempgraphicswidth\textwidth]{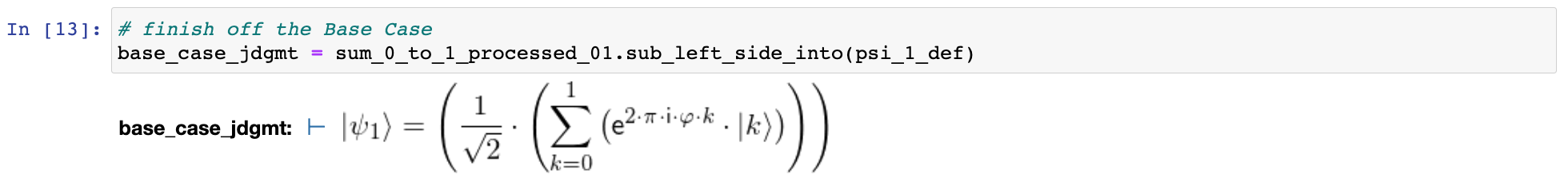}\\
  \includegraphics[width = \tempgraphicswidth\textwidth]{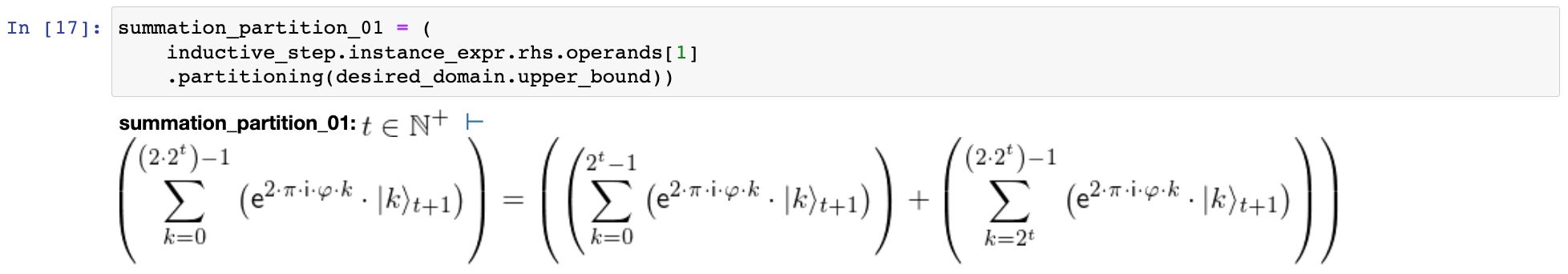}\\
  \includegraphics[width = \tempgraphicswidth\textwidth]{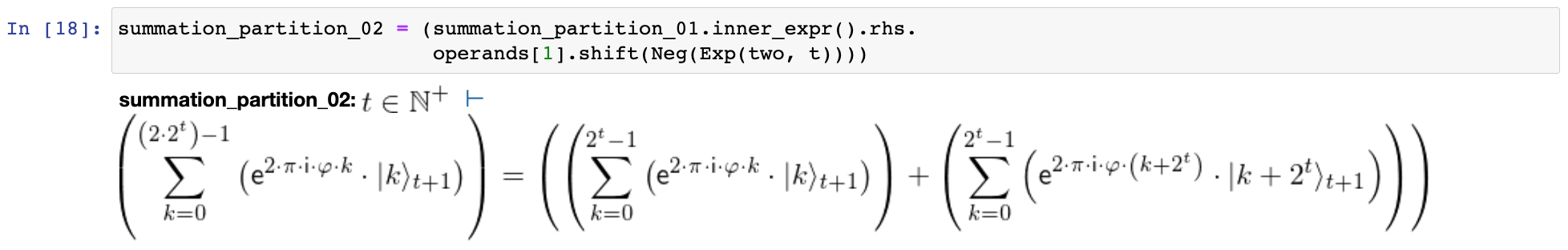}\\
  \includegraphics[width = \tempgraphicswidth\textwidth]{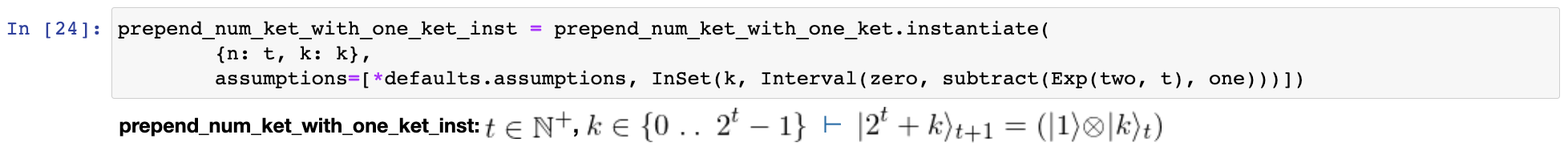}\\
  \includegraphics[width = \tempgraphicswidth\textwidth]{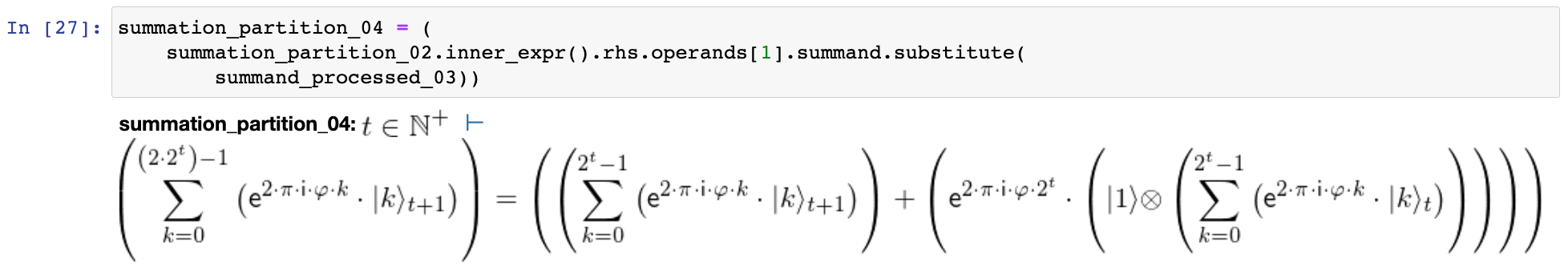}\\
  \includegraphics[width = \tempgraphicswidth\textwidth]{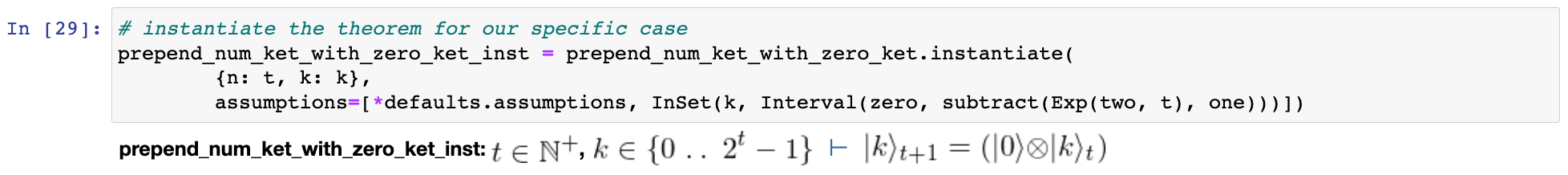}\\
  \includegraphics[width = \tempgraphicswidth\textwidth]{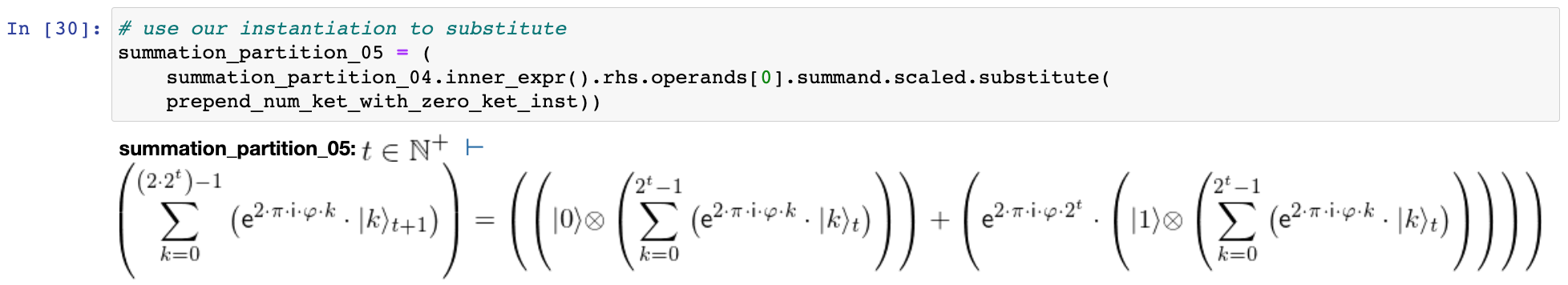}\\
  \includegraphics[width = \tempgraphicswidth\textwidth]{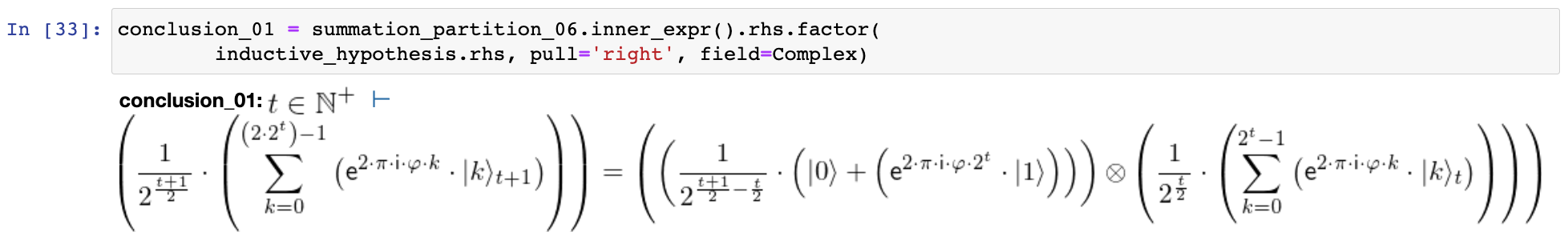}\\
  \includegraphics[width = \tempgraphicswidth\textwidth]{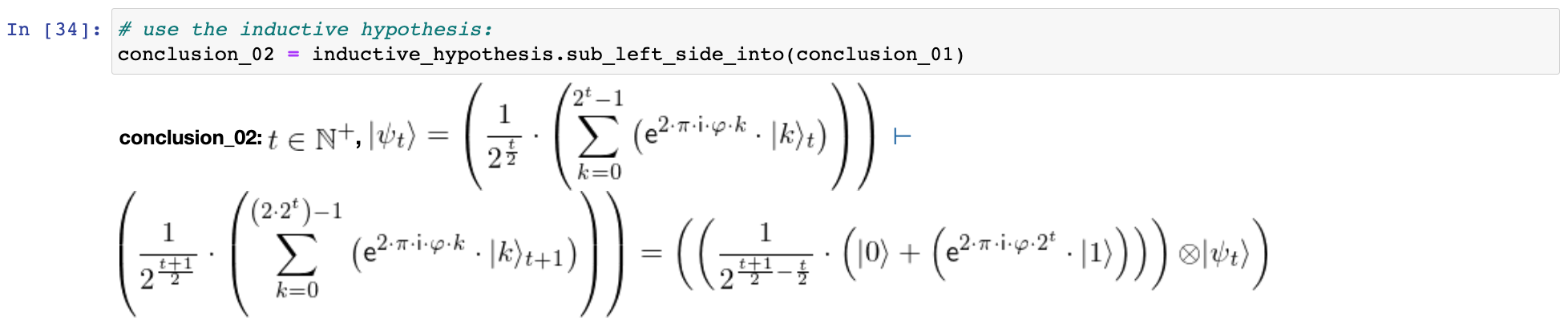}\\
  \includegraphics[width = \tempgraphicswidth\textwidth]{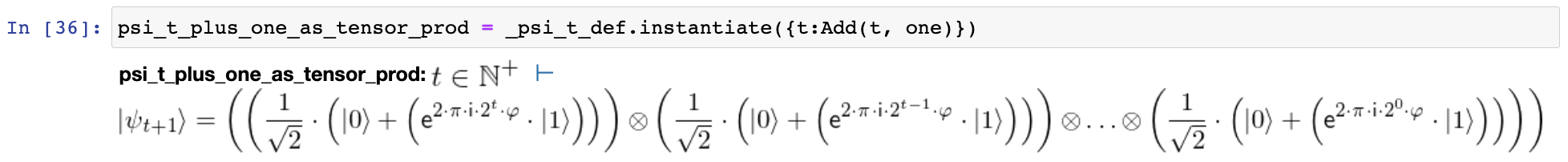}\\
  \includegraphics[width = \tempgraphicswidth\textwidth]{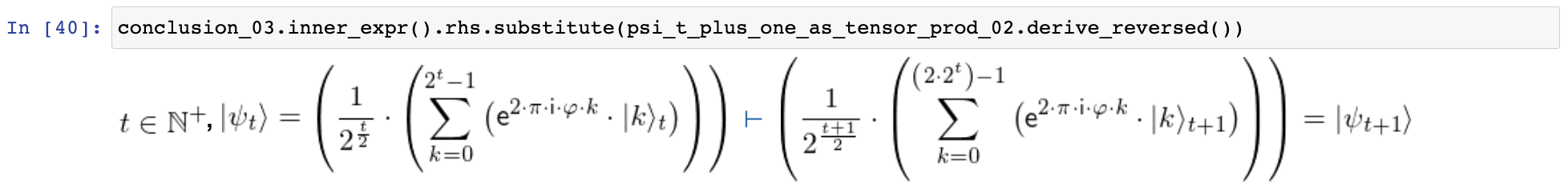}
  \caption{A subset (13 of 23) of the interactive steps that construct the proof of \texttt{\_psi\_t\_formula}~(\ref{eq:thm_psi_t_formula}) chosen to show the basic flow of this inductive proof and demonstrate some useful derivation commands in action such as \texttt{partitioning}, \texttt{shift}, \texttt{substitute}, and \texttt{factor}.  Appendix~\ref{appendix:structured_proof} compares this to a structured proof format intended to communicate a proof among experts.
  }
  \label{fig:example_interactive_steps}
\end{figure*}

\begin{figure*}%[tp]
\captionsetup{font=footnotesize}
  \includegraphics[width = 0.75\textwidth]{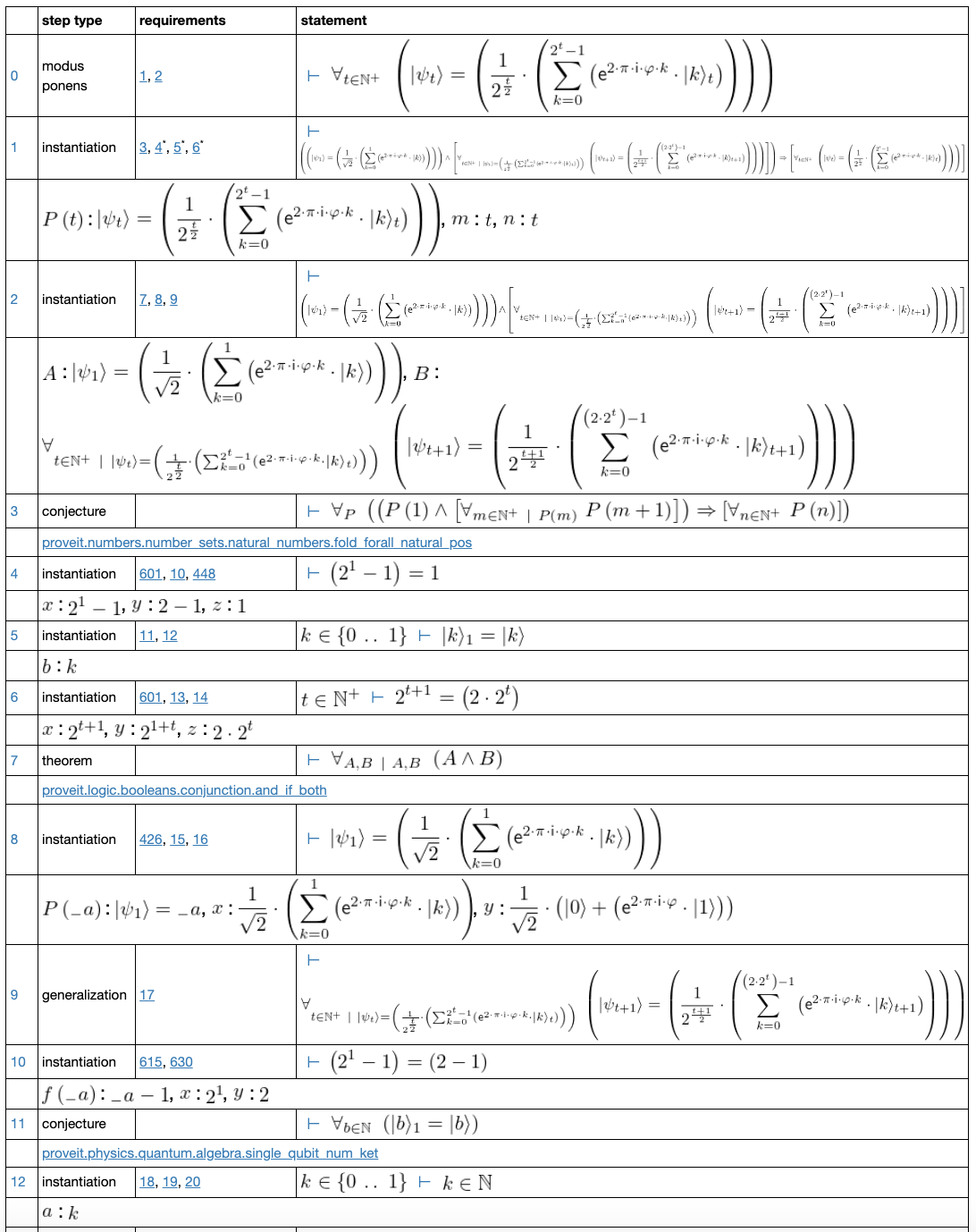}
  \caption{The first few lines of \ProveIt{}'s formal proof output for \texttt{\_psi\_t\_formula}~(\ref{eq:thm_psi_t_formula}), illustrating both the general structure of a formal proof in \ProveIt{} and the rigorous and explicitly human-checkable details involved in each proof step (the full proof has 684 steps at this time). The steps appear in reverse order with the proven statement at the top as the root of the proof DAG. Each numbered step (along the left-hand side) includes a ``step type'' and an explicit listing of all previous (higher-numbered) steps on which the current step depends (note the underlined numbers in the ``requirements'' column, always higher-numbered than the current step to indicate they occur further down the proof tree with no circular dependency). The underlined numbers are hyperlinks that will take the user to the corresponding step in the formal proof. Instantiation steps (such as \texttt{Step 4}) explicitly list the substitutions used in the derivation. Steps that introduce an assumption, axiom, theorem, or conjecture (such as \texttt{Step 7}) show the detailed formula being introduced, as well as its name in the \ProveIt{} database hyperlinked to its proof (or the placeholder for the proof if one does not yet exist).
  }
  \label{fig:example_formal_proof_12_steps_of_psi_t_formula}
\end{figure*}

\begin{figure}[htb]
\captionsetup{font=footnotesize}
  \includegraphics[width = 0.48\textwidth]{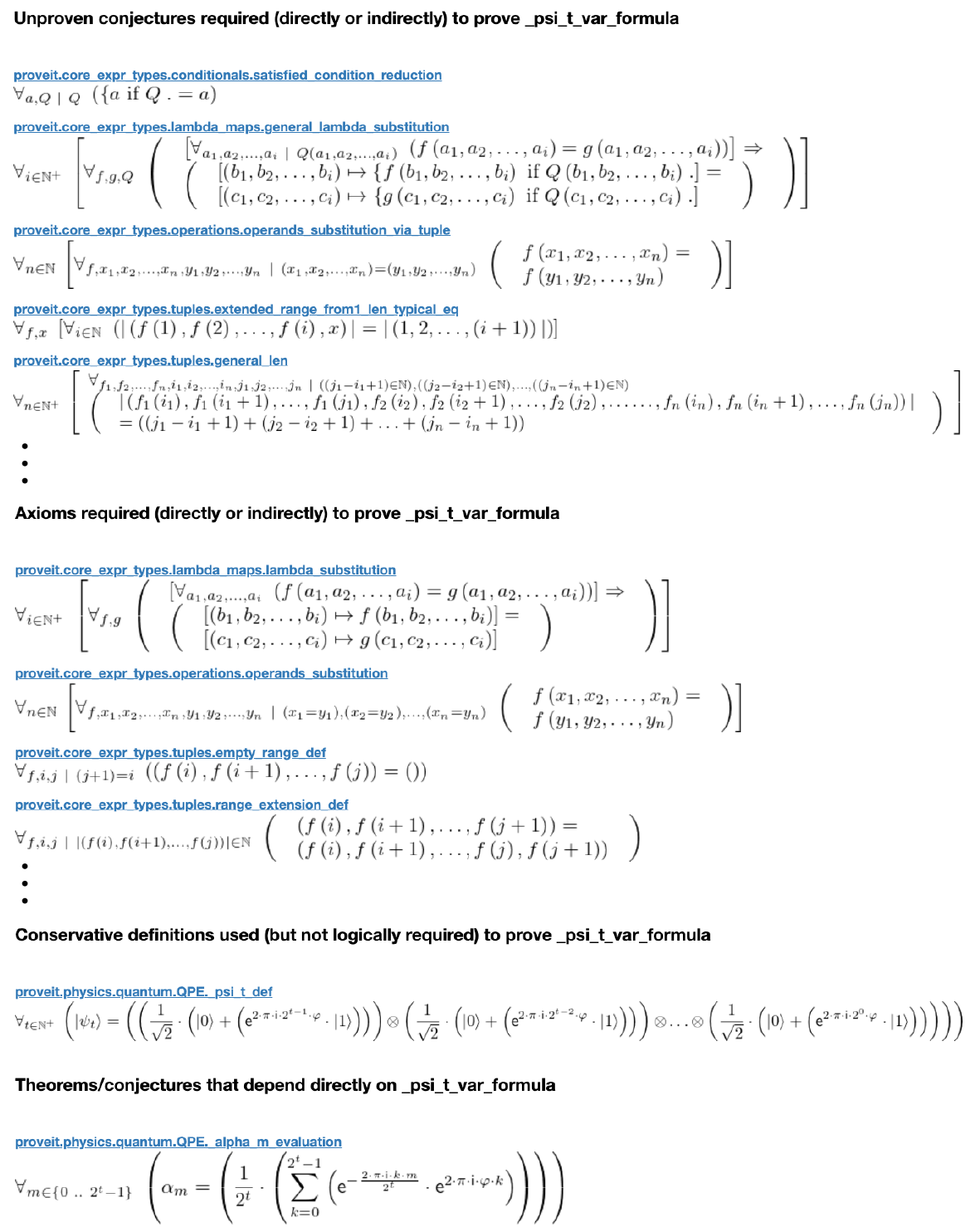}
\caption{Excerpt from the dependencies page \texttt{\_psi\_t\_formula}~(\ref{eq:thm_psi_t_formula}), showing some of the unproven theorems, axioms, and conservative definitions on which the theorem depends, and \texttt{\_alpha\_m\_evaluation}~(\ref{eq:thm_alpha_m_evaluation}) which depends directly on the \texttt{\_psi\_t\_formula} theorem itself. Such dependencies are all explicitly tracked in \ProveIt{} and readily accessible for review by the user.
}
\label{fig:example_psi_t_formula_dependencies}
\end{figure}

Finally, as discussed in \S\ref{subsec:theorem_dependencies}, the user has easy access to a complete summary list of all conjectures, axioms, and conservative definitions required in the formal proof (as well as a list of all theorems that depend on the theorem in question) through its dependencies page. An excerpt of the dependencies page for \texttt{\_psi\_t\_formula}~(\ref{eq:thm_psi_t_formula}) is shown in Fig.~\ref{fig:example_psi_t_formula_dependencies}.

\section{Comparisons with other approaches}
\label{sec:comparisons}

\ProveIt{} is designed to enable robust theorem-proving via a user-interactive process that can mimic informal proof styles.
It is not intended as a black-box automated theorem-proving system.
The automation in \ProveIt{} is mainly intended as a means to skip obvious steps that would normally be skipped in an informal proof.
Furthermore, \ProveIt{} generates human-readable formal proofs --- they can be quite long, but any given derivation step can be understood with a small amount of training.
To our knowledge, \ProveIt{} is unique in these respects.

We are not the first to provide a formal verification of QPE-related claims --- related recent work on this problem can be found in Liu, \textit{et al.} \cite{Liu_et_al:2019_quantum_hoare_logic_proof_document}, Hietala, \textit{et al.} \cite{hietala_et_al_proving_quantum_programs:LIPIcs.ITP.2021.21}, Chareton, \textit{et al.} \cite{Chareton_et_al:2021_auto_deductive_verif_framework_Qbricks}, and Bauer-Marquart, \textit{et al.} \cite{Bauer-Marquart_et_al:2022_Symbolic_Verification_Quantum_Programs}.
Our work is unique in providing a verification for Nielsen \& Chuang's \cite{Nielsen_Chuang:2010} QPE-related precision guarantee, \texttt{qpe\_precision\_guarantee}~(\ref{eq:thm_qpe_precision_guarantee}), and in providing not just \textit{verifications} but detailed formal human-readable proofs of the exact, general, and precision-guarantee cases. This aspect of our proof attempt is elaborated further below. 

Liu, \textit{et al.} \cite{Liu_Theorem_Prover_Hoare_Logic:2016} implemented quantum Hoare logic (QHL) \cite{Ying:2011_Floyd_Hoare_logic_for_quantum_programs} in Isabel/HOL and Python and reported having the first ``mechanized proofs'' (\textit{i.e.}, proofs of correctness utilizing a theorem prover) of both Grover's quantum search algorithm and the QPE algorithm.
Two conceptual challenges arise in the work: First, the quantum algorithms need to be manually translated into QHL, which seems a significant effort, the result of which can be obfuscating of the original sense of the algorithm; and second, the formal verification appears to rely on extensive calculations of functions of matrices, with the calculations being off-loaded to Python's \texttt{Numpy} and \texttt{Sympy} packages.
The proof outputs also do not seem to be human readable.
Quoting \cite{hietala_et_al_proving_quantum_programs:LIPIcs.ITP.2021.21}, this is ``then only a partial verification effort.''
Also, they do not prove any precision guarantees on the phase estimation in QPE.

Using \textsc{sqir} (``small quantum intermediate representation''), which they describe as ``a simple quantum language deeply embedded in the Coq proof assistant,'' Hietala, \textit{et al.} \cite{hietala_et_al_proving_quantum_programs:LIPIcs.ITP.2021.21} provide proofs of correctness for a number of quantum algorithms, such as quantum teleportation, the Deutsch-Jozsa algorithm [10], Simon’s algorithm [32], the quantum
Fourier transform (QFT), quantum phase estimation (QPE), and Grover’s algorithm [16].

\begin{figure}[htb]
  \captionsetup{font=footnotesize, justification=justified}
  \centering
  \begin{minipage}[]{0.01\linewidth}
      \subcaption{}\label{fig:subfig_sqir_qpe_specification}
  \end{minipage}
  \begin{adjustbox}{minipage={0.9\linewidth}}
  \setlength{\fboxsep}{1pt}
  \setlength{\fboxrule}{0.5pt}
  \fbox{
      \includegraphics[width = 0.9\textwidth]{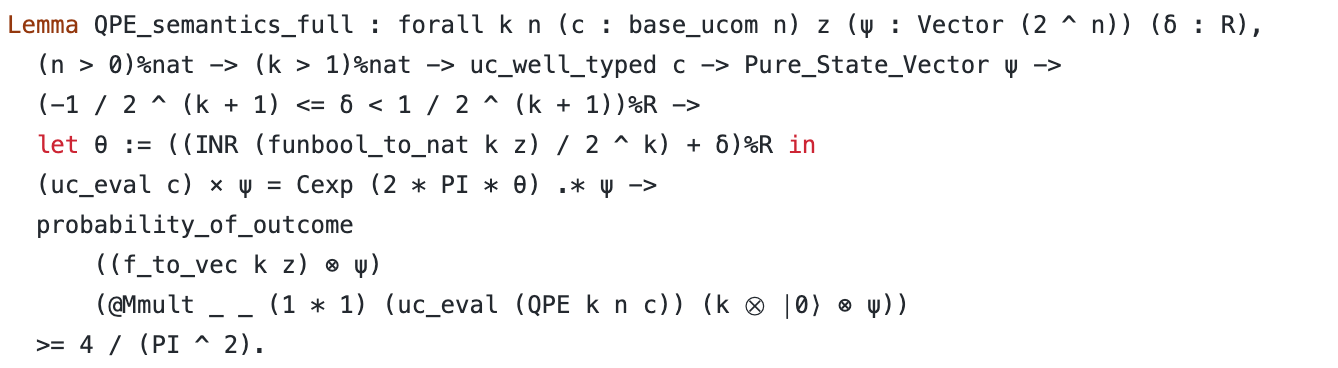}}
  \end{adjustbox}\\[1ex]
  
  \begin{minipage}[]{0.01\linewidth}
      \subcaption{}\label{fig:subfig_qbricks_qpe_specification}
  \end{minipage}
  \begin{adjustbox}{minipage={0.9\linewidth}}
  \setlength{\fboxsep}{1pt}
  \setlength{\fboxrule}{0.5pt}
  \fbox{
      \includegraphics[width = 0.9\textwidth]{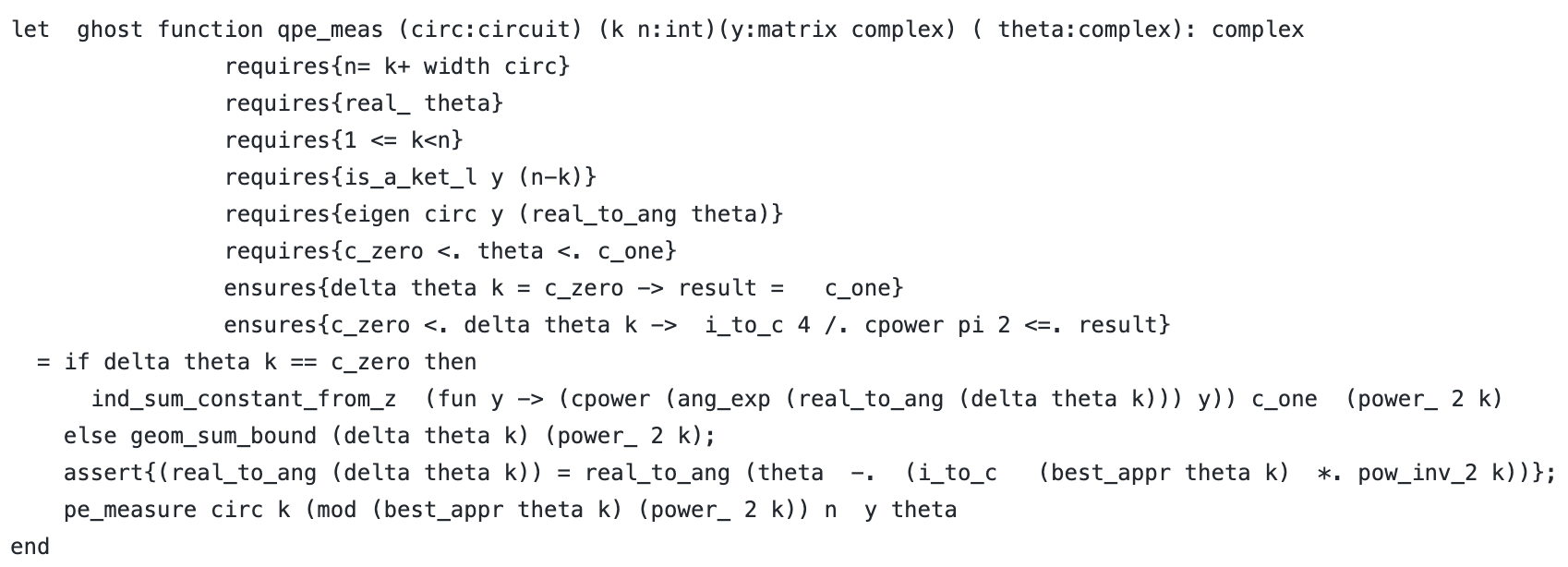}}
  \end{adjustbox}\\[1ex]

  \begin{minipage}[]{0.01\linewidth}
      \subcaption{}\label{fig:subfig_proveit_qpe_specification}
  \end{minipage}
  \begin{adjustbox}{minipage={0.9\linewidth}}
  \setlength{\fboxsep}{1pt}
  \setlength{\fboxrule}{0.5pt}
  \fbox{
      \includegraphics[width = 0.9\textwidth]{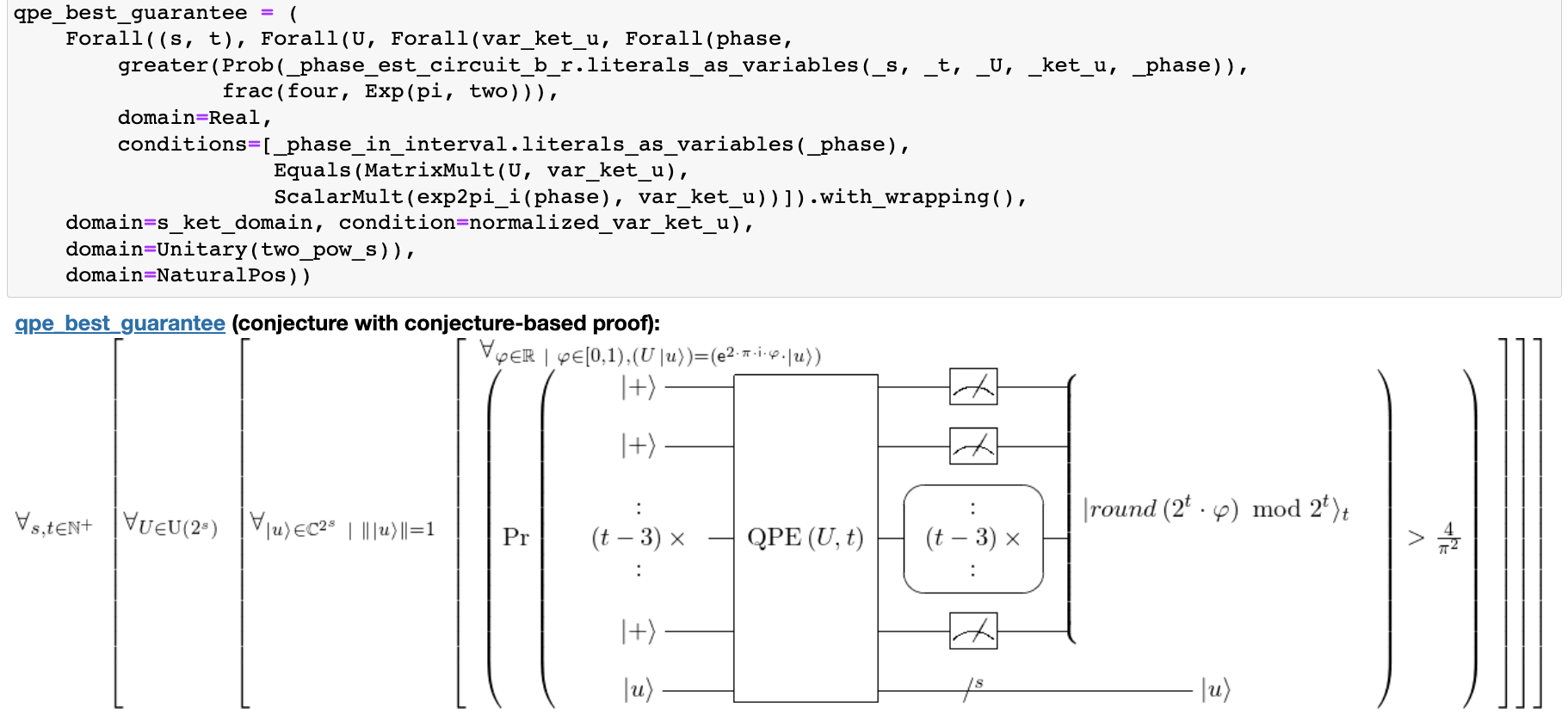}}
  \end{adjustbox}
\caption{Specifications of \texttt{qpe\_best\_guarantee}~(\ref{eq:thm_qpe_best_guarantee}) in (\textit{a}) SQIR \cite{hietala_et_al_proving_quantum_programs:LIPIcs.ITP.2021.21} \textit{vs.} (\textit{b}) QBricks \cite{Chareton_et_al:2022_formal_methods_for_quantum_programs} \textit{vs.} (\textit{c}) \ProveIt{}. The specification in \ProveIt{} automatically generates the accompanying circuit representation, contributing to clarity of presentation and human-readability. 
}
\label{fig:compare_qpe_best_guarantee_specifications}
\end{figure}

It is possible to make direct comparisons with Coq proofs for the first two of our high-level statements about the QPE output.  Both \texttt{qpe\_exact}~(\ref{eq:thm_qpe_exact}) and \texttt{qpe\_best\_guarantee}~(\ref{eq:thm_qpe_best_guarantee}) were  essentially proven in Ref.~\cite{hietala_et_al_proving_quantum_programs:LIPIcs.ITP.2021.21} and can be found in their SQIR repository. 
There are immediate differences in the way these statements are presented.
Their quantum algorithm definition and statements, by necessity, are expressed as computer code which requires knowledge of Coq to fully parse and understand.
And while their paper presents quantum circuits to describe algorithms, these are human translations that are not directly tied to or produced by their definitions in Coq. 
In contrast, our QPE specifications in terms of quantum circuits are generated directly from our definition in \ProveIt{}.  
Our high-level QPE output statements, as expressed above, were generated directly from the proven representations in \ProveIt{}.
There is an essentially direct correspondence between the statements that are certified by \ProveIt{} and their automatically-generated \LaTeX{} representations.
The specification of \texttt{qpe\_best\_guarantee}~(\ref{eq:thm_qpe_best_guarantee}) in SQIR appears in the top panel of Fig.~\ref{fig:compare_qpe_best_guarantee_specifications}; the specification in \ProveIt{}, including the automatically-generated graphical circuit representation, appears in the bottom panel.
\ProveIt{}'s graphical representation of the specification makes the theorem more readily comprehensible.

We can also compare the level of effort required to generate the proofs in Coq versus \ProveIt{}. 
Although \ProveIt{} does not have Coq's maturity and thus a significant part of the current effort in \ProveIt{} was devoted to its infrastructure, one helpful comparison to make is in the number of user-invoked derivation commands in \ProveIt{} versus Coq commands (\textit{e.g.}, tactic invocations) necessary to generate the proofs at the high level (\textit{i.e.}, taking more fundamental packages for granted).\footnote{Coq commands were enumerated for Hietala, \textit{et al.}'s \texttt{QPE\_simplify}, \texttt{QPE\_semantics\_simplified}, and \texttt{QPE\_full\_semantics} lemmas \cite{hietala_et_al_proving_quantum_programs:LIPIcs.ITP.2021.21}, the most appropriate comparison to the analogous QPE-specific proofs in \ProveIt{}. The number of Coq commands was estimated by simply counting all commands that ended in a period or semi-colon.  The Prove-It derivation commands tally the numbers in Table~\ref{tab:proof_steps} and~\ref{tab:proof_steps} according to the dependencies of the three high-level statements respectively as shown in Fig.~\ref{fig:qpe_theorems_dependencies_tree}.}
This comparison is presented in Table~\ref{tab:deriv_steps_comparison_ProveIt_Coq}.  
We cannot make a comparison for the most challenging proof since this particular form was not proven in Ref.~\cite{hietala_et_al_proving_quantum_programs:LIPIcs.ITP.2021.21}. 
Such numbers demonstrate that the interactive \ProveIt{} process is quite competitive with the effort required in the Coq proof assistant, in addition to \ProveIt{} also producing a human-readable formal proof for one's effort: 87 \ProveIt{} derivation commands versus 137 Coq commands for the \texttt{qpe\_exact} theorem, and 213 versus 313 for the \texttt{qpe\_best\_guarantee} theorem.

\begin{table*}[tb]
    \captionsetup{font=footnotesize}
    \centering
    \begin{tabular}{|l||c|c|c||c|c|c|}
    \hline
    & \multicolumn{3}{c||}{\thead{\ProveIt{}\\derivation commands}}
    & \multicolumn{3}{c|}{\thead{\textit{Coq} commands}}\\
    \thead{Theorem} & \thead{total} & \thead{reused} & \thead{add'l} & \thead{total} &
    \thead{reused} &
    \thead{add'l}\\
    \hline
    \texttt{qpe\_exact}
    & 87
    & ---
    & ---
    & 137
    & ---
    & ---
    \\ \hline
    \texttt{qpe\_best\_guarantee}
    & 213
    & 72
    & 141
    & 313
    & 73
    & 240
    \\ \hline
    \texttt{qpe\_precision\_guarantee}
    & 437
    & 145
    & 292
    & ---
    & ---
    & ---
    \\ \hline
    \end{tabular}
    \vspace{0.1in}
    \caption{Number of \ProveIt{} derivation command steps vs. Coq commands \cite{hietala_et_al_proving_quantum_programs:LIPIcs.ITP.2021.21} for the main QPE theorems. The ``reused'' and ``additional'' steps add to the left-column total, providing a measure of how many derivation steps in later theorems are reused from earlier theorems.}
    \label{tab:deriv_steps_comparison_ProveIt_Coq}
\end{table*}

In his Ph.D. thesis, Amy used the Feynman path integral model of quantum circuits for their optimization, synthesis and functional verification \cite{Amy_PhD_Thesis:2019_Quantum_Circuit_Design}. It particularly focused on the functional verification of quantum circuits built using Clifford+$R_k$ gates, where $R_k = {R_z}({\frac{2\pi}{2^k}})$, and $R_z$ performs rotation around $z$-axis. A semantic model of such Clifford+$R_z$ circuits as path sums is developed along with a calculus of rewrite rules; these circuits are verified against a human-readable input/output specification. It is shown there that path sums associated with Clifford+$R_z$ circuits can be represented by a collection of path variables, a pseudo-Boolean phase function and an affine basis transformation, whereas a pseudo-Boolean phase function has a unique representation as a multilinear polynomial over variables. A calculus of rewrite rules over polynomial representations is used to compute
a normal form polynomial associated with a quantum circuit that can be computed in polynomial time; these normal forms are however not unique. Quantum algorithms/circuits for reversible functions, particularly implementations of Toffoli gates and adders using Clifford$+T$ gates were verified; more nontrivial algorithms including quantum Fourier transform and quantum hidden-shift algorithms were also verified. Verification thus amounts to checking for circuit equivalence performed by computing normal forms of circuits.
The method is implemented in a quantum compiler backend infrastructure FEYNMAN for quantum circuits, similar to LLVM for programming languages \cite{Lattner_Vikram:2004_LLVM, website:LLVM}. Since Amy's work does not consider measurement gates, precision guarantees on phase estimation are omitted. 

Using their Qbricks quantum circuit-building program environment \cite{website:Qbricks} within the Why3 deductive verification environment \cite{website:Why3}, Chareton, \textit{et al.} \cite{Chareton_et_al:2021_auto_deductive_verif_framework_Qbricks} encode and verify implementations of Grover’s quantum search algorithm, the Quantum Fourier Transform (QFT), Quantum Phase Estimation (QPE), and Shor’s algorithm.
They used Amy's Feynman path sum model of quantum circuits as building blocks/program constructs which allows other quantum algorithms to be written as finite branching straight-line programs in a Floyd-Hoare axiomatic framework for which Why3 is well-suited for generating the associated verification conditions for input-output behavioral verification.
They show that their approach produces verifications with ``high automaticity'' and high efficiency (in terms of the size of the algorithm specification and the number of commands required in the verification effort) compared to similar proofs in SQIR and QHL.
Like other approaches, they have built a collection of theories which enable reasoning about sets, algebra, arithmetic, iterators, quantum data, Kronecker product and unity circle, etc.
They also generalized Amy's work to implement parameterized path sums for modeling quantum circuits with arbitrary numbers of qbits.

The Qbricks approach provides formal verification with high automaticity, whereas \ProveIt{} facilitates the generation of a human-readable formal proof based on user-interactive informal proof steps. 
As with Hietala, \textit{et al.}'s \cite{hietala_et_al_proving_quantum_programs:LIPIcs.ITP.2021.21} use of SQIR, the algorithm specification in Qbricks is more difficult to grasp compared to the specification in \ProveIt{}. The Qbricks specification for the exact and general QPE results appears in the middle panel of Fig.~\ref{fig:compare_qpe_best_guarantee_specifications}.
Moreover, the interactive steps in \ProveIt{}, understandably missing from a system such as Qbricks dedicated to high automation, and rather obscure in Coq code, can provide insights, as evidenced by this process uncovering a number of minor errors in the classic Nielsen \& Chuang presentation \cite{Nielsen_Chuang:2010} which we describe in \S\ref{Sec:Discussion_Conclusion}. More generally, we would argue that \ProveIt{} offers advantages in flexibility and relative ease of use over the Qbricks and SQIR/Coq approaches, and compares favorably in terms of time and effort even as it produces a human-readable formal proof largely lacking in the other systems.

\section{Discussion \& Conclusions}
\label{Sec:Discussion_Conclusion} 

Current quantum computers are not powerful enough, in size or fidelity, to test ideas beyond the smallest demonstrations.  As they become more powerful, debugging quantum programs through direct probing will be an increasing challenge due to the complex nature of quantum information: an arbitrary state of $n$ qubits must be described using $2^n$ complex numbers, and probing of circuit states results in probabilistic outcomes. Formal program verification offers an alternative approach to helping ensure correct quantum programs. 

In this paper, we demonstrated a new approach to formal theorem proving applied to the verification of an important quantum algorithm with an exponential speed advantage in comparison to the best known classical algorithm for accomplishing the same feat, quantum phase estimation (QPE).
There is significant value in being able to formally verify quantum algorithms and their implementations in order to guide investments and prevent hard-to-find bugs in quantum programs. 
The Jupyter notebook-based interface and underlying Python code make the \ProveIt{} system accessible to a wide audience, moderately easy to learn and use, and relatively easy to make contributions to its expandable knowledge base of axioms, conjectures, theorems, proofs, and proof tactics.

\ProveIt{}'s approach is unique by presenting mathematical/technical statements, allowing a user to derive new statements in a fashion that mimics informal theorem proving, and finally presenting a full, airtight proof that is human readable and machine check-able. 
In this paper, we demonstrated these features as applied to proving three successively sophisticated high-level statements about the output distribution of the QPE algorithm.
We showed our decomposition of these proofs into several supporting theorems (lemmas) that parallel a textbook proof~\cite{Nielsen_Chuang:2010}.
We then explained the guarantees that \ProveIt{} provides, by tracking dependencies, ensuring there is no circular logic, and certifying each proof with understandable, complete, and machine check-able derivations. 
After describing how our proofs were generated with the use of automation features to fill in obvious steps, making formal theorem proving nearly as straightforward as constructing an informal but rigorous proof, we focused on our proof for one of the QPE-specific ``lemmas'' as a demonstration.
Finally, we made comparisons with QPE proofs in other theorem-proving systems and noted that \ProveIt{} is uniquely versatile in using broadly-understood and convenient mathematical/diagrammatic notation for human-readable theorems and proofs.  We also noted the we compare favorably against a Coq implementation for proving QPE in the number of derivation/tactic commands that were used in our respective proofs.

It is interesting to note that the interactive proving process within \ProveIt{}'s theorem notebooks helped uncover several typographical errors in the original text \cite{Nielsen_Chuang:2010} and eventually led to a better bound on the precision guarantee. For example, in their section \S5.2.1 on ``Performance \& Requirements,'' Nielsen \& Chuang cite the constraint $0 \le \delta \le 2^{-t}$ whereas $0 \le \delta < 2^{-t}$ (with the strict inequality on the right) is actually needed, and they use $-\pi \le 2 \pi (\delta - \ell/2^{t}) \le \pi$ where $-\pi \le 2 \pi (\delta - \ell/2^{t}) < \pi$ (again with the strict inequality on the right) is actually needed.
There is also an absolute value missing on the right-hand side of Nielsen \& Chuang's Eq. 5.29, and the equal sign in their Eq 5.34 should be a less-than.
We were also able to process the summations slightly differently in their Eq. 5.31 to eventually arrive at a better bound on the probability of a measurement ``failure,'' obtaining the tighter $p\big(|m-b|>e\big)< \frac{1}{2e}+\frac{1}{4e^2}$ instead of $\frac{1}{2(1-e)}$ and allowing $e \ge 1$ instead of just $e \ge 2$.

There is much work remaining to be done in developing the \ProveIt{} system to meet its potential as a powerful and valuable theorem proving system.
This current work demonstrates what is possible with this approach.
But, in this instance, it did require a concerted effort by a small team to implement specific capabilities. 
Ultimately, we envision a system able to incorporate features developed by a diverse range of contributors to support the needs of an even broader audience of users for a variety of applications.
With more contributions, the system would become more powerful and garner further contributions in a positive feedback loop.
Before reaching that critical transition, there is more work to be done by a small, dedicated team. 
Tutorials and package ``demonstration'' pages need to be completed so people can learn how to use the system and make contributions, a number of known software issues need to be addressed, and processes need to be improved for handling higher though-puts.

\section{Acknowledgments}
\label{Sec:Acknowledgments}

We gratefully acknowledge Kenny Rudinger and Mohan Sarovar who contributed substantially to an earlier, proof-of-concept demonstration of using \ProveIt{} to verify the QPE algorithm~\cite{ProveItQPE} and for their continued support and encouragement.  We acknowledge Joaquín~E.~Madrid~Larrañaga who contributed to a number of \ProveIt{} features (including the generation of quantum circuit \LaTeX, basic arithmetic, etc.), testing, and documentation.
We also thank Jon~Aytac, Geoffrey~Hulette, Denis~Bueno, and Geoffrey Reedy for valuable discussions and contributions to the development of \ProveIt{}, Kesha~Hietala for generous help in accessing and understanding SQIR and related Coq material, and Christophe~Chareton for generous help in accessing and understanding Qbricks.

This work was supported by the U.S. Department of
Energy, Office of Science, Office of Advanced Scientific Computing Research under the Quantum Computing Applications Team (QCAT) and Quantum Systems Accelerator (QSA) programs, and the Laboratory Directed Research and Development program at Sandia National Laboratories.
Sandia National Laboratories is a multi-program
laboratory managed and operated by National Technology and Engineering Solutions of Sandia, LLC., a wholly owned subsidiary of Honeywell International, Inc., for the U.S. Department of Energy’s National Nuclear Security Administration under contract DE-NA-0003525.
This paper describes objective technical results and analysis.
Any subjective views or opinions that might be expressed in the paper do not necessarily represent the views of the U.S. Department of Energy or the United States Government

\appendix

\section{Structured Proof Example}
\label{appendix:structured_proof}

% The following used for
% continuing enumerated item
% lists with intercollary text
\newcounter{tempCounter}

Writing structured proofs that clearly and concisely communicate concepts and proof strategies to a community of experts is an important goal in its own right which does not require a sophisticated formal theorem proving software system~\cite{Lamport:2012, Lamport:1995_How_to_write_a_proof}.
Ideally, a proof is verified in a sophisticated software system like \ProveIt{} \emph{and is also} presentable for clearly communicating concepts.  \ProveIt{}'s origins drew inspiration from work by one of our authors, Robert~Carr, in developing a system to format structured proofs by hand. His ``proof via dependencies'' format first lists a collection of accepted, relevant mathematical facts in a familiar notation as ``Assumptions.'' Then each proof step begins with a conjunction of explicit assumptions and results from previous proof steps and ends with the keyword ``Thus'' and a conclusion based on that specific collection of assumptions and previous step results.
The following is a demonstration of using this format to prove \texttt{\_psi\_t\_formula} ((\ref{eq:thm_psi_t_formula}) and the focus of \S\ref{sec:theorem_proof_demo}).

\vspace{0.1in}

{\footnotesize
\noindent\textbf{\textit{Theorem:}} \\[0.5ex]
\begin{tabular}{ll}
    If $\forall_{t\in\mathbb{N}^{+}}\psi_{t} =$ &  $\frac{1}{2^{t/2}}\Bigl((\ket{0} + e^{2\pi i 2^{t-1} \varphi }\ket{1}) \otimes (\ket{0} + e^{2\pi i 2^{t-2} \varphi }\ket{1})\otimes$\\
     & \hspace{0.1in}$(\ket{0} + e^{2\pi i 2^{t-2} \varphi }\ket{1})\otimes\cdots\otimes(\ket{0} + e^{2\pi i 2^{0} \varphi }\ket{1})\Bigr)$,
\end{tabular}\\
then $\forall_{t\in\mathbb{N}^{+}} \psi_{t} = \frac{1}{2^{t/2}}\sum_{k=0}^{2^{t} -1}\; e^{2\pi i \varphi k} \ket{k}_{t}$.

\vspace{0.1in}

\noindent\textbf{\textit{Assumptions}}

{\footnotesize

{\raggedright
\begin{enumerate}
\itemindent=-10pt%
\item $\varphi \in [0, 1)$

\item $\forall_{k\in]t[}\; p_k = \frac{1}{\sqrt{2}}(\ket{0} + e^{2\pi i 2^{k} \varphi }\ket{1})$\\ (where $]t[ = \{0, 1, 2, \ldots, t-1\}$)

\item $\psi_1 = p_0$ and $\forall_{k\in[t-1]}\; \psi_{k+1} = p_{k} \otimes \psi_{k}$\\(\textit{e.g.}, $\psi_{t} = p_{t-1} \otimes p_{t-2} \otimes \ldots \otimes p_0$.)

% A14
\item $\forall_{k\in]t[}\;p_k^{\prime} = (\ket{0} + e^{2\pi i 2^{k} \varphi }\ket{1})$

% A15
\item $\psi_1^{\prime} = p_0^{\prime}$, and $\forall_{k\in[t-1]}\; \psi_{k+1}^{\prime} = p_{k}^{\prime} \otimes \psi_{k}^{\prime}$

% A16
\item
$\forall_{t\in\mathbb{N}^{+}}\forall_{k\in ]2^{t}[}\;\ket{0}\otimes \ket{k}_t = \ket{k}_{t+1}$

% A17
\item
$\forall_{t\in\mathbb{N}^{+}}\forall_{k\in ]2^{t}[}\;\ket{1}\otimes \ket{k}_t = \ket{2^{t} + k}_{t+1}$

% A18
\item\label{axiom:linearity_of_tensor_product}
$\forall_{a}\,\forall_b\,\forall_{J}\; \sum_{j\in J}^{} a\otimes b_j = a \otimes \sum_{j\in J}^{} b_j$

% A19
\item $\sum_{k=0}^{k=t} a \ket{k}_{t} = a \sum_{k=0}^{k=t} \ket{k}_{t}$

% A19
\item $\psi_{t-1} = \frac{1}{2^{t/2}}\sum_{k=0}^{2^{t} -1}\; e^{2\pi i \varphi k} \ket{k}_{t}$ iff $\psi_{t-1}^{\prime} = \sum_{k=0}^{2^{t} -1}\; e^{2\pi i \varphi k} \ket{k}_{t}$.
    
  \item Principle of Mathematical Induction:\\If $P(1)$ and $\forall_{t\in\mathbb{N}^+} \left[P(t) \Rightarrow P(t+1)\right]$, then $\forall_{t\in\mathbb{N}^+} P(t)$.
  
  \item $\forall_{t\in\mathbb{N}^+}\Bigl[
  \psi_{t} = \frac{1}{2^{t/2}}\Bigl((\ket{0} + e^{2\pi i 2^{t-1} \varphi }\ket{1}) \otimes$ \\ \hspace{0.1in}$
  (\ket{0} + e^{2\pi i 2^{t-2} \varphi }\ket{1})
  \otimes
  \cdots
  \otimes
  (\ket{0} + e^{2\pi i 2^{0} \varphi }\ket{1})
  \Bigr)
  \Bigr]$.
  
  \item $\forall_{t\in\mathbb{N}^{+}}
  \left[f(t) = \frac{1}{2^{t/2}}\sum_{k=0}^{2^{t} -1}\; e^{2\pi i \varphi k} \ket{k}_{t}
  \right]$.
    
  \item $\forall_{t \in \mathbb{N}^{+}}\left[
  P(t) \Leftrightarrow \psi_{t} = f(t)
  \right]$
  
  \item $\psi_{1} = \frac{1}{2^{1/2}}
  \left(
  \ket{0} +
  e^{2\pi i \varphi} \ket{1}
  \right)$
  
  \item $f(1) = \frac{1}{2^{1/2}}
  \left(
  \ket{0}_{1} +
  e^{2\pi i \varphi} \ket{1}_{1}
  \right)
  $
  
  \item $\forall_{a, b, c\in \mathbb{Z}}\, \forall_{\alpha}\left(
  \sum_{k = a}^{c} \alpha_k = \sum_{k=a}^{b} \alpha_k
  +
  \sum_{k=b+1}^{c }\alpha_k\right)$
  
  \item $\forall_{k\in \{2^{t^{\prime}}, 2^{t^{\prime}}+1, \ldots, 2^{t^{*}}-1\}}$\\$
  \left[e^{2\pi i \varphi k} \ket{k}_{t^{*}}= 
  e^{2\pi i \varphi (k - 2^{t^{\prime}} + 2^{t^{\prime}})} \ket{k - 2^{t^{\prime}} + 2^{t^{\prime}}}_{t^{*}}\right]$
    
\end{enumerate}
} % end raggedright
} % end footnotesize

\noindent \textbf{\textit{Structured Proof (by induction on $t$):}}

\vspace{0.1in}

\noindent\underline{Base Case}.

{\raggedright
\begin{enumerate}[labelindent=5pt, itemindent=0pt, leftmargin=*]
  % \itemindent=-10pt%
  \item $1 \in \mathbb{N}^{+}$ and  $P(1) \Leftrightarrow \psi_{1} = f(1)$\\ and
  $\psi_1 = \frac{1}{2^{1/2}}\left((\ket{0} + e^{2\pi i \varphi }\ket{1})
  \right)$\\
  and
  $f(1) =
  \frac{1}{2^{1/2}}\left((\ket{0}_{1} + e^{2\pi i \varphi }\ket{1}_{1})
  \right)$\\ and
  $\ket{0} = \ket{0}_{1}$ and $\ket{1} = \ket{1}_{1}$.\\[0.05in]
  Thus $P(1)$.
  
  \setcounter{tempCounter}{\value{enumi}}
\end{enumerate}
} % end raggedright

\noindent\underline{Inductive Step}.

{\raggedright

\begin{enumerate}[labelindent=5pt, itemindent=0pt, leftmargin=*]
\setcounter{enumi}{\value{tempCounter}}
  
  % STEP 2
  \item Let $t^{\prime} \in \mathbb{N}^{+}$ such that $P(t^{\prime})$ and $\forall_{t \in \mathbb{N}^{+}}\left[
  P(t) \Leftrightarrow \psi_{t} = f(t)
  \right]$.\\[0.05in]
  Thus, $\psi_{t^{\prime}} = f(t^{\prime})$.
	
  % STEP 3
  \item $t^{*} = t^{\prime} + 1$ 
  and $\forall_{a, b, c\in \mathbb{Z}}\, \forall_{\alpha}\;
  \sum_{k = a}^{c} \alpha_k = \sum_{k=a}^{b} \alpha_k
  +
  \sum_{k=b+1}^{c }\alpha_k$.\\[0.05in]
  Thus $\sum_{k=0}^{2^{t^{*}}\!-1} e^{2\pi i \varphi k} \ket{k}_{t^{*}}$\\
  \hspace{0.2in}$=
  \sum_{k=0}^{2^{t^{\prime}} -1}\; e^{2\pi i \varphi k} \ket{k}_{t^{*}}
  +
  \sum_{k=2^{t^{\prime}}}^{2^{t^{*}} - 1}\; e^{2\pi i \varphi k} \ket{k}_{t^{*}}$
	
  % STEP 4
  \item
  $\forall_{k\in \{2^{t^{\prime}}, 2^{t^{\prime}}+1, \ldots, 2^{t^{*}}-1\}}$\\
  $\left[e^{2\pi i \varphi k} \ket{k}_{t^{*}}= 
  e^{2\pi i \varphi (k - 2^{t^{\prime}} + 2^{t^{\prime}})} \ket{k - 2^{t^{\prime}} + 2^{t^{\prime}}}_{t^{*}}\right]$.\\[0.05in]
  Thus $\sum_{k=2^{t^{\prime}}}^{2^{t^{*}} - 1}\; e^{2\pi i \varphi k} \ket{k}_{t^{*}}=
  \sum_{k=0}^{2^{t^{\prime}} -1}\;e^{2\pi i \varphi (k + 2^{t^{\prime}})} \ket{k + 2^{t^{\prime}}}_{t^{*}}$.
	
  % STEP 5
  \item
  $\sum_{k=0}^{2^{t^{*}} -1}\; e^{2\pi i \varphi k} \ket{k}_{t^{*}}
	=
	\sum_{k=0}^{2^{t^{\prime}} -1}\; e^{2\pi i \varphi k} \ket{k}_{t^{*}}
	+
	\sum_{k=2^{t^{\prime}}}^{2^{t^{*}} - 1}\; e^{2\pi i \varphi k} \ket{k}_{t^{*}}$\\
    and
    $\sum_{k=2^{t^{\prime}}}^{2^{t^{*}} - 1}\; e^{2\pi i \varphi k} \ket{k}_{t^{*}}=
	\sum_{k=0}^{2^{t^{\prime}} -1}\;e^{2\pi i \varphi (k + 2^{t^{\prime}})} \ket{k + 2^{t^{\prime}}}_{t^{*}}
	$.\\[0.05in]
	Thus $\sum_{k=0}^{2^{t^{*}} \!-1} e^{2\pi i \varphi k} \ket{k}_{t^{*}}$\\
     \hspace{0.2in}$= 
	\sum_{k=0}^{2^{t^{\prime}} -1}\; e^{2\pi i \varphi k} \ket{k}_{t^{*}}+ \sum_{k=0}^{2^{t^{\prime}} -1}\;e^{2\pi i \varphi (k + 2^{t^{\prime}})} \ket{k + 2^{t^{\prime}}}_{t^{*}}$.
	
  % STEP 6
  \item $\forall_{a}\,\forall_b\,\forall_{J}\; \sum_{j\in J}^{} a\otimes b_j = a \otimes \sum_{j\in J}^{} b_j$\\ and $\forall_{t\in\mathbb{N}^{+}}\forall_{k\in ]2^{t}[}\;\ket{1}\otimes \ket{k}_t = \ket{2^{t} + k}_{t+1}$.\\[0.05in]
	Thus 
	$ \sum_{k=0}^{2^{t^{\prime}} \!-1}\;e^{2\pi i \varphi (2^{t^{\prime}} + k)} \ket{2^{t^{\prime}} + k}_{t^{*}}$\\\hspace{0.2in}
	$=
	e^{2\pi i \varphi (2^{t^{\prime}})} \sum_{k=0}^{2^{t^{\prime}} -1}\;e^{2\pi i \varphi (k)}\; \ket{1} \otimes \ket{k}_{t^{\prime}}$\\\hspace{0.2in}
	$=
	\ket{1} \otimes \left( e^{2\pi i \varphi (2^{t^{\prime}})} \sum_{k=0}^{2^{t^{\prime}} -1}\;e^{2\pi i \varphi (k)} \ket{k}_{t^{\prime}} \right)
	$
	
    % STEP 7
    \item
    $\forall_{a}\,\forall_b\,\forall_{J}\; \sum_{j\in J}^{} a\otimes b_j = a \otimes \sum_{j\in J}^{} b_j$\\and $\forall_{t\in\mathbb{N}^{+}}\forall_{k\in [2^{t-1}]}\;\ket{0}\otimes \ket{k}_t = \ket{k}_{t+1}$.\\[0.05in]
    Thus
	$\sum_{k=0}^{2^{t^{\prime}} -1}\; e^{2\pi i \varphi k} \ket{k}_{t^{*}}$\\ \hspace{0.2in}$ = \sum_{k=0}^{2^{t^{\prime}} -1}\; e^{2\pi i \varphi k} \ket{0} \otimes \ket{k}_{t^{\prime}}
	=
	\ket{0}\otimes
	\sum_{k=0}^{2^{t^{\prime}} -1}\; e^{2\pi i \varphi k}  \ket{k}_{t^{\prime}}$.
	
    % STEP 8
    \item $\sum_{k=0}^{2^{t^{*}} -1}\; e^{2\pi i \varphi k} \ket{k}_{t^{*}}$\\
    \hspace{0.2in}$ = 
	\sum_{k=0}^{2^{t^{\prime}} -1}\; e^{2\pi i \varphi k} \ket{k}_{t^{*}}+ \sum_{k=0}^{2^{t^{\prime}} -1}\;e^{2\pi i \varphi (k + 2^{t^{\prime}})} \ket{k + 2^{t^{\prime}}}_{t^{*}}$\\
	and $ \sum_{k=0}^{2^{t^{\prime}} -1}\;e^{2\pi i \varphi (2^{t^{\prime}} + k)} \ket{2^{t^{\prime}} + k}_{t^{*}}$\\
    \hspace{0.2in}$
	=
	\ket{1} \otimes e^{2\pi i \varphi (2^{t^{\prime}})} \sum_{k=0}^{2^{t^{\prime}} -1}\;e^{2\pi i \varphi (k)} \ket{k}_{t^{\prime}}
	$\\
	and $\sum_{k=0}^{2^{t^{\prime}} -1}\; e^{2\pi i \varphi k} \ket{k}_{t^{*}} = \ket{0}\otimes
	\sum_{k=0}^{2^{t^{\prime}} -1}\; e^{2\pi i \varphi k}  \ket{k}_{t^{\prime}}$.\\[0.05in]
    Thus $\sum_{k=0}^{2^{t^{*}} -1}\; e^{2\pi i \varphi k} \ket{k}_{t^{*}}$\\
    \hspace{0.2in}$ = 
	\ket{0}\otimes
	\sum_{k=0}^{2^{t^{\prime}} -1}\; e^{2\pi i \varphi k}  \ket{k}_{t^{\prime}} + \ket{1}$\\
    \hspace{0.4in}$ \otimes e^{2\pi i \varphi (2^{t^{\prime}})} \sum_{k=0}^{2^{t^{\prime}} -1}\;e^{2\pi i \varphi (k)} \ket{k}_{t^{\prime}}$\\
    \hspace{0.2in}$= 
	\left(\ket{0} + \ket{1} e^{2\pi i \varphi (2^{t^{\prime}})}\right)
	\otimes
	\sum_{k=0}^{2^{t^{\prime}} -1}\;e^{2\pi i \varphi (k)} \ket{k}_{t^{\prime}}$.
	
    % STEP 9
    \item $\sum_{k=0}^{2^{t^{*}} -1}\; e^{2\pi i \varphi k} \ket{k}_{t^{*}}$\\
    \hspace{0.2in}$= 
	\left(\ket{0} + \ket{1} e^{2\pi i \varphi (2^{t^{\prime}})}\right)
	\otimes
	\sum_{k=0}^{2^{t^{\prime}} -1}\;e^{2\pi i \varphi (k)} \ket{k}_{t^{\prime}}$\\
	and $\forall_{k\in]t[}\; p_k^{\prime} = (\ket{0} + e^{2\pi i 2^{k} \varphi }\ket{1})$\\ and $\psi_{t^{\prime}}^{\prime} = \sum_{k=0}^{2^{t^{\prime}} -1}\; e^{2\pi i \varphi k} \ket{k}_{t^{\prime}}$.
	\\[0.05in]
	Thus
	$\sum_{k=0}^{2^{t^{*}} -1}\; e^{2\pi i \varphi k} \ket{k}_{t^{*}}= 
	p_{t^{\prime}}^{\prime}
	\otimes
	\psi_{t^{\prime}}^{\prime}$
	
    % STEP 10
    \item $\sum_{k=0}^{2^{t^{*}} -1}\; e^{2\pi i \varphi k} \ket{k}_{t^{*}}= 
	p_{t^{\prime}}^{\prime}
	\otimes
	\psi_{t^{\prime}}^{\prime}$ and $\psi_1^{\prime} = p_0^{\prime}$\\[0.05in] and $\forall_{k\in[t-1]}\; \psi_{k+1} = p_{k}^{\prime} \otimes \psi_{k}^{\prime}$.\\[0.05in]
	Thus
	$\sum_{k=0}^{2^{t^{*}} -1}\; e^{2\pi i \varphi k} \ket{k}_{t^{*}}= 
	\psi_{t^{*}}^{\prime}$
	
    % STEP 11
    \item $\sum_{k=0}^{2^{t^{*}} -1}\; e^{2\pi i \varphi k} \ket{k}_{t^{*}}= 
	\psi_{t^{*}}^{\prime}$ and [$\forall_{a, b, \alpha} \;\text{if}\; a = b \;\text{then}\; \alpha a = \alpha b$]\\
	and $a = \sum_{k=0}^{2^{t^{*}} -1}\; e^{2\pi i \varphi k} \ket{k}_{t^{*}} $ and $b = \psi_{t^*}^{\prime}$ and $\alpha = \frac{1}{2^{t^* / 2}}$.\\[0.05in]
	Thus $\frac{1}{2^{t^* / 2}}\psi_{t^{*}}^{\prime} = \frac{1}{2^{t^{*}/2}}\sum_{k=0}^{2^{t^{*}} -1}\; e^{2\pi i \varphi k} \ket{k}_{t^{*}}$
	
    % STEP 12
    \item $\frac{1}{2^{t^* / 2}}\psi_{t^{*}}^{\prime} = \frac{1}{2^{t^{*}/2}}\sum_{k=0}^{2^{t^{*}} -1}\; e^{2\pi i \varphi k} \ket{k}_{t^{*}}$ and $\frac{1}{2^{t^* / 2}}\psi_{t^{*}}^{\prime} = \psi_{t^*}$. \\[0.05in]
	Thus
	$\psi_{t^{*}} = \frac{1}{2^{t^{*}/2}}\sum_{k=0}^{2^{t^{*}} -1}\; e^{2\pi i \varphi k} \ket{k}_{t^{*}}$.
	
    % STEP 13
    \item Assuming $\psi_{t^{\prime}} = \frac{1}{2^{t^{\prime}/2}}\sum_{k=0}^{2^{t^{\prime}} -1}\; e^{2\pi i \varphi k} \ket{k}_{t^{\prime}}$\\
	and $t^{\prime}$ was arbitrary,\\
	we obtain $\psi_{t^{*}} = \frac{1}{2^{t^{*}/2}}\sum_{k=0}^{2^{t^{*}} -1}\; e^{2\pi i \varphi k} \ket{k}_{t^{*}}$.\\[0.05in]
	Thus, $\forall_{t\in\mathbb{N}^+} \left[P(t) \Rightarrow P(t+1)\right]$.
	
    % STEP 13
    \item $P(1)$ and $\forall_{t\in\mathbb{N}^+} \left[P(t) \Rightarrow P(t+1)\right]$\\ and the Principle of Mathematical Induction.\\[0.05in]
	Thus, $\forall_{t\in\mathbb{N}^+}
	P(t)$.
	
\end{enumerate}

} % end raggedright
} % end smaller size 

% ================== %
% PROOF SKETCH TABLE %
% ================== %

\begin{table*}
\scriptsize
\captionsetup{font=footnotesize}
\begin{tabular}{l|l|l|l}
\thead[l]{step}
& \thead[l]{judgment} & \thead[l]{reason(s)} 
& \thead[l]{notebook\\cell}
\\
\hline
\hline
% == STEP   0 == %
0 & {$ \vdash \forall_{t \in \mathbb{N}^+}~\left(\lvert \psi_{t} \rangle = \left(\frac{1}{2^{\frac{t}{2}}} \cdot \left(\sum_{k=0}^{2^{t} - 1} \left(\mathsf{e}^{2 \cdot \pi \cdot \mathsf{i} \cdot \varphi \cdot k} \cdot \lvert k \rangle_{t}\right)\right)\right)\right) $ } & 3, 8, 17\\
\hline
% == STEP   3 == %
3 & {$ \vdash \forall_{P}~\left(\left(P\left(1\right) \land \left[\forall_{m \in \mathbb{N}^+~|~P\left(m\right)}~P\left(m + 1\right)\right]\right) \Rightarrow \left[\forall_{n \in \mathbb{N}^+}~P\left(n\right)\right]\right) $ } & 
\begin{minipage}[t]{1.75cm}{induction\\theorem} \end{minipage} &
4
\\ [0.1in]
\hline
% == STEP   8 == %
8 & {$ \vdash \lvert \psi_{1} \rangle = \frac{1}{\sqrt{2}} \sum_{k=0}^{1}\left(e^{2 \pi i \varphi k}\ket{k}_{1}\right) $ } & 101 & 13\\
\hline
% == STEP 17 == %
17 & {$\begin{array}{c}
{t \in \mathbb{N}^+, \ket{\psi_{t}}  = \left(\frac{1}{2^{t/2}}\left(\sum_{k=0}^{2^{t} - 1} \left(\mathsf{e}^{2 \pi \mathsf{i}  \varphi k} \ket{k}_{t}\right)\right)\right)} \vdash \\
\lvert \psi_{t + 1} \rangle = \left(\frac{1}{2^{\frac{t + 1}{2}}} \left(\sum_{k=0}^{\left(2 \cdot 2^{t}\right) - 1} \left(\mathsf{e}^{2 \pi \mathsf{i} \varphi k} \lvert k \rangle_{t + 1}\right)\right)\right) 
\end{array}$ } & 47, 84, 101 & 40\\
\hline
% == STEP 47 == %
47 & {$
\begin{array}{l}
{t \in \mathbb{N}^+, \ket{\psi_{t}} = \left(\frac{1}{2^{t/2}}
\left(\sum_{k=0}^{2^{t} - 1} \left(\mathsf{e}^{2 \pi \mathsf{i} \varphi k} \ket{k}_{t}\right)\right)\right)} \vdash \\
\begin{array}{l}
\left(\frac{1}{2^{\frac{t + 1}{2}}} \left(\sum_{k=0}^{\left(2 \cdot 2^{t}\right) - 1} \left(\mathsf{e}^{2 \pi \mathsf{i} \varphi k} \ket{k}_{t + 1}\right)\right)\right) = 
\left(\left(\frac{1}{\sqrt{2}} \left(\lvert 0 \rangle + \left(\mathsf{e}^{2 \pi \mathsf{i} \varphi 2^{t}} \lvert 1 \rangle\right)\right)\right) {\otimes} \lvert \psi_{t} \rangle\right) 
\end{array}
\end{array}$ } & 60 & 34\\
\hline
% == STEP 60 == %
60 & {$ {t \in \mathbb{N}^+} \vdash 
\begin{array}{l}
\left(\frac{1}{2^{(t + 1)/2}} \cdot \left(\sum_{k=0}^{\left(2 \cdot 2^{t}\right) - 1} \left(\mathsf{e}^{2 \pi \mathsf{i} \varphi k} \ket{k}_{t + 1}\right)\right)\right) = \\
\left(
\begin{array}{l}
\left(\frac{1}{\sqrt{2}} \left(\lvert 0 \rangle + \left(\mathsf{e}^{2 \pi \mathsf{i} \varphi 2^{t}}  \ket{1}\right)\right)\right) {\otimes}
\left(\frac{1}{2^{t/2}} \cdot \left(\sum_{k=0}^{2^{t} - 1} \left(\mathsf{e}^{2 \pi \mathsf{i} \varphi k} \ket{k}_{t}\right)\right)\right)
\end{array}
\right)
\end{array}$ } & 92 & 33\\
\hline
% == STEP 84 == %
84 & {$ {t \in \mathbb{N}^+} \vdash \ket{\psi_{t + 1}} = \left(
\begin{array}{c}
\left(\frac{1}{\sqrt{2}} \cdot \left(\lvert 0 \rangle + \left(\mathsf{e}^{2 \cdot \pi \cdot \mathsf{i} \cdot 2^{t} \cdot \varphi} \cdot \lvert 1 \rangle\right)\right)\right) {\otimes} \\
\left(\frac{1}{\sqrt{2}} \cdot \left(\lvert 0 \rangle + \left(\mathsf{e}^{2 \cdot \pi \cdot \mathsf{i} \cdot 2^{t - 1} \cdot \varphi} \cdot \lvert 1 \rangle\right)\right)\right) {\otimes} \\
\ldots {\otimes}  \left(\frac{1}{\sqrt{2}} \cdot \left(\lvert 0 \rangle + \left(\mathsf{e}^{2 \cdot \pi \cdot \mathsf{i} \cdot 2^{0} \cdot \varphi} \cdot \lvert 1 \rangle\right)\right)\right)
\end{array}\right) $ } & 101 & 36\\
\hline
% == STEP 92 == %
92 & {$ {t \in \mathbb{N}^+} \vdash 
\begin{array}{c}
\left(\sum_{k=0}^{\left(2 \cdot 2^{t}\right) - 1} \left(\mathsf{e}^{2 \cdot \pi \cdot \mathsf{i} \cdot \varphi \cdot k} \cdot \lvert k \rangle_{t + 1}\right)\right) = \\
\left(
\begin{array}{c}
\left(\lvert 0 \rangle {\otimes} \left(\sum_{k=0}^{2^{t} - 1} \left(\mathsf{e}^{2 \cdot \pi \cdot \mathsf{i} \cdot \varphi \cdot k} \cdot \lvert k \rangle_{t}\right)\right)\right) + \\
\left(\left(\mathsf{e}^{2 \cdot \pi \cdot \mathsf{i} \cdot \varphi \cdot 2^{t}} \cdot \lvert 1 \rangle\right) {\otimes} \left(\sum_{k=0}^{2^{t} - 1} \left(\mathsf{e}^{2 \cdot \pi \cdot \mathsf{i} \cdot \varphi \cdot k} \cdot \lvert k \rangle_{t}\right)\right)\right)
\end{array}\right)
\end{array}$ } & \begin{minipage}[t]{1.75cm}{142, 178, 181,\\218, 350} \end{minipage} & 30\\
\hline
% == STEP 101 == %
101 & {$ \vdash \forall_{t \in \mathbb{N}^+}~\left(\ket{\psi_{t}} = 
\left(
\begin{array}{c}
\left(\frac{1}{\sqrt{2}} \cdot \left(\lvert 0 \rangle + \left(\mathsf{e}^{2 \cdot \pi \cdot \mathsf{i} \cdot 2^{t - 1} \cdot \varphi} \cdot \lvert 1 \rangle\right)\right)\right) {\otimes} \\
\left(\frac{1}{\sqrt{2}} \cdot \left(\lvert 0 \rangle + \left(\mathsf{e}^{2 \cdot \pi \cdot \mathsf{i} \cdot 2^{t - 2} \cdot \varphi} \cdot \lvert 1 \rangle\right)\right)\right) {\otimes} \\ 
\ldots {\otimes}  \left(\frac{1}{\sqrt{2}} \cdot \left(\lvert 0 \rangle + \left(\mathsf{e}^{2 \cdot \pi \cdot \mathsf{i} \cdot 2^{0} \cdot \varphi} \cdot \lvert 1 \rangle\right)\right)\right)
\end{array}\right)\right)$ } & definition & 8\\
\hline
% == STEP 142 == %
142 & {
$ {t \in \mathbb{N}^+} \vdash 
\begin{array}{c}
\left(\sum_{k=0}^{\left(2 \cdot 2^{t}\right) - 1} \left(\mathsf{e}^{2 \cdot \pi \cdot \mathsf{i} \cdot \varphi \cdot k} \cdot \lvert k \rangle_{t + 1} \right) \right) = \\
\left(
\begin{array}{c}
\left(\sum_{k=0}^{2^{t} - 1} \left(\mathsf{e}^{2 \cdot \pi \cdot \mathsf{i} \cdot \varphi \cdot k} \cdot \lvert k \rangle_{t + 1}\right)\right) + \\ \left(\sum_{k=0}^{2^{t} - 1} \left(\mathsf{e}^{2 \cdot \pi \cdot \mathsf{i} \cdot \varphi \cdot \left(k + 2^{t}\right)} \cdot \lvert k + 2^{t} \rangle_{t + 1}\right)\right)
\end{array}\right) 
\end{array}
$ }
& 173, 174 & 18\\
\hline
% == STEP 173 == %
173 & {
$ {t \in \mathbb{N}^+} \vdash 
% \begin{array}{c}
\left(\sum_{k=0}^{\left(2 \cdot 2^{t}\right) - 1} \left(\mathsf{e}^{2 \pi \mathsf{i} \varphi k} \lvert k \rangle_{t + 1}\right)\right) =  %\\
\left(
\begin{array}{c}
\left(\sum_{k=0}^{2^{t} - 1} \left(\mathsf{e}^{2 \pi \mathsf{i} \varphi \cdot k} \lvert k \rangle_{t + 1}\right)\right) + \\ \left(\sum_{k=2^{t}}^{\left(2 \cdot 2^{t}\right) - 1} \left(\mathsf{e}^{2 \pi \mathsf{i} \varphi \cdot k} \lvert k \rangle_{t + 1}\right)\right)
\end{array}
\right) 
% \end{array}
$ } & \begin{tabular}{c}summation\\split\end{tabular} & 17\\
\hline
% == STEP  174 == %
174 & {$ {t \in \mathbb{N}^+} \vdash 
%\begin{array}{c}
\left(\sum_{k=2^{t}}^{\left(2 \cdot 2^{t}\right) - 1} \left(\mathsf{e}^{2 \pi \mathsf{i} \varphi k} \lvert k \rangle_{t + 1}\right)\right) =%\\
\left(\sum_{k=0}^{2^{t} - 1} \left(\mathsf{e}^{2 \pi \mathsf{i} \varphi \left(k + 2^{t}\right)} \lvert k + 2^{t} \rangle_{t + 1}\right)\right) 
%\end{array}
$ } & \begin{tabular}{c}summation\\shift\end{tabular} & 18\\
\hline
% == STEP 178 == %
178 & {$ {t \in \mathbb{N}^+} \vdash 
\begin{array}{c}
\left(\sum_{k=0}^{2^{t} - 1} \left(\left(\mathsf{e}^{2 \cdot \pi \cdot \mathsf{i} \cdot \varphi \cdot k} \cdot \mathsf{e}^{2 \cdot \pi \cdot \mathsf{i} \cdot \varphi \cdot 2^{t}}\right) \cdot \left(\lvert 1 \rangle {\otimes} \lvert k \rangle_{t}\right)\right)\right) \\
= \left(\mathsf{e}^{2 \cdot \pi \cdot \mathsf{i} \cdot \varphi \cdot 2^{t}} \cdot \left(\lvert 1 \rangle {\otimes} \left(\sum_{k=0}^{2^{t} - 1} \left(\mathsf{e}^{2 \cdot \pi \cdot \mathsf{i} \cdot \varphi \cdot k} \cdot \lvert k \rangle_{t}\right)\right)\right)\right) 
\end{array}$ }
& \begin{minipage}[t]{1.75cm}{linearity of\\ tensor product} \end{minipage} & 27\\
\hline
% == STEP 181 == %
181 & {$ {t \in \mathbb{N}^+} \vdash \begin{array}{c} \begin{array}{l} \left(\lvert 0 \rangle {\otimes} \left(\sum_{k=0}^{2^{t} - 1} \left(\mathsf{e}^{2 \pi \mathsf{i} \varphi k} \ket{k}_{t}\right)\right)\right)  = \left(\sum_{k=0}^{2^{t} - 1} \left(\mathsf{e}^{2 \pi \mathsf{i} \varphi k} \left(\lvert 0 \rangle {\otimes} \lvert k \rangle_{t}\right)\right)\right) \end{array} \end{array} $ } & \begin{minipage}[t]{1.75cm}{linearity of\\ tensor product} \end{minipage} & 30\\
\hline
% == STEP 218 == %
218 & {$ {t \in \mathbb{N}^+, k \in \{0~\ldotp \ldotp~2^{t} - 1\}} \vdash \lvert k \rangle_{t + 1} = \left(\lvert 0 \rangle {\otimes} \lvert k \rangle_{t}\right) $ } & ket definition & 29\\
\hline
% == STEP 350 == %
350 & {$ {t \in \mathbb{N}^+, k \in \{0~\ldotp \ldotp~2^{t} - 1\}} \vdash \lvert 2^{t} + k \rangle_{t + 1} = \left(\lvert 1 \rangle {\otimes} \lvert k \rangle_{t}\right) $ } & ket definition & 24
\\
\hline
\end{tabular}
\vspace{0.1in}
\caption{
A condensed, annotated version of \ProveIt{}'s formal proof of \texttt{\_psi\_t\_formula}~(\ref{eq:thm_psi_t_formula}), capturing the most substantive derivation steps.  Each step specifies previous steps on which it depends (listed in the ``reason(s)'' column), except where dependent steps have been omitted (in which case the ``reason'' is articulated in words).  The ``notebook cell'' numbers refer to the corresponding user-interactive inputs shown in Fig.~\ref{fig:example_interactive_steps} leading to the formal proof.}
\label{tab:abstracted_proof_psi_t_formula}
\end{table*}
% ============= %
% END NEW TABLE %
% ============= %

\vspace{0.1in} 

Could we use \ProveIt{}'s output to generate a structured proof format to convey a proof clearly and concisely?
The full formal proof output of \texttt{\_psi\_t\_formula}, for example, contains a large number of details that are trivial and distracting to any experts.
All of the steps of a good structured proof are contained in the formal proof output, but the excessive details should be stripped away.
One idea for doing this would be to present the proof output in an interactively expandable tree structure (much like the way that directories may be navigated in an operating system) and have further options for the user to reorder the steps as desired.
Table~\ref{tab:abstracted_proof_psi_t_formula} illustrates the potential for building a more readily communicable ``proof sketch'' from \ProveIt{}'s output, similar to the ``proof via dependencies'' shown above.

\vspace{0.25in}

%%
%% Bibliography
%%

% \bibliographystyle{apsrev4-2}
\bibliography{bibliography}

\providecommand{\noopsort}[1]{}\providecommand{\singleletter}[1]{#1}%
\begin{thebibliography}{52}%
\makeatletter
\providecommand \@ifxundefined [1]{%
 \@ifx{#1\undefined}
}%
\providecommand \@ifnum [1]{%
 \ifnum #1\expandafter \@firstoftwo
 \else \expandafter \@secondoftwo
 \fi
}%
\providecommand \@ifx [1]{%
 \ifx #1\expandafter \@firstoftwo
 \else \expandafter \@secondoftwo
 \fi
}%
\providecommand \natexlab [1]{#1}%
\providecommand \enquote  [1]{``#1''}%
\providecommand \bibnamefont  [1]{#1}%
\providecommand \bibfnamefont [1]{#1}%
\providecommand \citenamefont [1]{#1}%
\providecommand \href@noop [0]{\@secondoftwo}%
\providecommand \href [0]{\begingroup \@sanitize@url \@href}%
\providecommand \@href[1]{\@@startlink{#1}\@@href}%
\providecommand \@@href[1]{\endgroup#1\@@endlink}%
\providecommand \@sanitize@url [0]{\catcode `\\12\catcode `\$12\catcode
  `\&12\catcode `\#12\catcode `\^12\catcode `\_12\catcode `\%12\relax}%
\providecommand \@@startlink[1]{}%
\providecommand \@@endlink[0]{}%
\providecommand \url  [0]{\begingroup\@sanitize@url \@url }%
\providecommand \@url [1]{\endgroup\@href {#1}{\urlprefix }}%
\providecommand \urlprefix  [0]{URL }%
\providecommand \Eprint [0]{\href }%
\providecommand \doibase [0]{https://doi.org/}%
\providecommand \selectlanguage [0]{\@gobble}%
\providecommand \bibinfo  [0]{\@secondoftwo}%
\providecommand \bibfield  [0]{\@secondoftwo}%
\providecommand \translation [1]{[#1]}%
\providecommand \BibitemOpen [0]{}%
\providecommand \bibitemStop [0]{}%
\providecommand \bibitemNoStop [0]{.\EOS\space}%
\providecommand \EOS [0]{\spacefactor3000\relax}%
\providecommand \BibitemShut  [1]{\csname bibitem#1\endcsname}%
\let\auto@bib@innerbib\@empty
%</preamble>
\bibitem [{\citenamefont {{IBM}}(nd)}]{website:IBM_Quantum}%
  \BibitemOpen
  \bibfield  {author} {\bibinfo {author} {\bibnamefont {{IBM}}},\ }\href@noop
  {} {\bibinfo {title} {{IBM Quantum}}} (\bibinfo {year} {n.d.}),\ \bibinfo
  {note} {\url{https://quantum-computing.ibm.com/}. Last accessed on
  3/04/2023.}\BibitemShut {Stop}%
\bibitem [{\citenamefont {{Rigetti Computing}}(2023)}]{website:Rigetti}%
  \BibitemOpen
  \bibfield  {author} {\bibinfo {author} {\bibnamefont {{Rigetti Computing}}},\
  }\href@noop {} {\bibinfo {title} {{IBM Q}}} (\bibinfo {year} {2023}),\
  \bibinfo {note} {\url{https://www.rigetti.com/}. Last accessed on
  3/04/2023.}\BibitemShut {Stop}%
\bibitem [{\citenamefont {{Oxford Quantum
  Circuits}}(2023)}]{website:OxfordQuantumCircuits}%
  \BibitemOpen
  \bibfield  {author} {\bibinfo {author} {\bibnamefont {{Oxford Quantum
  Circuits}}},\ }\href@noop {} {} (\bibinfo {year} {2023}),\ \bibinfo {note}
  {\url{https://oxfordquantumcircuits.com/technology/}. Last accessed on
  3/04/2023.}\BibitemShut {Stop}%
\bibitem [{\citenamefont {{Quantum
  Inspire}}(2023{\natexlab{a}})}]{website:QuantumInspire_Starmon5}%
  \BibitemOpen
  \bibfield  {author} {\bibinfo {author} {\bibnamefont {{Quantum Inspire}}},\
  }\href@noop {} {\bibinfo {title} {Starmon-5}} (\bibinfo {year}
  {2023}{\natexlab{a}}),\ \bibinfo {note}
  {\url{https://www.quantum-inspire.com/backends/starmon-5/}. Last accessed on
  3/04/2023.}\BibitemShut {Stop}%
\bibitem [{\citenamefont {{IonQ}}(2023)}]{website:IonQ}%
  \BibitemOpen
  \bibfield  {author} {\bibinfo {author} {\bibnamefont {{IonQ}}},\ }\href@noop
  {} {\bibinfo {title} {{Trapped Ion Quantum Computing}}} (\bibinfo {year}
  {2023}),\ \bibinfo {note} {\url{https://ionq.com/}. Last accessed on
  3/04/2023.}\BibitemShut {Stop}%
\bibitem [{\citenamefont {{Sandia National
  Laboratories}}(2023)}]{website:Qscout}%
  \BibitemOpen
  \bibfield  {author} {\bibinfo {author} {\bibnamefont {{Sandia National
  Laboratories}}},\ }\href@noop {} {\bibinfo {title} {{QSCOUT}}} (\bibinfo
  {year} {2023}),\ \bibinfo {note}
  {\url{https://www.sandia.gov/quantum/quantum-information-sciences/projects/qscout/}.
  Last accessed on 3/04/2023.}\BibitemShut {Stop}%
\bibitem [{\citenamefont {{QuEra}}(2022)}]{website:QuEra}%
  \BibitemOpen
  \bibfield  {author} {\bibinfo {author} {\bibnamefont {{QuEra}}},\ }\href@noop
  {} {\bibinfo {title} {{Neutral Atom Platform}}} (\bibinfo {year} {2022}),\
  \bibinfo {note} {\url{https://www.quera.com/neutral-atom-platform}. Last
  accessed on 3/04/2023.}\BibitemShut {Stop}%
\bibitem [{\citenamefont {{Xanadu}}(2023)}]{website:Xanadu}%
  \BibitemOpen
  \bibfield  {author} {\bibinfo {author} {\bibnamefont {{Xanadu}}},\
  }\href@noop {} {\bibinfo {title} {Photonics}} (\bibinfo {year} {2023}),\
  \bibinfo {note} {\url{https://xanadu.ai/photonics/}. Last accessed on
  3/04/2023.}\BibitemShut {Stop}%
\bibitem [{\citenamefont {{Quandela}}(2022)}]{website:Quandela}%
  \BibitemOpen
  \bibfield  {author} {\bibinfo {author} {\bibnamefont {{Quandela}}},\
  }\href@noop {} {\bibinfo {title} {{Quandela Cloud}}} (\bibinfo {year}
  {2022}),\ \bibinfo {note} {\url{https://cloud.quandela.com/}. Last accessed
  on 3/04/2023.}\BibitemShut {Stop}%
\bibitem [{\citenamefont {{Quantum
  Inspire}}(2023{\natexlab{b}})}]{website:QuantumInspire_Spin2}%
  \BibitemOpen
  \bibfield  {author} {\bibinfo {author} {\bibnamefont {{Quantum Inspire}}},\
  }\href@noop {} {\bibinfo {title} {Spin2}} (\bibinfo {year}
  {2023}{\natexlab{b}}),\ \bibinfo {note}
  {\url{https://www.quantum-inspire.com/backends/spin-2/}. Last accessed on
  3/04/2023.}\BibitemShut {Stop}%
\bibitem [{\citenamefont {Shor}(1994)}]{Shor:1994_shor_factoring_algorithm}%
  \BibitemOpen
  \bibfield  {author} {\bibinfo {author} {\bibfnamefont {P.}~\bibnamefont
  {Shor}},\ }\bibfield  {title} {\bibinfo {title} {Algorithms for quantum
  computation: discrete logarithms and factoring},\ }in\ \href
  {https://doi.org/10.1109/SFCS.1994.365700} {\emph {\bibinfo {booktitle}
  {Proceedings 35th Annual Symposium on Foundations of Computer Science}}}\
  (\bibinfo {year} {1994})\ pp.\ \bibinfo {pages} {124--134}\BibitemShut
  {NoStop}%
\bibitem [{\citenamefont {Boneh}\ and\ \citenamefont
  {Lipton}(1995)}]{Boneh1995}%
  \BibitemOpen
  \bibfield  {author} {\bibinfo {author} {\bibfnamefont {D.}~\bibnamefont
  {Boneh}}\ and\ \bibinfo {author} {\bibfnamefont {R.~J.}\ \bibnamefont
  {Lipton}},\ }\bibfield  {title} {\bibinfo {title} {Quantum cryptanalysis of
  hidden linear functions},\ }in\ \href
  {https://doi.org/10.1007/3-540-44750-4_34} {\emph {\bibinfo {booktitle}
  {Advances in Cryptology {\textemdash} {CRYPT}0' 95}}}\ (\bibinfo  {publisher}
  {Springer Berlin Heidelberg},\ \bibinfo {year} {1995})\ pp.\ \bibinfo {pages}
  {424--437}\BibitemShut {NoStop}%
\bibitem [{\citenamefont {Harrow}\ \emph {et~al.}(2009)\citenamefont {Harrow},
  \citenamefont {Hassidim},\ and\ \citenamefont
  {Lloyd}}]{Harrow_Hassidim_Lloyd:2009_quantum_alg_linear_eqs}%
  \BibitemOpen
  \bibfield  {author} {\bibinfo {author} {\bibfnamefont {A.~W.}\ \bibnamefont
  {Harrow}}, \bibinfo {author} {\bibfnamefont {A.}~\bibnamefont {Hassidim}},\
  and\ \bibinfo {author} {\bibfnamefont {S.}~\bibnamefont {Lloyd}},\ }\bibfield
   {title} {\bibinfo {title} {Quantum algorithm for linear systems of
  equations},\ }\href {https://doi.org/10.1103/PhysRevLett.103.150502}
  {\bibfield  {journal} {\bibinfo  {journal} {Phys. Rev. Lett.}\ }\textbf
  {\bibinfo {volume} {103}},\ \bibinfo {pages} {150502} (\bibinfo {year}
  {2009})}\BibitemShut {NoStop}%
\bibitem [{\citenamefont {Motta}\ and\ \citenamefont
  {Rice}(2022)}]{Motta_Rice:2021_quantum_algs_quantum_chem}%
  \BibitemOpen
  \bibfield  {author} {\bibinfo {author} {\bibfnamefont {M.}~\bibnamefont
  {Motta}}\ and\ \bibinfo {author} {\bibfnamefont {J.~E.}\ \bibnamefont
  {Rice}},\ }\bibfield  {title} {\bibinfo {title} {Emerging quantum computing
  algorithms for quantum chemistry},\ }\href
  {https://doi.org/https://doi.org/10.1002/wcms.1580} {\bibfield  {journal}
  {\bibinfo  {journal} {WIREs Computational Molecular Science}\ }\textbf
  {\bibinfo {volume} {12}},\ \bibinfo {pages} {e1580} (\bibinfo {year}
  {2022})},\ \bibinfo {note} {available at
  \url{https://wires.onlinelibrary.wiley.com/doi/pdf/10.1002/wcms.1580}}\BibitemShut
  {NoStop}%
\bibitem [{\citenamefont {Chareton}\ \emph {et~al.}(2022)\citenamefont
  {Chareton}, \citenamefont {Bardin}, \citenamefont {Lee}, \citenamefont
  {Valiron}, \citenamefont {Vilmart},\ and\ \citenamefont
  {Xu}}]{Chareton_et_al:2022_formal_methods_for_quantum_programs}%
  \BibitemOpen
  \bibfield  {author} {\bibinfo {author} {\bibfnamefont {C.}~\bibnamefont
  {Chareton}}, \bibinfo {author} {\bibfnamefont {S.}~\bibnamefont {Bardin}},
  \bibinfo {author} {\bibfnamefont {D.}~\bibnamefont {Lee}}, \bibinfo {author}
  {\bibfnamefont {B.}~\bibnamefont {Valiron}}, \bibinfo {author} {\bibfnamefont
  {R.}~\bibnamefont {Vilmart}},\ and\ \bibinfo {author} {\bibfnamefont
  {Z.}~\bibnamefont {Xu}},\ }\href {https://doi.org/10.48550/ARXIV.2109.06493}
  {\bibinfo {title} {Formal methods for quantum programs: A survey}} (\bibinfo
  {year} {2022}),\ \bibinfo {note} {available at
  \url{https://arxiv.org/abs/2109.06493}}\BibitemShut {NoStop}%
\bibitem [{\citenamefont {Hietala}\ \emph {et~al.}(2019)\citenamefont
  {Hietala}, \citenamefont {Rand}, \citenamefont {Hung}, \citenamefont {Wu},\
  and\ \citenamefont {Hicks}}]{Hietala_et_al_verified_optimization:2019}%
  \BibitemOpen
  \bibfield  {author} {\bibinfo {author} {\bibfnamefont {K.}~\bibnamefont
  {Hietala}}, \bibinfo {author} {\bibfnamefont {R.}~\bibnamefont {Rand}},
  \bibinfo {author} {\bibfnamefont {S.-H.}\ \bibnamefont {Hung}}, \bibinfo
  {author} {\bibfnamefont {X.}~\bibnamefont {Wu}},\ and\ \bibinfo {author}
  {\bibfnamefont {M.}~\bibnamefont {Hicks}},\ }\href@noop {} {\bibinfo {title}
  {Verified optimization in a quantum intermediate representation}} (\bibinfo
  {year} {2019}),\ \Eprint {https://arxiv.org/abs/1904.06319} {arXiv:1904.06319
  [cs.LO]} \BibitemShut {NoStop}%
\bibitem [{\citenamefont {Hietala}\ \emph
  {et~al.}(2021{\natexlab{a}})\citenamefont {Hietala}, \citenamefont {Rand},
  \citenamefont {Hung}, \citenamefont {Wu},\ and\ \citenamefont
  {Hicks}}]{Hietala_et_al_verified_optimizer:2021}%
  \BibitemOpen
  \bibfield  {author} {\bibinfo {author} {\bibfnamefont {K.}~\bibnamefont
  {Hietala}}, \bibinfo {author} {\bibfnamefont {R.}~\bibnamefont {Rand}},
  \bibinfo {author} {\bibfnamefont {S.-H.}\ \bibnamefont {Hung}}, \bibinfo
  {author} {\bibfnamefont {X.}~\bibnamefont {Wu}},\ and\ \bibinfo {author}
  {\bibfnamefont {M.}~\bibnamefont {Hicks}},\ }\bibfield  {title} {\bibinfo
  {title} {A verified optimizer for quantum circuits},\ }\href
  {https://doi.org/10.1145/3434318} {\bibfield  {journal} {\bibinfo  {journal}
  {Proceedings of the {ACM} on Programming Languages}\ }\textbf {\bibinfo
  {volume} {5}},\ \bibinfo {pages} {1} (\bibinfo {year}
  {2021}{\natexlab{a}})}\BibitemShut {NoStop}%
\bibitem [{\citenamefont {Paykin}\ \emph {et~al.}(2017)\citenamefont {Paykin},
  \citenamefont {Rand},\ and\ \citenamefont
  {Zdancewic}}]{Paykin_Rand_Zdancewic_QWIRE:2017}%
  \BibitemOpen
  \bibfield  {author} {\bibinfo {author} {\bibfnamefont {J.}~\bibnamefont
  {Paykin}}, \bibinfo {author} {\bibfnamefont {R.}~\bibnamefont {Rand}},\ and\
  \bibinfo {author} {\bibfnamefont {S.}~\bibnamefont {Zdancewic}},\ }\bibfield
  {title} {\bibinfo {title} {Qwire: A core language for quantum circuits},\
  }in\ \href {https://doi.org/10.1145/3009837.3009894} {\emph {\bibinfo
  {booktitle} {Proceedings of the 44th ACM SIGPLAN Symposium on Principles of
  Programming Languages}}},\ \bibinfo {series and number} {POPL 2017}\
  (\bibinfo  {publisher} {Association for Computing Machinery},\ \bibinfo
  {address} {New York, NY, USA},\ \bibinfo {year} {2017})\ p.\ \bibinfo {pages}
  {846–858}\BibitemShut {NoStop}%
\bibitem [{\citenamefont {Rand}\ \emph
  {et~al.}(2018{\natexlab{a}})\citenamefont {Rand}, \citenamefont {Paykin},\
  and\ \citenamefont
  {Zdancewic}}]{Rand_Paykin_Zdancewic_Phantom_Types_for_Quantum_Programs:2018}%
  \BibitemOpen
  \bibfield  {author} {\bibinfo {author} {\bibfnamefont {R.}~\bibnamefont
  {Rand}}, \bibinfo {author} {\bibfnamefont {J.}~\bibnamefont {Paykin}},\ and\
  \bibinfo {author} {\bibfnamefont {S.}~\bibnamefont {Zdancewic}},\ }\href
  {https://doi.org/10.48550/ARXIV.2204.07112} {\bibinfo {title} {Phantom types
  for quantum programs}} (\bibinfo {year} {2018}{\natexlab{a}}),\ \bibinfo
  {note} {available at \url{https://arxiv.org/abs/2204.07112}}\BibitemShut
  {NoStop}%
\bibitem [{\citenamefont {Rand}\ \emph
  {et~al.}(2018{\natexlab{b}})\citenamefont {Rand}, \citenamefont {Paykin},\
  and\ \citenamefont {Zdancewic}}]{Rand_Paykin_ZDancewic_QWIRE_Practice:2018}%
  \BibitemOpen
  \bibfield  {author} {\bibinfo {author} {\bibfnamefont {R.}~\bibnamefont
  {Rand}}, \bibinfo {author} {\bibfnamefont {J.}~\bibnamefont {Paykin}},\ and\
  \bibinfo {author} {\bibfnamefont {S.}~\bibnamefont {Zdancewic}},\ }\bibfield
  {title} {\bibinfo {title} {{QWIRE} practice: Formal verification of quantum
  circuits in coq},\ }\href {https://doi.org/10.4204/eptcs.266.8} {\bibfield
  {journal} {\bibinfo  {journal} {Electronic Proceedings in Theoretical
  Computer Science}\ }\textbf {\bibinfo {volume} {266}},\ \bibinfo {pages}
  {119} (\bibinfo {year} {2018}{\natexlab{b}})}\BibitemShut {NoStop}%
\bibitem [{\citenamefont {Team}(2022)}]{Coq:2022}%
  \BibitemOpen
  \bibfield  {author} {\bibinfo {author} {\bibfnamefont {T.~C.~D.}\
  \bibnamefont {Team}},\ }\href {https://doi.org/10.5281/zenodo.7313584}
  {\bibinfo {title} {The coq proof assistant}} (\bibinfo {year} {2022}),\
  \bibinfo {note} {available at
  \url{https://zenodo.org/record/7313584\#.Y36kuOzMKu4}}\BibitemShut {NoStop}%
\bibitem [{\citenamefont {{The Coq Proof Assistant}}(nd)}]{website:Coq}%
  \BibitemOpen
  \bibfield  {author} {\bibinfo {author} {\bibnamefont {{The Coq Proof
  Assistant}}},\ }\href@noop {} {} (\bibinfo {year} {n.d.}),\ \bibinfo {note}
  {last accessed on 11/21/2022 at \url{https://coq.inria.fr/}}\BibitemShut
  {NoStop}%
\bibitem [{\citenamefont {Chareton}\ \emph {et~al.}(2021)\citenamefont
  {Chareton}, \citenamefont {Bardin}, \citenamefont {Bobot}, \citenamefont
  {Perrelle},\ and\ \citenamefont
  {Valiron}}]{Chareton_et_al:2021_auto_deductive_verif_framework_Qbricks}%
  \BibitemOpen
  \bibfield  {author} {\bibinfo {author} {\bibfnamefont {C.}~\bibnamefont
  {Chareton}}, \bibinfo {author} {\bibfnamefont {S.}~\bibnamefont {Bardin}},
  \bibinfo {author} {\bibfnamefont {F.}~\bibnamefont {Bobot}}, \bibinfo
  {author} {\bibfnamefont {V.}~\bibnamefont {Perrelle}},\ and\ \bibinfo
  {author} {\bibfnamefont {B.}~\bibnamefont {Valiron}},\ }\bibfield  {title}
  {\bibinfo {title} {An automated deductive verification framework for
  circuit-building quantum programs.},\ }in\ \href@noop {} {\emph {\bibinfo
  {booktitle} {Programming Languages and Systems : 30th European Symposium on
  Programming (ESOP 2021)}}},\ \bibinfo {series} {Lecture Notes in Computer
  Science. 12648}, Vol.\ \bibinfo {volume} {12648}\ (\bibinfo  {publisher}
  {Springer International Publishing},\ \bibinfo {year} {2021})\ pp.\ \bibinfo
  {pages} {12648:148--177},\ \bibinfo {note} {available at
  \url{https://link.springer.com/chapter/10.1007/978-3-030-72019-3_6}}\BibitemShut
  {NoStop}%
\bibitem [{\citenamefont {{Qbricks}}(nd)}]{website:Qbricks}%
  \BibitemOpen
  \bibfield  {author} {\bibinfo {author} {\bibnamefont {{Qbricks}}},\
  }\href@noop {} {\bibinfo {title} {{Qbricks}}} (\bibinfo {year} {n.d.}),\
  \bibinfo {note} {\url{https://qbricks.github.io/}. Last accessed on
  2/01/2023.}\BibitemShut {Stop}%
\bibitem [{\citenamefont {Liu}\ \emph {et~al.}(2016)\citenamefont {Liu},
  \citenamefont {Li}, \citenamefont {Wang}, \citenamefont {Ying},\ and\
  \citenamefont {Zhan}}]{Liu_Theorem_Prover_Hoare_Logic:2016}%
  \BibitemOpen
  \bibfield  {author} {\bibinfo {author} {\bibfnamefont {T.}~\bibnamefont
  {Liu}}, \bibinfo {author} {\bibfnamefont {Y.}~\bibnamefont {Li}}, \bibinfo
  {author} {\bibfnamefont {S.}~\bibnamefont {Wang}}, \bibinfo {author}
  {\bibfnamefont {M.}~\bibnamefont {Ying}},\ and\ \bibinfo {author}
  {\bibfnamefont {N.}~\bibnamefont {Zhan}},\ }\href@noop {} {\bibinfo {title}
  {A theorem prover for quantum {H}oare logic and its applications}} (\bibinfo
  {year} {2016}),\ \Eprint {https://arxiv.org/abs/1601.03835} {arXiv:1601.03835
  [cs.LO]} \BibitemShut {NoStop}%
\bibitem [{\citenamefont {Liu}\ \emph {et~al.}(2019{\natexlab{a}})\citenamefont
  {Liu}, \citenamefont {Zhan}, \citenamefont {Wang}, \citenamefont {Ying},
  \citenamefont {Liu}, \citenamefont {Li}, \citenamefont {Ying},\ and\
  \citenamefont {Zhan}}]{Liu_Formal_Verification_Hoare_Logic:2019}%
  \BibitemOpen
  \bibfield  {author} {\bibinfo {author} {\bibfnamefont {J.}~\bibnamefont
  {Liu}}, \bibinfo {author} {\bibfnamefont {B.}~\bibnamefont {Zhan}}, \bibinfo
  {author} {\bibfnamefont {S.}~\bibnamefont {Wang}}, \bibinfo {author}
  {\bibfnamefont {S.}~\bibnamefont {Ying}}, \bibinfo {author} {\bibfnamefont
  {T.}~\bibnamefont {Liu}}, \bibinfo {author} {\bibfnamefont {Y.}~\bibnamefont
  {Li}}, \bibinfo {author} {\bibfnamefont {M.}~\bibnamefont {Ying}},\ and\
  \bibinfo {author} {\bibfnamefont {N.}~\bibnamefont {Zhan}},\ }\bibfield
  {title} {\bibinfo {title} {Formal verification of quantum algorithms using
  quantum {H}oare logic},\ }in\ \href@noop {} {\emph {\bibinfo {booktitle}
  {Computer Aided Verification}}},\ \bibinfo {editor} {edited by\ \bibinfo
  {editor} {\bibfnamefont {I.}~\bibnamefont {Dillig}}\ and\ \bibinfo {editor}
  {\bibfnamefont {S.}~\bibnamefont {Tasiran}}}\ (\bibinfo  {publisher}
  {Springer International Publishing},\ \bibinfo {year} {2019})\ pp.\ \bibinfo
  {pages} {187--207}\BibitemShut {NoStop}%
\bibitem [{\citenamefont {Liu}\ \emph {et~al.}(2019{\natexlab{b}})\citenamefont
  {Liu}, \citenamefont {Zhan}, \citenamefont {Wang}, \citenamefont {Ying},
  \citenamefont {Liu}, \citenamefont {Li}, \citenamefont {Ying},\ and\
  \citenamefont {Zhan}}]{Liu_et_al:2019_quantum_hoare_logic_proof_document}%
  \BibitemOpen
  \bibfield  {author} {\bibinfo {author} {\bibfnamefont {J.}~\bibnamefont
  {Liu}}, \bibinfo {author} {\bibfnamefont {B.}~\bibnamefont {Zhan}}, \bibinfo
  {author} {\bibfnamefont {S.}~\bibnamefont {Wang}}, \bibinfo {author}
  {\bibfnamefont {S.}~\bibnamefont {Ying}}, \bibinfo {author} {\bibfnamefont
  {T.}~\bibnamefont {Liu}}, \bibinfo {author} {\bibfnamefont {Y.}~\bibnamefont
  {Li}}, \bibinfo {author} {\bibfnamefont {M.}~\bibnamefont {Ying}},\ and\
  \bibinfo {author} {\bibfnamefont {N.}~\bibnamefont {Zhan}},\ }\bibfield
  {title} {\bibinfo {title} {Quantum {H}oare logic},\ }\href@noop {} {\bibfield
   {journal} {\bibinfo  {journal} {Archive of Formal Proofs}\ } (\bibinfo
  {year} {2019}{\natexlab{b}})},\ \bibinfo {note}
  {\url{https://isa-afp.org/entries/QHLProver.html}, Formal proof
  development}\BibitemShut {NoStop}%
\bibitem [{\citenamefont
  {Ying}(2011)}]{Ying:2011_Floyd_Hoare_logic_for_quantum_programs}%
  \BibitemOpen
  \bibfield  {author} {\bibinfo {author} {\bibfnamefont {M.}~\bibnamefont
  {Ying}},\ }\bibfield  {title} {\bibinfo {title} {Floyd-hoare logic for
  quantum programs.},\ }\href
  {https://libproxy.unm.edu/login?url=https://search.ebscohost.com/login.aspx?direct=true&db=edsgao&AN=edsgcl.278595644&site=eds-live&scope=site}
  {\bibfield  {journal} {\bibinfo  {journal} {ACM Transactions on Programming
  Languages \& Systems}\ }\textbf {\bibinfo {volume} {33}},\ \bibinfo {pages}
  {19} (\bibinfo {year} {2011})}\BibitemShut {NoStop}%
\bibitem [{\citenamefont {Ying}(2018)}]{Ying_Auto_Verif_Quantum_Programs:2019}%
  \BibitemOpen
  \bibfield  {author} {\bibinfo {author} {\bibfnamefont {M.}~\bibnamefont
  {Ying}},\ }\bibfield  {title} {\bibinfo {title} {Toward automatic
  verification of quantum programs},\ }\href
  {https://doi.org/10.1007/s00165-018-0465-3} {\bibfield  {journal} {\bibinfo
  {journal} {Formal Aspects of Computing}\ }\textbf {\bibinfo {volume} {31}},\
  \bibinfo {pages} {3} (\bibinfo {year} {2018})}\BibitemShut {NoStop}%
\bibitem [{\citenamefont {Nielsen}\ and\ \citenamefont
  {Chuang}(2010)}]{Nielsen_Chuang:2010}%
  \BibitemOpen
  \bibfield  {author} {\bibinfo {author} {\bibfnamefont {M.~A.}\ \bibnamefont
  {Nielsen}}\ and\ \bibinfo {author} {\bibfnamefont {I.~L.}\ \bibnamefont
  {Chuang}},\ }\href@noop {} {\emph {\bibinfo {title} {Quantum Computation and
  Quantum Information (10th Anniversary Edition)}}}\ (\bibinfo  {publisher}
  {Cambridge University Press},\ \bibinfo {address} {New York, NY},\ \bibinfo
  {year} {2010})\BibitemShut {NoStop}%
\bibitem [{\citenamefont {Witzel}\ \emph {et~al.}(2020)\citenamefont {Witzel},
  \citenamefont {Craft}, \citenamefont {Carr},\ and\ \citenamefont
  {Larrañaga}}]{Preprint:ProveIt}%
  \BibitemOpen
  \bibfield  {author} {\bibinfo {author} {\bibfnamefont {W.~M.}\ \bibnamefont
  {Witzel}}, \bibinfo {author} {\bibfnamefont {W.~D.}\ \bibnamefont {Craft}},
  \bibinfo {author} {\bibfnamefont {R.~D.}\ \bibnamefont {Carr}},\ and\
  \bibinfo {author} {\bibfnamefont {J.~E.~M.}\ \bibnamefont {Larrañaga}},\
  }\href@noop {} {\bibinfo {title} {Prove-it: A proof assistant for organizing
  and verifying general mathematical knowledge}} (\bibinfo {year} {2020}),\
  \Eprint {https://arxiv.org/abs/2012.10987} {arXiv:2012.10987 [cs.LO]}
  \BibitemShut {NoStop}%
\bibitem [{\citenamefont {{Prove-It}}(2023)}]{website:Prove-It}%
  \BibitemOpen
  \bibfield  {author} {\bibinfo {author} {\bibnamefont {{Prove-It}}},\
  }\href@noop {} {\bibinfo {title} {Welcome to the expanding {P}rove-{I}t
  library of proofs}} (\bibinfo {year} {2023}),\ \bibinfo {note}
  {\url{www.pyproveit.org}}\BibitemShut {NoStop}%
\bibitem [{\citenamefont {Cleve}\ \emph {et~al.}(1998)\citenamefont {Cleve},
  \citenamefont {Ekert}, \citenamefont {Macchiavello},\ and\ \citenamefont
  {Mosca}}]{Cleve_et_al:1998_quantum_algs_revisited}%
  \BibitemOpen
  \bibfield  {author} {\bibinfo {author} {\bibfnamefont {R.}~\bibnamefont
  {Cleve}}, \bibinfo {author} {\bibfnamefont {A.}~\bibnamefont {Ekert}},
  \bibinfo {author} {\bibfnamefont {C.}~\bibnamefont {Macchiavello}},\ and\
  \bibinfo {author} {\bibfnamefont {M.}~\bibnamefont {Mosca}},\ }\bibfield
  {title} {\bibinfo {title} {Quantum algorithms revisited},\ }\href@noop {}
  {\bibfield  {journal} {\bibinfo  {journal} {Proceedings: Mathematical,
  Physical and Engineering Sciences}\ }\textbf {\bibinfo {volume} {454}},\
  \bibinfo {pages} {339} (\bibinfo {year} {1998})},\ \bibinfo {note} {stable
  URL: https://www.jstor.org/stable/53169}\BibitemShut {NoStop}%
\bibitem [{\citenamefont
  {Mermin}(2007)}]{Mermin_Quantum_Computer_Science:2007}%
  \BibitemOpen
  \bibfield  {author} {\bibinfo {author} {\bibfnamefont {N.~D.}\ \bibnamefont
  {Mermin}},\ }\href
  {https://libproxy.unm.edu/login?url=https://search.ebscohost.com/login.aspx?direct=true&db=cat05987a&AN=unm.137221653&site=eds-live&scope=site}
  {\emph {\bibinfo {title} {Quantum computer science : an introduction.}}}\
  (\bibinfo  {publisher} {Cambridge University Press},\ \bibinfo {year}
  {2007})\BibitemShut {NoStop}%
\bibitem [{\citenamefont {Benenti}\ \emph {et~al.}(2004)\citenamefont
  {Benenti}, \citenamefont {Casati},\ and\ \citenamefont
  {Strini}}]{benenti:2004_principles_QCI}%
  \BibitemOpen
  \bibfield  {author} {\bibinfo {author} {\bibfnamefont {G.}~\bibnamefont
  {Benenti}}, \bibinfo {author} {\bibfnamefont {G.}~\bibnamefont {Casati}},\
  and\ \bibinfo {author} {\bibfnamefont {G.}~\bibnamefont {Strini}},\ }\href
  {https://books.google.com/books?id=5gmH-DLuIlMC} {\emph {\bibinfo {title}
  {Principles of Quantum Computation and Information}}},\ \bibinfo {series}
  {Principles of Quantum Computation and Information}\ No.\ \bibinfo {number}
  {v. 1}\ (\bibinfo  {publisher} {World Scientific},\ \bibinfo {year}
  {2004})\BibitemShut {NoStop}%
\bibitem [{\citenamefont {Hietala}\ \emph
  {et~al.}(2021{\natexlab{b}})\citenamefont {Hietala}, \citenamefont {Rand},
  \citenamefont {Hung}, \citenamefont {Li},\ and\ \citenamefont
  {Hicks}}]{hietala_et_al_proving_quantum_programs:LIPIcs.ITP.2021.21}%
  \BibitemOpen
  \bibfield  {author} {\bibinfo {author} {\bibfnamefont {K.}~\bibnamefont
  {Hietala}}, \bibinfo {author} {\bibfnamefont {R.}~\bibnamefont {Rand}},
  \bibinfo {author} {\bibfnamefont {S.-H.}\ \bibnamefont {Hung}}, \bibinfo
  {author} {\bibfnamefont {L.}~\bibnamefont {Li}},\ and\ \bibinfo {author}
  {\bibfnamefont {M.}~\bibnamefont {Hicks}},\ }\bibfield  {title} {\bibinfo
  {title} {{Proving Quantum Programs Correct}},\ }in\ \href
  {https://doi.org/10.4230/LIPIcs.ITP.2021.21} {\emph {\bibinfo {booktitle}
  {12th International Conference on Interactive Theorem Proving (ITP 2021)}}},\
  \bibinfo {series} {Leibniz International Proceedings in Informatics
  (LIPIcs)}, Vol.\ \bibinfo {volume} {193},\ \bibinfo {editor} {edited by\
  \bibinfo {editor} {\bibfnamefont {L.}~\bibnamefont {Cohen}}\ and\ \bibinfo
  {editor} {\bibfnamefont {C.}~\bibnamefont {Kaliszyk}}}\ (\bibinfo
  {publisher} {Schloss Dagstuhl -- Leibniz-Zentrum f{\"u}r Informatik},\
  \bibinfo {address} {Dagstuhl, Germany},\ \bibinfo {year} {2021})\ pp.\
  \bibinfo {pages} {21:1--21:19}\BibitemShut {NoStop}%
\bibitem [{\citenamefont {Hodel}(2020)}]{Hoedel:1995_Intro_Math_Logic}%
  \BibitemOpen
  \bibfield  {author} {\bibinfo {author} {\bibfnamefont {R.~E.}\ \bibnamefont
  {Hodel}},\ }\href@noop {} {\emph {\bibinfo {title} {An Introduction to
  Mathematical Logic}}}\ (\bibinfo  {publisher} {Dover Publications, Inc.},\
  \bibinfo {address} {Mineola, NY},\ \bibinfo {year} {1995/2020})\BibitemShut
  {NoStop}%
\bibitem [{\citenamefont {Mendelson}(2015)}]{Mendelson:2015}%
  \BibitemOpen
  \bibfield  {author} {\bibinfo {author} {\bibfnamefont {E.}~\bibnamefont
  {Mendelson}},\ }\href@noop {} {\emph {\bibinfo {title} {Introduction to
  Mathematical Logic (6th Ed.)}}},\ \bibinfo {edition} {6th}\ ed.,\ Textbooks
  in Mathematics\ (\bibinfo  {publisher} {CRC Press/Taylor \& Francis Group},\
  \bibinfo {address} {Boca Raton, Fla.},\ \bibinfo {year} {2015})\BibitemShut
  {NoStop}%
\bibitem [{\citenamefont {Lamport}(1995)}]{Lamport:1995_How_to_write_a_proof}%
  \BibitemOpen
  \bibfield  {author} {\bibinfo {author} {\bibfnamefont {L.}~\bibnamefont
  {Lamport}},\ }\bibfield  {title} {\bibinfo {title} {How to write a proof},\
  }\href@noop {} {\bibfield  {journal} {\bibinfo  {journal} {American
  Mathematical Monthly}\ }\textbf {\bibinfo {volume} {102}},\ \bibinfo {pages}
  {600} (\bibinfo {year} {August-September 1995})},\ \bibinfo {note} {available
  at
  \url{http://lamport.azurewebsites.net/pubs/lamport-how-to-write.pdf}}\BibitemShut
  {NoStop}%
\bibitem [{\citenamefont {Lamport}(2012)}]{Lamport:2012}%
  \BibitemOpen
  \bibfield  {author} {\bibinfo {author} {\bibfnamefont {L.}~\bibnamefont
  {Lamport}},\ }\bibfield  {title} {\bibinfo {title} {How to write a 21st
  century proof},\ }\href {https://doi.org/10.1007/s11784-012-0071-6}
  {\bibfield  {journal} {\bibinfo  {journal} {Journal of Fixed Point Theory and
  Applications}\ }\textbf {\bibinfo {volume} {11}},\ \bibinfo {pages} {43}
  (\bibinfo {year} {2012})},\ \bibinfo {note} {available at
  \url{http://lamport.azurewebsites.net/pubs/proof.pdf}}\BibitemShut {NoStop}%
\bibitem [{\citenamefont {Barendregt}\ and\ \citenamefont
  {Barendsen}(2002)}]{Barendregt_Barendsen:2002_Autarkic_Computations_in_Formal_Proofs}%
  \BibitemOpen
  \bibfield  {author} {\bibinfo {author} {\bibfnamefont {H.}~\bibnamefont
  {Barendregt}}\ and\ \bibinfo {author} {\bibfnamefont {E.}~\bibnamefont
  {Barendsen}},\ }\bibfield  {title} {\bibinfo {title} {Autarkic computations
  in formal proofs},\ }\href {https://doi.org/10.1023/A:1015761529444}
  {\bibfield  {journal} {\bibinfo  {journal} {Journal of Automated Reasoning}\
  }\textbf {\bibinfo {volume} {28}},\ \bibinfo {pages} {321} (\bibinfo {year}
  {2002})}\BibitemShut {NoStop}%
\bibitem [{\citenamefont {Barendregt}\ and\ \citenamefont
  {Freek}(2002)}]{Barendregt_Wiedijk:2005_challenge_of_computer_mathematics}%
  \BibitemOpen
  \bibfield  {author} {\bibinfo {author} {\bibfnamefont {H.}~\bibnamefont
  {Barendregt}}\ and\ \bibinfo {author} {\bibfnamefont {W.}~\bibnamefont
  {Freek}},\ }\bibfield  {title} {\bibinfo {title} {The challenge of computer
  mathematics},\ }\href {https://doi.org/10.1098/rsta.2005.1650} {\bibfield
  {journal} {\bibinfo  {journal} {Philosophical Transactions of the Royal
  Society A}\ }\textbf {\bibinfo {volume} {363}},\ \bibinfo {pages} {2351}
  (\bibinfo {year} {2002})}\BibitemShut {NoStop}%
\bibitem [{\citenamefont {Shoenfield}(1967)}]{Schoenfield_Math_Logic:1967}%
  \BibitemOpen
  \bibfield  {author} {\bibinfo {author} {\bibfnamefont {J.~R.}\ \bibnamefont
  {Shoenfield}},\ }\href@noop {} {\emph {\bibinfo {title} {Mathematical
  Logic}}},\ Addison-Wesley Series in Logic\ (\bibinfo  {publisher}
  {Addison-Wesley Publishing Co.},\ \bibinfo {address} {Reading, MA},\ \bibinfo
  {year} {1967})\BibitemShut {NoStop}%
\bibitem [{\citenamefont {Eastin}\ and\ \citenamefont
  {Flammia}(2004)}]{Eastin_Fammia:2004_Q_Circuit}%
  \BibitemOpen
  \bibfield  {author} {\bibinfo {author} {\bibfnamefont {B.}~\bibnamefont
  {Eastin}}\ and\ \bibinfo {author} {\bibfnamefont {S.~T.}\ \bibnamefont
  {Flammia}},\ }\href {https://doi.org/10.48550/ARXIV.QUANT-PH/0406003}
  {\bibinfo {title} {Q-circuit tutorial}} (\bibinfo {year} {2004}),\ \bibinfo
  {note} {available at
  \url{https://arxiv.org/abs/quant-ph/0406003}}\BibitemShut {NoStop}%
\bibitem [{\citenamefont {de~Bruijn}(1980)}]{deBruijn:1980_automath}%
  \BibitemOpen
  \bibfield  {author} {\bibinfo {author} {\bibfnamefont {N.~G.}\ \bibnamefont
  {de~Bruijn}},\ }\bibfield  {title} {\bibinfo {title} {A survey of the project
  automath},\ }in\ \href@noop {} {\emph {\bibinfo {booktitle} {To H.B. Curry:
  Essays on Combinatory Logic, Lambda Calculus and Formalism}}},\ \bibinfo
  {editor} {edited by\ \bibinfo {editor} {\bibfnamefont {J.}~\bibnamefont
  {Seldin}}\ and\ \bibinfo {editor} {\bibfnamefont {J.}~\bibnamefont
  {Hindley}}}\ (\bibinfo  {publisher} {Academic Press},\ \bibinfo {year}
  {1980})\ pp.\ \bibinfo {pages} {579--606},\ \bibinfo {note} {available at
  \url{https://pure.tue.nl/ws/files/2092478/597605.pdf}}\BibitemShut {NoStop}%
\bibitem [{\citenamefont {Wiedijk}(nd)}]{Wiedijk:nd_De_Bruijn_Factor}%
  \BibitemOpen
  \bibfield  {author} {\bibinfo {author} {\bibfnamefont {F.}~\bibnamefont
  {Wiedijk}},\ }\href@noop {} {\bibinfo {title} {The {D}e {B}ruijn factor}}
  (\bibinfo {year} {n.d.}),\ \bibinfo {note} {available at
  \url{https://www.cs.ru.nl/~freek/factor/factor.pdf}}\BibitemShut {NoStop}%
\bibitem [{\citenamefont {Bauer-Marquart}\ \emph {et~al.}(2022)\citenamefont
  {Bauer-Marquart}, \citenamefont {Leue},\ and\ \citenamefont
  {Schilling}}]{Bauer-Marquart_et_al:2022_Symbolic_Verification_Quantum_Programs}%
  \BibitemOpen
  \bibfield  {author} {\bibinfo {author} {\bibfnamefont {F.}~\bibnamefont
  {Bauer-Marquart}}, \bibinfo {author} {\bibfnamefont {S.}~\bibnamefont
  {Leue}},\ and\ \bibinfo {author} {\bibfnamefont {C.}~\bibnamefont
  {Schilling}},\ }\href {https://doi.org/10.48550/ARXIV.2212.02267} {\bibinfo
  {title} {symqv: Automated symbolic verification of quantum programs}}
  (\bibinfo {year} {2022}),\ \bibinfo {note} {available at
  \url{https://arxiv.org/abs/2212.02267}}\BibitemShut {NoStop}%
\bibitem [{\citenamefont
  {Amy}(2019)}]{Amy_PhD_Thesis:2019_Quantum_Circuit_Design}%
  \BibitemOpen
  \bibfield  {author} {\bibinfo {author} {\bibfnamefont {M.}~\bibnamefont
  {Amy}},\ }\emph {\bibinfo {title} {Formal Methods in Quantum Circuit
  Design}},\ \href@noop {} {Ph.D. thesis},\ \bibinfo  {school} {University of
  Waterloo} (\bibinfo {year} {2019}),\ \bibinfo {note} {available at
  \url{https://uwspace.uwaterloo.ca/bitstream/handle/10012/14480/Amy_Matthew.pdf}}\BibitemShut
  {NoStop}%
\bibitem [{\citenamefont {Lattner}\ and\ \citenamefont
  {Adve}(2004)}]{Lattner_Vikram:2004_LLVM}%
  \BibitemOpen
  \bibfield  {author} {\bibinfo {author} {\bibfnamefont {C.}~\bibnamefont
  {Lattner}}\ and\ \bibinfo {author} {\bibfnamefont {V.}~\bibnamefont {Adve}},\
  }\bibfield  {title} {\bibinfo {title} {{LLVM: A Compilation Framework for
  Lifelong Program Analysis \& Transformation}},\ }in\ \href@noop {} {\emph
  {\bibinfo {booktitle} {{Proceedings of the 2004 International Symposium on
  Code Generation and Optimization (CGO'04)}}}}\ (\bibinfo {address} {Palo
  Alto, California},\ \bibinfo {year} {2004})\BibitemShut {NoStop}%
\bibitem [{\citenamefont {{LLVM}}(nd)}]{website:LLVM}%
  \BibitemOpen
  \bibfield  {author} {\bibinfo {author} {\bibnamefont {{LLVM}}},\ }\href@noop
  {} {\bibinfo {title} {The {LLVM} {C}ompiler {I}nfrastructure project}}
  (\bibinfo {year} {n.d.}),\ \bibinfo {note} {\url{llvm.org}}\BibitemShut
  {NoStop}%
\bibitem [{\citenamefont {{Why3}}(2023)}]{website:Why3}%
  \BibitemOpen
  \bibfield  {author} {\bibinfo {author} {\bibnamefont {{Why3}}},\ }\href@noop
  {} {} (\bibinfo {year} {2023}),\ \bibinfo {note} {last accessed on 2/01/2023
  at \url{https://why3.lri.fr/}}\BibitemShut {NoStop}%
\bibitem [{\citenamefont {Witzel}\ \emph {et~al.}(2015)\citenamefont {Witzel},
  \citenamefont {Sarovar},\ and\ \citenamefont {Rudinger}}]{ProveItQPE}%
  \BibitemOpen
  \bibfield  {author} {\bibinfo {author} {\bibfnamefont {W.}~\bibnamefont
  {Witzel}}, \bibinfo {author} {\bibfnamefont {M.}~\bibnamefont {Sarovar}},\
  and\ \bibinfo {author} {\bibfnamefont {K.}~\bibnamefont {Rudinger}},\
  }\href@noop {} {\bibinfo {title} {{Versatile Formal Methods Applied to
  Quantum Information}}} (\bibinfo {year} {2015}),\ \bibinfo {note}
  {\url{prod.sandia.gov/techlib/access-control.cgi/2015/159617r.pdf}}\BibitemShut
  {NoStop}%
\end{thebibliography}%

\end{document}